\documentclass[aps,prd,amsmath,floats,floatfix,twocolumn,nofootinbib,superscriptaddress,showpacs]{revtex4-1}
\usepackage{graphicx}
\usepackage{multirow}
\usepackage{color}
\usepackage{latexsym}
\usepackage{amsfonts}
\usepackage{amsmath}
\usepackage{mathrsfs}
\usepackage[colorlinks]{hyperref}
\hypersetup{
  colorlinks=true,        
  linkcolor=blue,         
  citecolor=cyan,         
}
\graphicspath{{figures/}} 

\newcommand{\be}{\begin{eqnarray}}
\newcommand{\ee}{\end{eqnarray}}

\begin{document}
	
\title{Observing the eye of the storm I: testing regular black holes with LVK and EHT observations}

\author{Carlos~A.~Benavides-Gallego}
\email[corresponding author: ]{ cabenavidesg@sjtu.edu.cn}
\affiliation{School of Aeronautics and Astronautics, Shanghai Jiao Tong University, Shanghai 200240, PRC.}
\affiliation{School of Physics and Astronomy, Shanghai Jiao Tong University, 800 Dongchuan Road, Minhang, Shanghai 200240, PRC.}
\affiliation{Shanghai Frontiers Science Center of Gravitational Wave Detection, 800 Dongchuan Road, Minhang, Shanghai 200240, People's Republic of China}

\author{Swarnim Shashank}
\email{swarnim@fudan.edu.cn}
\affiliation{Center for Astronomy and Astrophysics, Center for Field Theory and Particle Physics and Department of Physics, Fudan University, 200438 Shanghai, China}

\author{Haiguang Xu}
\email{hrm@sjtu.edu.cn}
\affiliation{School of Physics and Astronomy, Shanghai Jiao Tong University, 800 Dongchuan Road, Minhang, Shanghai 200240, PRC.}
\affiliation{Shanghai Frontiers Science Center of Gravitational Wave Detection, 800 Dongchuan Road, Minhang, Shanghai 200240, People's Republic of China}

\date{\today}

\begin{abstract}
According to the celebrated singularity theorems, space-time singularities in general relativity are inevitable. However, it is generally believed that singularities do not exist in nature, and their existence suggests the necessity of a new theory of gravity. In this paper, we investigated a regular astrophysically viable space-time (regular in the sense that it is singularity-free) from the observational point of view using observations from the LIGO, Virgo, and KAGRA (LVK), and the event horizon telescope (EHT) collaborations. This black hole solution depends on a free parameter $\ell$ in addition to the mass, $M$, and the spin, $a$, violating, in this way, the non-hair theorem/conjecture. In the case of gravitational wave observations, we use the catalogs GWTC-1, 2, and 3 to constrain the free parameter. In the case of the EHT, we use the values of the angular diameter reported for SgrA* and M87*. We also investigated the photon ring structure by considering scenarios such as static spherical accretion, infalling spherical accretion, and thin accretion disk. Our results show that the EHT observations constrain the free parameter $\ell$ to the intervals $0\leq \ell \leq 0.148$ and $0\leq \ell \leq 0.212$ obtained for SgrA* and M87*, respectively. On the other hand, GW observations constrain the free parameter with values that satisfy the theoretical limit, particularly those events for which $\ell<<1$. Our results show that the most stringent constraints on $\ell$ correspond to the events GW191204-171526 ($\ell=0.041^{+0.106}_{-0.041}$) and GW190924-021846 ($\ell=0.050^{+0.165}_{-0.050}$) for the SEOB model.
\end{abstract}
	
\maketitle
\section{Introduction}
      
      	The theorems demonstrated by Penrose and Hawking show that space-time singularities in general relativity (GR) are an inherent consequence of the theory itself~\cite{Hawking:1973uf}. One of the earliest examples is the well-known Schwarzschild black hole (BH) solution, which contains a singularity within the event horizon~\cite{Schwarzschild:1916uq}. Nevertheless, the scientific community maintains that singularities do not exist in nature. In a recent paper, for example, Roy Kerr claims that there is no proof that black holes (BH) contain singularities when generated by real physical bodies~\cite{Kerr:2023rpn}. For him, ``\textit{gravitational clumping leads inevitably to black holes in our universe, confirming what is observed, but this does not lead to singularities},'' criticizing, in this way, the work of Penrose and Hawking. The fact that Einstein's theory predicts singularities implies the existence of a region in space-time where the laws of physics break down, revealing the necessity of a new theory of gravity, e.g., a quantum theory of gravity.
	
	From the perspective of quantum mechanics, it is possible to circumvent the issue of space-time singularities by assuming a de Sitter core at the center of the space-time. This idea was initially proposed by Sakharov and Gliner~\cite{Sakharov:1966aja, Gliner:1966aja} and later utilized by Bardeen to develop the first regular black hole (RBH) solution during the 60s~\cite{Bardeen:1968aja}. In contrast to classical BHs, RBH solutions are free from singularities, and their metrics and curvature invariants remain finite throughout space-time. In this sense, there is no infinite increase in the space-time curvature during a collapse if the quantum fluctuations dominate the process, imposing a limit on the curvature and leading to the formation of the central core. Since Bardeen's proposal, several RBH solutions have been proposed~\cite{Borde:1994ai, Barrabes:1995nk, Bogojevic:1998ma, Cabo:1997rm, Hayward:2005gi, Bambi:2013ufa, Ghosh:2014pba, Ghosh:2014hea, Toshmatov:2014nya, Azreg-Ainou:2014pra, Dymnikova:2015hka, Simpson:2021dyo, Simpson:2021zfl}. Moreover, thanks to the work of Ayón-Beato y Gracía, we know that the physical source of the RHBs could be a nonlinear electromagnetic field produced by a nonlinear electric field or a magnetic monopole~\cite{Ayon-Beato:1998hmi, Ayon-Beato:1999qin, Ayon-Beato:1999kuh, Ayon-Beato:2000mjt, Ayon-Beato:2004ywd}. Hence, while a specific quantum theory of gravity remains elusive, the prospect of RBH solutions offers a compelling avenue for investigating phenomena analogous to those in classical BH theory.
 
      On the other hand, it is well-known that BHs predicted by GR are characterized uniquely by the mass, the angular momentum, and the electric charge. This is the famous no-hair conjecture~\cite{Israel:1967wq, Israel:1967za, Carter:1971zc, Hawking:1972qk, Robinson:1975bv}. The conjecture considers the existence of an event horizon that encloses the singularity~\cite{Penrose:1969pc} and the absence of closed time-like loops in the exterior domain of the BH~\cite{Johannsen:2010ru}. However, in the case of astrophysical BHs, the electric charge is usually neglected because the plasma (e.g., accretion disk) around the BH quickly discharges them~\cite{Zajacek:2019kla}. Therefore, scientists use the Kerr space-time to model astrophysical BHs~\cite{Kerr:1963ud}, which only considers the mass and the angular momentum, i.e., the Kerr hypothesis~\cite{Bambi:2011mj}.
      
      Following the Kerr hypothesis, theorists have constructed several BH hole solutions that resemble the Kerr space-time~\cite{Collins:2004ex, Glampedakis:2005cf, Vigeland:2009pr, Vigeland:2011ji, Johannsen:2013asa, Johannsen:2013szh, Konoplya:2016jvv, Ghasemi-Nodehi:2016wao, Canate:2022gpy}. However, when proposing Kerr-like BH solutions, it is crucial to ensure their astrophysical viability. According to Simpson and Visser,``\textit{one should impose physics constraints which speak directly to the observational community}''~\cite{Simpson:2021zfl}. Consequently, an astrophysical viable space-time must satisfy certain physical assumptions~\cite{Simpson:2021zfl}: 
\begin{enumerate}
\item{}  Given that most astrophysical objects have spin, the space-time must be axially symmetric. 
\item{} Similarly to the Kerr geometry, one must impose asymptotic flatness at spatial infinity.
\item{} To compare theoretical results with observations, it is necessary to impose the separability of the Hamilton-Jacobi (HJ) equations. According to Simpson and Visser, in the case of axisymmetry, a sufficient condition for this is the existence of a nontrivial Killing tensor $K_{\mu\nu}$. 
\item{} In order to analyze the permitted quasinormal modes for spin-one electromagnetism and spin-two GR perturbation using standard numerical techniques, it is necessary to impose the separability of Maxwell's equations and the equations governing spin-two axial and polar modes.
\item{} All RBH solutions violate the strong and the weak energy conditions (the latter in the case of rotating RHB), thus violating the singularity theorems~\cite{Fan:2016hvf}. For this reason, it is imperative to impose constraints on the satisfaction or violation of the classical energy conditions, at least in the region outside the outer horizon. Empirical evidence indicates that violations of the energy conditions should occur only at the quantum scale. Apart from strong energy condition (SEC) violations due to positive cosmological constant, there is no observed evidence of exotic matter in astrophysical contexts. 
\end{enumerate}      

	From the observational point of view, any BH model, whether classical/regular or from an alternative theory of gravity, must be tested experimentally. Recent experimental accomplishments, including the direct detection of gravitational waves (GWs) from astrophysical sources in the LIGO/Virgo merger events~\cite{LIGOScientific:2016aoc, LIGOScientific:2018mvr, LIGOScientific:2020ibl, LIGOScientific:2019fpa, LIGOScientific:2020tif, LIGOScientific:2021sio} and the groundbreaking image of the BH M87 by the event horizon telescope (EHT)~\cite{EventHorizonTelescope:2019dse, EventHorizonTelescope:2019uob,  EventHorizonTelescope:2019jan, EventHorizonTelescope:2019ths, EventHorizonTelescope:2019pgp, EventHorizonTelescope:2019ggy}, have increased the capability of the scientific community and theoreticians to obtain phenomenological evidence of the different BH models based on their astrophysical signature. In this sense, testing BH models via observation will help to enlighten our understanding of space-time singularities and provide experimental clues to think about possible modifications of Einstein's theory to construct a quantum theory of gravity.
     
     Two possible approaches to test BH models are electromagnetic and gravitational wave observations. For example, in recent years, numerous studies have employed X-ray data from NuSTAR, RXTE, Suzaku, and XMM-Newton~\cite{Cao:2017kdq, Tripathi:2018lhx, Tripathi:2020qco, Tripathi:2020dni, Tripathi:2020yts, Zhang:2021ymo}, radio data from the Event Horizon Telescope experiment~\cite{Psaltis:2020ctj, Bambi:2019tjh, EventHorizonTelescope:2020qrl, Volkel:2020xlc}, and gravitational wave data from LIGO and Virgo~\cite{Psaltis:2020ctj, Cardenas-Avendano:2019zxd, Carson:2020iik, Das:2024mjq, Riaz:2022rlx} to test the Kerr hypothesis. In this paper, we use the RBH solution proposed by Ghosh and Simpson-Visser~\cite{Ghosh:2014pba, Simpson:2021dyo, Simpson:2021zfl} to constrain the free parameter $\ell$ using GW and EHT observations. In the former case, we use the binary black hole (BBH) data from LIGO, Virgo and KAGRA (LVK)  collaboration published in their first, second and third gravitational-wave transient catalog GWTC-1, 2, and 3~\cite{LIGOScientific:2018mvr, LIGOScientific:2020ibl, LIGOScientific:2019fpa, LIGOScientific:2020tif, LIGOScientific:2021sio}. In a subsequent work\footnote{This work is already in progress.}, we plan to use X-ray observations to constrain the free parameter $\ell$. 
     
	We organize the manuscript as follows. In Sec.\ref{SecII}, we revisit the regular BH solution proposed by Ghosh and Simpson-Visser, known as the eye of the storm (EOS), and discuss the equations of motion in Sec.~\ref{SecIII}. Then, in Sec.~\ref{SecIV}, we discuss the methodology for testing RBHs using LVK observation using the GWTC-1, 2, and 3 observations. In Sec.~\ref{SecV}, we focus on constraining EOS BH using the ETH observations. We investigate the angular diameter of the shadow and the ring structure using different scenarios, such as static spherical accretion, infalling spherical accretion, and disk accretion, where we assume an optical and geometrically thin accretion disk. In Sec.~\ref{SecVI}, we solve the equations of motion using the Hamiltonian formalism to obtain the image cast by EOS BH. Finally, in Sec.~\ref{SecVII}, we conclude and discuss our results. Along the manuscript, we use geometrized units with $G=c=1$ and denote the dimensionless free parameter using $\ell$. In Sec.\ref{SecIV}, we use $p^\mu$ to denote the four-momentum of massive particles and $V_\text{eff}$ its efective potential. In Sec.~\ref{SecV}, on the other hand, we use $k^\mu$ and $\mathcal{V}_\text{eff}$, respectively, to denote the four-momentum and the effective potential for photons.
   

\section{The Ghosh-Simpson-Visser space-time \label{SecII}}
	In this section, we review and discuss some aspects of the RBH proposed (independently) by Ghosh and Simpson-Visser. As mentioned above, the motivation to propose RBH solutions stems from the fact that, classically, singularities in GR occur at distance scales where the theory is no longer applicable. Therefore, it becomes crucial to excise these singularities by considering appropriate astrophysical regimes to investigate and test quantities that could be observationally falsifiable or verifiable.

\begin{figure}[t]
\centering
\includegraphics[scale=0.55]{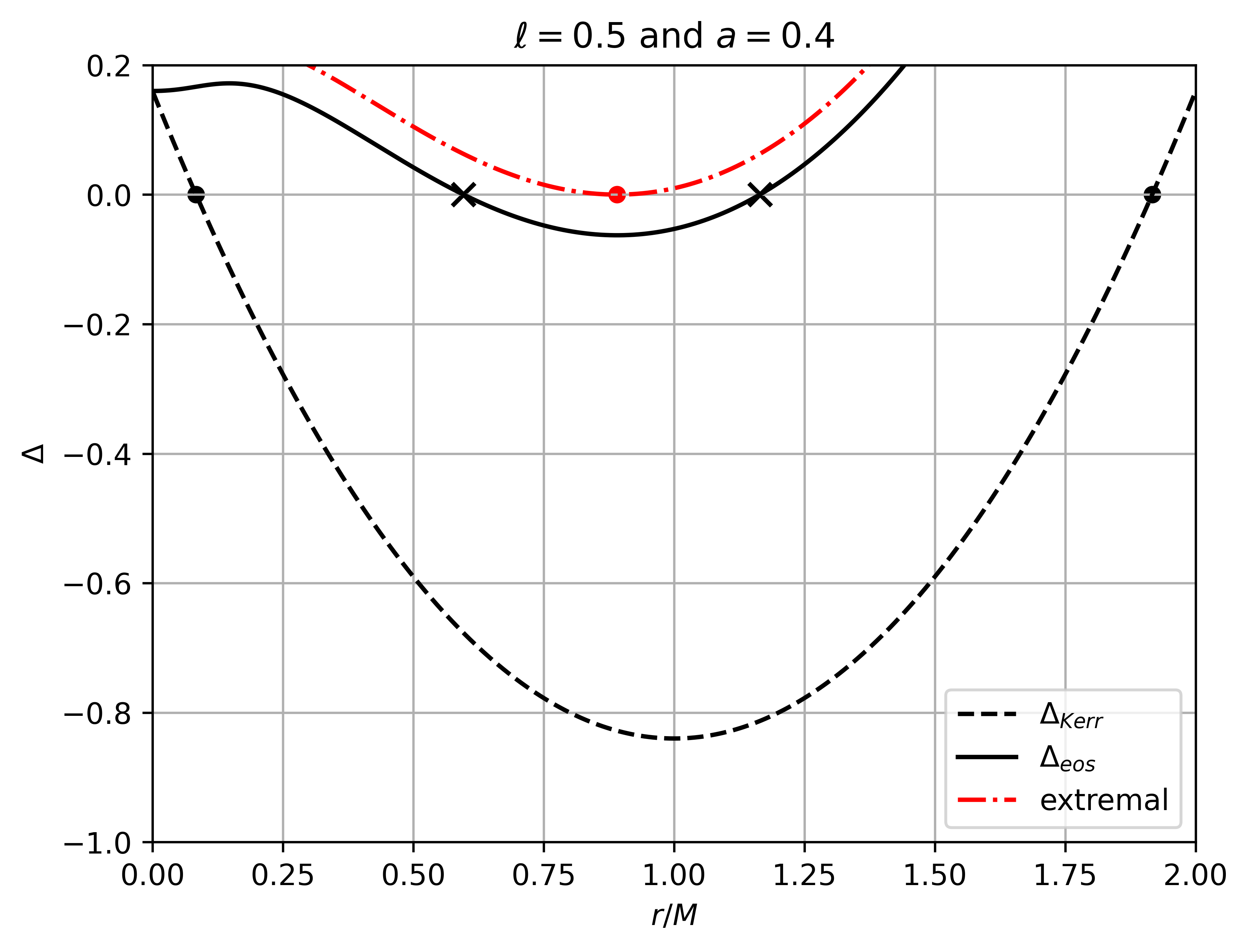}
\caption{Plot of $\Delta_{Kerr}$ and $\Delta_{eos}$ for $a=0.4$, and $\ell=0.5$. The Inner and outer horizons for Kerr and EOS BHs are shown using dots and crosses, respectively. The figure also shows the extremal case (red dashed-dot line) for the Kerr-like black hole.  We use $M=1$.}
\label{Fig1}
\end{figure}
\begin{figure}[t]
\centering
\includegraphics[scale=0.55]{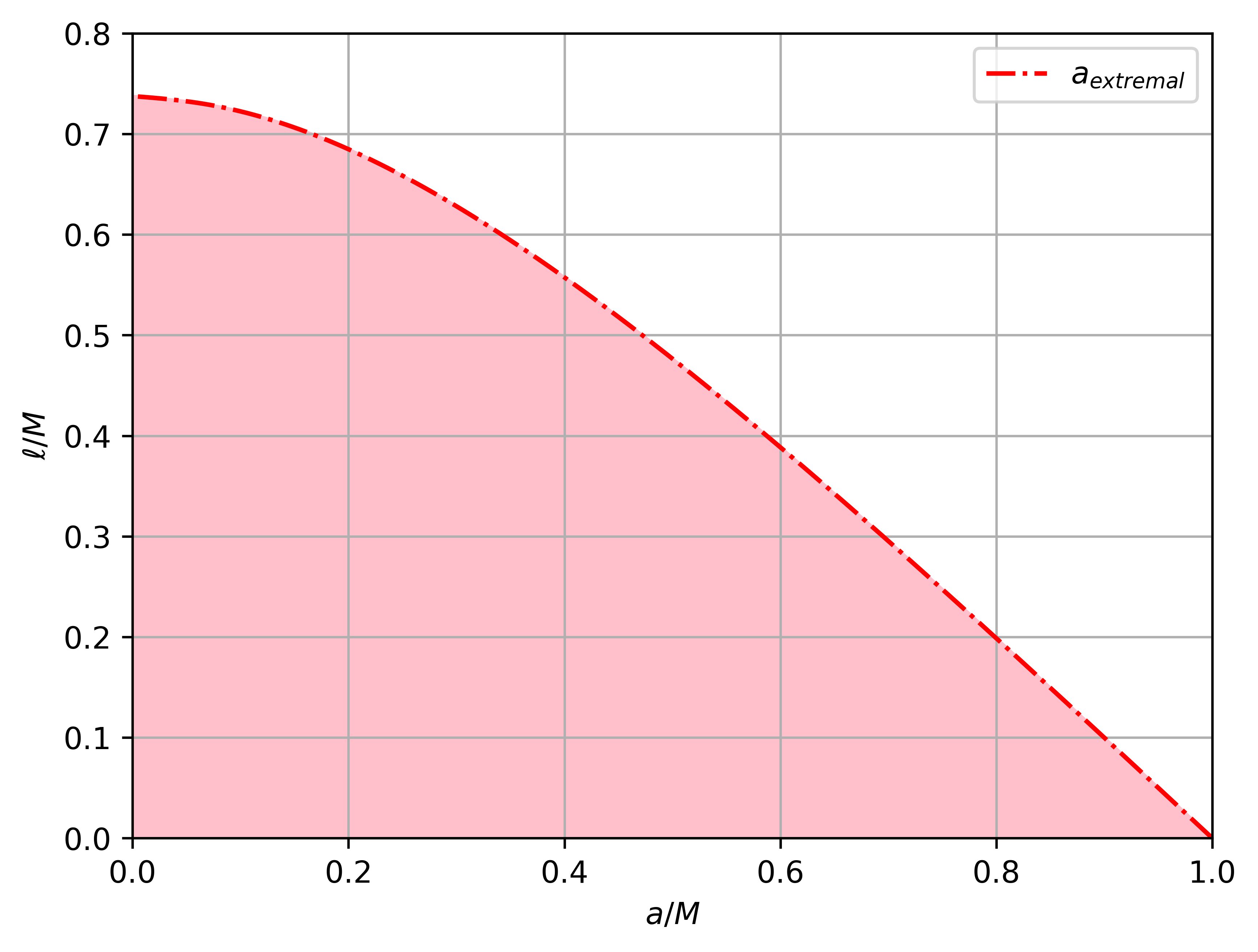}
\caption{$\ell$ vs. $a$ for the EOS space-time. We use $M=1$.}
\label{Fig2}
\end{figure}
	
	The line element describing the  Ghosh-Simpson-Visser space-time is given by~\cite{Ghosh:2014pba, Simpson:2021dyo, Simpson:2021zfl}
\begin{equation}
\label{SIe1}
\begin{aligned}
ds^2&=-\left(1-\frac{2Mr e^{-\ell M/r}}{\Sigma}\right)dt^2+\frac{\Sigma}{\Delta_\text{eos}}dr^2+\Sigma d\theta^2\\
&-\frac{4aMre^{-\ell M/r}}{\Sigma}\sin^2\theta dt d\varphi \\
&+\left[r^2+a^2+\frac{2M r a^2 e^{-\ell M/r}}{\Sigma}\sin^2\theta\right]\sin^2 \theta d\varphi^2,
\end{aligned}
\end{equation}
where 
\begin{equation}
\label{SIe2}
\begin{array}{ccc}
\Sigma=r^2+a^2\cos^2\theta,&\text{and}&
\Delta_\text{eos}=r^2+a^2-2Mr\text{e}^{-\frac{\ell M}{r}}.
\end{array}
\end{equation}	
In Eqs.~\eqref{SIe1} and \eqref{SIe2}, $M$ and $a$ correspond, respectively, to the black hole's mass and spin; $\ell$ is a dimensionless free parameter; i.e., a deviation from the Kerr space-time. When $\ell=0$, the line element reduces to that of Kerr BH, and for $\ell=0$ and $a=0$, the metric reduces to Schwarzschild. The parameters $M$, $a$, and $\ell$ are assumed to be positive~\cite{Ghosh:2014pba}.

	The Ghosh-Simpson-Visser (from now on EOS\footnote{The metric is also known as the eye of the storm (EOS) by Simpson and Visser~\cite{Simpson:2021dyo, Simpson:2021zfl}.}) space-time is inspired by the RBH with an asymptotically Minkowski core analyzed in Ref.~\cite{Simpson:2019mud}, obtained by changing $M\rightarrow M(r)=Me^{-\ell M/r}$ in the Schwarzschild solution. According to Simpson and Visser, the line element in Eq.~\eqref{SIe1}  is an astrophysical viable space-time in the sense described above. It is axially symmetric and preserves asymptotic flatness as $r\rightarrow \infty$. Moreover, as shown in Refs.~\cite{Ghosh:2014pba, Simpson:2021dyo, Simpson:2021zfl}, the physics is well-behaved for $r\geq0$ since the non-zero components of the Riemann curvature tensor are globally finite. 
	
	The separability of the HJ equations is guaranteed by the existence of a non-trivial Killing tensor, $K_{\mu\nu}$. In this sense, it is possible to solve the geodesic equations with the help of a Carter constant, as demonstrated in Refs.~\cite{Amir:2016cen, Kumar:2018ple}; this allows us to compare the theoretical model with GW, EHT or X-ray observations. On the other hand,  since the Ricci tensor and Killing tensor are diagonal in a tetrad basis, the Klein-Gordon equation is separable~\cite{Simpson:2021dyo, Simpson:2021zfl, Baines:2021qaw}. The Maxwell's equations are also separable. Hence,  the EOS space-time is amenable to a standard spin-zero and spin-one quasinormal modes analysis~\cite{Simpson:2021dyo, Simpson:2021zfl}.
	
	The analysis performed by Simpson and Visser in Ref.~\cite{Simpson:2021dyo} regarding the energy conditions demonstrates that the radial null energy condition (NEC) is satisfied globally, given that the density and radial pressure are in equilibrium, i.e., $\rho+p_r=0$. In the case of the transverse NEC, on the other hand, the condition is violated if $\Sigma/r^3>4/\ell$ and satisfied if  $\Sigma/r^3<4/\ell$. On the equatorial plane, the violation of the NEC occurs when $r<\ell/4$. Similarly, when considering the strong energy condition (SEC), the authors show that it is violated if $\Sigma/r^3>2/\ell$ and satisfied if $\Sigma/r^3<2/\ell$. On the equatorial plane, the SEC is unsatisfied whenever $r<\ell/2$. From the physical point of view, these results show the freedom in suppressing the free parameter $\ell$. In this sense, the violation of the energy conditions can be constrained to an arbitrarily small region in the deep core, making the line element \eqref{SIIe1} an astrophysical viable space-time. 	
	
	Although the EOS space-time is singularity-free, it possesses an event horizon characterized by a coordinate singularity: the roots of $\Delta_{eos}$. According to Simpson and Visser, there are two distinct roots, one double root or zero roots and, since $\Delta_{eos}>\Delta_{Kerr}$, one has that~\cite{Simpson:2021dyo, Simpson:2021zfl}
\begin{equation}
\label{SIe3}
M-\sqrt{M^2-a^2}<r^-_H<r^+_H<M+\sqrt{M^2-a^2},
\end{equation}
where $r^-_H$ and $r^-_H$ are the inner and outer horizons, respectively. In Fig.~\ref{Fig1}, we show the behavior of $\Delta_{Kerr}$ and $\Delta_{eos}$ as a function of the radial coordinate $r$. The figure also shows the location of the horizons for both solutions (obtained numerically in the case of the EOS BH) using dots and crosses. From the figure, it is clear that the horizons for the EOS space-time follow the relation stated in Eq.~\eqref{SIe3}. Furthermore, the figure also shows the existence of an extremal case for each value of $\ell$ and $a$, where $r^+_H$ and $r^-_H$ coincide in one horizon; i. e., when the equation $\Delta_{eos}=0$ has one double root. In this sense, the extremal case helps us constrain the values of $\ell$ and $a$ so that the EOS space-time describes a BH. Hence, for a given value of $a_\text{extremal}$, there is a corresponding $\ell_\text{extremal}$ for which the solution is a BH. If $\ell>\ell_\text{extremal}$, there is no horizon, and the line element \eqref{SIIe1} does not represent a BH. In Fig.~\ref{Fig2}, we show $\ell$ vs. $a$, where the red dashed-dot line represents $\ell_\text{extremal}$. Therefore, the pink region corresponds to those values of $\ell$ and $a$ for which the EOS space-time describes a BH while the white region corresponds to naked regular compact objects. It is important to point out that for $a=0$, $\ell_\text{extremal}\approx0.73576$. 

For a detailed analysis and discussion regarding different properties of the EOS regular BH solution, such as the static limit, horizons, and ergoregion, etc., the reader may refer to Refs.~\cite{Ghosh:2014pba, Simpson:2021dyo, Simpson:2021zfl} and references therein.


\section{Equations of motion\label{SecIII}}

In this section, following Chandrasekhar~\cite{Chandrasekhar:1998}, we obtain a system of differential equations necessary to investigate the motion of test particles and photons. In general, axisymmetric space-times allow three constants of motion: the energy, angular momentum, and the norm of the fourth velocity (conserved via parallel displacement). Nevertheless, these constants will not reduce the geodesic motion problem to one involving quadrature~\cite{Chandrasekhar:1998}. Therefore, it is necessary to obtain Carter's constant via the separability of the Hamilton-Jacobi equation~\cite{Carter:1971zc}.
	
	The Lagrangian describing the dynamics of massive/massless particles is given by
\begin{equation}
\label{SIIe1}
2\mathcal{L} = g_{\mu\nu}\dot{x}^\mu \dot{x}^\nu = -\epsilon,
\end{equation}
where an overdot denotes the partial derivative with respect to the affine parameter $\lambda$; $\epsilon=0$ and $\epsilon=1$ refer to massless and massive particles, respectively. 

	The Hamiltonian is defined by 
\begin{equation}
\label{SIIe2}
\mathcal{H}=p_\mu\dot{x}^\mu-\mathcal{L},
\end{equation}
with $p_\mu$ the canonical 4-momentum. It is straightforward to show the Hamiltonian and the Lagrangian are equal~\cite{Chandrasekhar:1998}. On the other hand, the Hamilton-Jacobi equation is given by 
\begin{equation}
\label{SIIe3}
\frac{\partial \mathcal{S}}{\partial \lambda}+\mathcal{H}=0,
\end{equation}
where $\mathcal{S}$ is the Jacobian action. Moreover, the relation between the action and the covariant component of the momentum, $p_\mu$, is 
\begin{equation}
\label{SIIe4}
p_\mu=\frac{\partial \mathcal{S}}{\partial x^\mu}.
\end{equation}
Hence, from Eqs.~\eqref{SIIe2}, and \eqref{SIIe4}, the Hamilton-Jacobi equation reduces to
\begin{widetext} 
\begin{equation}
\label{SIIe5}
\begin{aligned}
2\frac{\partial\mathcal{S}}{\partial\lambda}&=-g^{\alpha\beta}\frac{\partial \mathcal{S}}{\partial x^\alpha}\frac{\partial \mathcal{S}}{\partial x^\beta}=\frac{1}{\Sigma\Delta_\text{eos}}\left[(r^2+a^2)\frac{\partial\mathcal{S}}{\partial{t}}+a\frac{\partial\mathcal{S}}{\partial\varphi}\right]^2
-\frac{1}{\Sigma\sin^2\theta}\left[a\sin^2\theta\frac{\partial\mathcal{S}}{\partial t}+\frac{\partial\mathcal{S}}{\partial \varphi}\right]^2-\frac{\Delta_\text{eos}}{\Sigma}\left(\frac{\partial\mathcal{S}}{\partial r}\right)^2
-\frac{\Delta_\text{eos}}{\Sigma}\left(\frac{\partial\mathcal{S}}{\partial\theta}\right)^2,
\end{aligned}
\end{equation}
\end{widetext}
where we follow Refs.~\cite{Chandrasekhar:1998,Misner:1973prb}. Furthermore, since the metric components do not depend on $t$ and $\varphi$, we have two conserved quantities: the specific energy $E=-p_t$ and the specific angular momentum $L=p_\varphi$. Therefore, one can express the solution $\mathcal{S}$ with the ansatz
\begin{equation}
\label{SIIe6}
\mathcal{S}=\frac{1}{2}\epsilon\lambda-Et+L\varphi+\mathcal{S}_r(r)+\mathcal{S}_\theta(\theta).
\end{equation}
Using this ansatz in Eq.~\eqref{SIIe6}, it is possible to separate the equation through a constant $\pm\mathscr{Q}$ (the Carter's constant) obtaining two equations
\begin{equation}
\label{SIIe7}
\begin{aligned}
\left(\Delta_\text{eos}\frac{d\mathcal{S}}{dr}\right)^2&=\mathcal{R}(r)=[(a^2+r^2)E-aL]^2-\Delta_\text{eos}[\mathcal{K}+\epsilon r^2],\\
\left(\frac{d\mathcal{S}}{d\theta}\right)^2&=\Theta(\theta)=\mathscr{Q}-[L^2\csc^2\theta+a^2(\epsilon-E^2)]\cos^2\theta
\end{aligned}
\end{equation}
with
\begin{equation}
\label{SIIe7a}
\begin{aligned}
\mathcal{K}&=\mathscr{Q}+(L-aE)^2,\\
\mathscr{Q}&=p^2_\theta+[L^2\csc^2\theta+a^2(\epsilon-E^2)]\cos^2\theta.
\end{aligned}
\end{equation}
From which the ansatz \eqref{SIIe6}, takes the form
\begin{equation}
\label{SIIe8}
\mathcal{S}=\frac{1}{2}\epsilon\lambda-Et+L\varphi+\int^r\frac{\sqrt{\mathcal{R}(r')}}{\Delta_\text{eos}}dr'+\int^\theta\sqrt{\Theta(\theta')}d\theta'.
\end{equation}
Now, the basic equations governing the motion can be deduced from Eq.~\eqref{SIIe8}, by setting to zero the partial derivatives of $\mathcal{S}$ with respect to the different constants of motion; i.e,
\begin{equation}
\begin{array}{cccc}
\frac{\partial\mathcal{S}}{\partial\mathscr{Q}}=0,&\frac{\partial\mathcal{S}}{\partial\epsilon}=0,&\frac{\partial\mathcal{S}}{\partial E}=0,&\frac{\partial\mathcal{S}}{\partial L}=0.
\end{array}
\end{equation}
Thus, one obtains the following relations
\begin{equation}
\label{SIIe9a}
\begin{aligned}
0&=\int^\theta\frac{d\theta}{\sqrt{\Theta}}-\int^r\frac{dr}{\sqrt{\mathcal{R}}},\\\\
\lambda&=\int^r\frac{r^2}{\sqrt{\mathcal{R}}}dr+a^2\int^\theta\frac{\cos^2\theta}{\sqrt{\Theta}}d\theta,
\end{aligned}
\end{equation}
and 
\begin{equation}
\label{SIIe9b}
\begin{aligned}
\dot{t}&=\frac{[(r^2+a^2)E-aL](r^2+a^2)}{\Delta_\text{eos}\sqrt{\mathcal{R}}}\dot{r}+\frac{a[aE-L\csc^2\theta]}{\sqrt{\Theta}}\dot{\theta},\\\\
\dot{\varphi}&=\frac{a[(r^2+a^2)E-aL]}{\Delta_\text{eos}\sqrt{\mathcal{R}}}\dot{r}+\frac{[L\csc^2\theta-aE]}{\sqrt{\Theta}}\dot{\theta}.
\end{aligned}
\end{equation}
From Eqs.\eqref{SIIe4} and \eqref{SIIe8}, the differential equations for $r$ and $\theta$ are
\begin{equation}
\label{SIIe9c}
\begin{aligned}
\Sigma\dot{r}&=\pm\sqrt{\mathcal{R}(r)},\\
\Sigma\dot{\theta}&=\pm\sqrt{\Theta}(\theta).
\end{aligned}
\end{equation}
Finally, after substituting into Eq.~\eqref{SIIe9b}, one gets~\cite{Amir:2016cen}
\begin{equation}
\label{SIIe9d}
\begin{aligned}
\Sigma \dot{t}&=-a(aE\sin^2\theta-L)+\frac{(r^2+a^2)\left[(r^2+a^2)E-aL\right]}{r^2+a^2-2Mre^{-\ell M/r}},\\\\
\Sigma \dot{\varphi}&=-(aE-L\csc^2\theta)+\frac{a\left[(r^2+a^2)E^2-aL\right]}{r^2+a^2-2Mre^{-\ell M/r}}.
\end{aligned}
\end{equation}
Hence the geodesic motion in the EOS space-time is modeled by the Eqs.~\eqref{SIIe9c} and \eqref{SIIe9d}.


\section{Constraining $\ell$ using LVK observations \label{SecIV}}

In order to study the GWs for the EOS BH solution, we consider only the inspiral phase of the binary system (BS), which can be studied using the post-Newtonian (PN) formalism~\cite{Blanchet:2013haa}. In the context of inspirals, we consider equatorial circular obits. Thus, we assume $\theta=\pi/2$ and $\dot{\theta}=0$. On the other hand, it is well-known that the spin contributions enter at the 3.5PN level or higher~\cite{Cardenas-Avendano:2019zxd}. Accordingly, we focus on the leading-order approximation for the free parameter $\ell$ that enters at a smaller PN order than the spin, $a$. As we will demonstrate shortly, it turns out that the contribution of the free parameter $\ell$ enters at 1PN order. In this sense, we set $a=0$, from which the line element~\eqref{SIe1} becomes
\begin{equation}
\label{SIIIe1}
\begin{aligned}
ds^2&=-\left(1-\frac{2M(r)}{r}\right)dt^2+\left(1-\frac{2M(r)}{r}\right)^{-1}dr^2\\\\
&+r^2d\theta^2+r^2\sin^2 \theta d\varphi^2,
\end{aligned}
\end{equation}
with $M(r)=Me^{-\ell M/r}$. This space-time corresponds to a deformation of the Schwarzschild BH through the free parameter $\ell$.

	The production of GWs is modified when the space-time surrounding compact objects in a binary system is affected by a Schwarzschild deformation. As one derives the Hamiltonian/Lagrangian from the metric itself, any modification to the space-time near the compact objects will result in a corresponding modification to the Hamiltonian/Lagrangian. Consequently, this leads to changes in the equations of motion, influencing the evolution of the orbital phase and, ultimately, the emitted GWs.
	
	Following the above consequence, we compute the GWs emitted during the early inspiral of a BS composed of non-rotating EOS BHs. As mentioned before, we focus on the inspiral regime and, thus, work on the PN framework. The main idea is to map the two-body problem to an effective one-body (EOB) problem, controlled by an effective Hamiltonian derived from the deformed Schwarzschild space-time. Using this deformed Hamiltonian, we compute the binding energy of the BS, assuming the radiation-reaction force of GR. Then, we calculate the rate of change of the orbital frequency and the GWs emitted in the frequency domain. Finally, we map the result to the parameterized post-Einsteinian (PPE) framework and use the LIGO-Virgo-KAGRA catalog GWTC-1, 2, and 3 to constrain the leading-order metric deformation parameter $\ell$.

\subsection{Equatorial geodesics}

	According to Ref.~\cite{Cardenas-Avendano:2019zxd}, the two-body problem can be mapped to an effective one-body (EOB) problem using the NP formalism. From the physical point of view, this EOB problem corresponds to the physical setup in which a test particle of mass equal to the reduced mass\footnote{Where $m=m_1+m_2$ is the total mass of the BS, and $m_{1,2}$ are the component masses.} of the real BS $\mu=m_1m_2/m$ is moving in geodesic motion (circular motion) around a BH with mass equal to the total mass of the BS. In our case, the space-time background corresponds to the non-rotating EOS metric. Hence, the effective Hamiltonian~\cite{Buonanno:1998gg, Hinderer:2017jcs}, controlling the conservative sector of the orbital motion, can be constructed from the contraction of the EOS metric with the four-momenta of the test particle. Then, one maps back to the real two-body problem to compute the GWs emitted by such a system when the two BHs are deformed. 
	
	To begin with, we consider the motion of a massive test particle in circular motion around the non-rotating EOS BH with total mass $m$ and free parameter $\ell$. From the analysis performed in Sec~\ref{SecIII}, we know the existence of two conserved quantities: the energy $E$ and the z-component of the angular momentum $L$. In the case of massive particles with rest mass $\mu$, we define the energy/angular momentum per unit reduced mass as $\mathcal{E}=-p_t$ and $\mathscr{L}=p_\varphi$, respectively.
	
	The radial equation for a massive particle can be obtained from the normalization condition $g_{\mu\nu}\dot{x}^\mu\dot{x}^\nu=-1$, from which we obtain 
\begin{equation}
\label{SIIIe2}
\dot{r}^2=\mathcal{E}^2-1+\frac{2me^{-\ell m/r}}{r}+\frac{(2me^{-\ell m/r}-r)}{r^3}\mathscr{L}^2=V^\text{EOS}_\text{eff}.
\end{equation}
Note that $V^\text{EOS}_\text{eff}$ reduces to that of Schwarzschild for $\ell=0$, i. e.,
\begin{equation}
\label{SIIIe3}
V^\text{S}_\text{eff}=\mathcal{E}^2 -1+\frac{2m}{r}+\frac{(2m-r)}{r^3}\mathscr{L}^2.
\end{equation}
Furthermore, for small pertubation of the free parameter ($\ell<<1$), the radial equation can be expressed in the form
\begin{equation}
\label{SIIIe4}
\dot{r}^2=V^\text{S}_\text{eff}-\frac{2 m\left[\mathscr{L}^2+r^2\right]}{r^4}\ell+\mathcal{O}(\ell^2)=V^\text{EOS}_\text{eff}.
\end{equation}	
Therefore, the radial equation can be expressed as that of Schwarzschild plus a small pertubation; i. e. 
\begin{equation}
\label{SIIIe5}
V^\text{EOS}_\text{eff}= V^\text{S}_\text{eff}+V^\text{P}_\text{eff}+\mathcal{O}(\ell^2),
\end{equation}	
where $V^\text{P}_\text{eff}$ is the small perturbation depending on the free parameer $\ell$. 

	On the other hand, the energy and the angular momentum for circular orbits can be found using the condition
\begin{equation}
\label{SIIIe6}
V^\text{EOS}_\text{eff}=0=\frac{dV^\text{EOS}_\text{eff}}{dr},
\end{equation}
from which, after solving simultaneously for $\mathcal{E}^2$ and $\mathscr{L}^2$, we obtain 
\begin{equation}
\label{SIIIe7}
\begin{aligned}
\mathcal{E}&=\sqrt{\frac{\left[r(r-2m)+2m\ell\right]^2}{r^2\left[r(r-3m)+4m\ell\right]}},\\\\
\mathscr{L}&=\sqrt{\frac{mr^2(r-2\ell)}{r(r-3m)+4m\ell}}.
\end{aligned}
\end{equation}
Once again, note that the last expressions reduce to those of Schwarzschild when $\ell=0$; i.e.,
\begin{equation}
\label{SIIIe8}
\begin{aligned}
\mathcal{E}^\text{S}&=\sqrt{\frac{4m^2-4mr+r^2}{r(r-3m)}},\\
\mathscr{L}^\text{S}&=\sqrt{\frac{mr^2}{r-3m}}.
\end{aligned}
\end{equation}

	Similarly to the effective potential, $V^\text{EOS}_\text{eff}$, it is possible to represent the energy and the angular momentum in terms of small perturbations by considering the form of Eq.~\eqref{SIIIe5}. Hence, when $\ell<<1$, we obtain the following expressions:
\begin{equation}
\label{SIIIe9}
\begin{aligned}
\mathcal{E}&= \mathcal{E}^\text{S}+\delta \mathcal{E},\\
\mathscr{L}&=\mathscr{L}^\text{S}+\delta \mathscr{L},
\end{aligned}
\end{equation}   
where
\begin{equation}
\label{SIIIe10}
\begin{aligned}
\delta \mathcal{E}&=-\frac{2m^3}{\left[r(r-3m)\right]^{3/2}}\ell+\mathcal{O}(\ell^2),\\\\
\delta \mathscr{L}&=-\frac{m^{3/2}(r-m)}{(r-3m)^{3/2}}\ell+\mathcal{O}(\ell^2).
\end{aligned}
\end{equation}

	Using the above results, it is possible to obtain a modified Kepler law. To do so, we recall that in the in the far limit, $\mathscr{L}=r^2\dot{\varphi}\rightarrow r^2\Omega$, where $\Omega=d\varphi/dt$ is the angular velocity of the body as measured by the distant observer. In this sense, by considering $\ell<<1$ and expanding, we obtain the following expression 
\begin{equation}
\label{SIIIe11}
\Omega^2=\frac{m}{r^3}\left[1+\frac{3 m}{r}-\frac{2 m}{r}\ell+\frac{9 m^2}{r^2}+\mathcal{O}\left(\ell,\frac{m^2}{r^2}\right)\right].	
\end{equation}
Note that $\ell$ enters at order 1PN, see the power of $m/r$ in the $\ell$ term. This result is similar to that of Ref.~\cite{Carson:2020iik}.

\subsection{Gravitational wave constrain for $\ell$}

	In order to constrain the free parameter $\ell$ using GW observations, we need to use the total and effective energy to map our results back to the two-body problem~\cite{Cardenas-Avendano:2019zxd, Buonanno:1998gg, Damour:2000kk, Shashank:2021giy}. In the case of circular orbits, the total energy of the system, $E_T$, can be expressed in terms of the effective energy, which is the energy of the body in the rest frame of the other~\cite{Buonanno:1998gg, Damour:2000kk}
\begin{equation}
\label{SIIIe12}
E_T=m+E_b=m\left[1+2\eta(E_\text{eff}-1)\right],
\end{equation}
where~\cite{Damour:2000kk}
\begin{equation}
\label{SIIIe13}
E_\text{eff}=g_{tt}\left(1+\frac{\mathscr{L}^2}{r^2}\right)^{1/2}.
\end{equation}
In Eq.~\eqref{SIIIe12}, the rest-mass energy $m$ has been separated from the binding energy $E_b$; this allows us to express $E_b$ as its GR term plus a correction. Additionally, we have introduced the symmetric mass ratio, defined as $\eta=\mu/m$. Hence, by considering $\ell<<1$ and expanding, we obtain the following form for the binding energy:
\begin{equation}
\label{SIIIe14}
E_b=E^\text{GR}_b-\frac{\eta m^2}{2r}\left[\frac{4m^2}{r^2}\ell+\mathcal{O}\left(\ell,\frac{m^3}{r^3}\right)\right].
\end{equation}
Note that it is possible to rewrite Eq.~\eqref{SIIIe14} in terms of the orbital frequency $\nu=\Omega/2\pi$ since the angular frequencies of both EOB problem and the two-body problem are the same~\cite{Cardenas-Avendano:2019zxd, Shashank:2021giy}; we obtain the following expression: 
\begin{equation}
\label{SIIIe15}
\frac{E_b(\nu)}{\mu}=\frac{E^\text{GR}_b(\nu)}{\mu}-2\ell(2\pi m\nu)^2+\mathcal{O}\left(\ell^2,(2\pi m \nu)^{8/3}\right).
\end{equation}
Equations \eqref{SIIIe14} and \eqref{SIIIe15} correspond to the total energy of the real BS with a perturbation. From Eq.~\eqref{SIIIe14}, it is possible to conclude that the perturbation is proportional to the free parameter $\ell$ that appears in the second order term $(m/r)^2$; i.e., at 2PN order. Therefore, the perturbation term is of $\mathcal{O}(v^4)$ smaller than the leading PN order term $E^\text{GR}_b$~\cite{Cardenas-Avendano:2019zxd}. 

	The orbital phase for a BS in a circular orbit is given by~\cite{Cardenas-Avendano:2019zxd, Shashank:2021giy}
\begin{equation}
\label{SIIIe16}
\phi(\nu)=\int^\nu\Omega dt=\int^\nu\frac{1}{\dot{E}}\left(\frac{dE}{d\Omega}\right)\Omega d\Omega,
\end{equation}
where $\dot{E}$ is the rate of change of the binding energy of the system due to gravitational wave emission. Note  that the gravitational wave phase depends both on the conservative (time-symmetric) dynamics represented here in the binding energy, as well as on the dissipative (time-asymmetric) dynamics, represented here in the energy loss rate~\cite{Cardenas-Avendano:2019zxd}. Hence, to find $\phi(\nu)$, we need to obtain $dE/d\Omega$ and $\dot{E}$ as functions of $\Omega$, and then as functions of $\nu$. 

	In a future work work, we want to compare the GW and X-ray constrains. Nevertheless, in contrast to former, which are sensitive to both the conservative and the dissipative sectors, X-ray constrains are only sensitive to the conservative sector. In this sense, we only consider modifications on the conservative dynamics and assume dissipative dynamics to be the same as GR, following the ideas in Refs.~\cite{Cardenas-Avendano:2019zxd, Shashank:2021giy}. Hence, with this assumption in mind, we only need the quadrupole fromula to the leading PN order (0PN) for the change in the binding energy~\cite{Blanchet:2013haa}; i.e.
\begin{equation}
\label{SIIIe19}
\dot{E}^\text{0PN}_\text{GR}=-\frac{32}{5}\eta^2m^2r^4\Omega^6.
\end{equation}
From which, after replacing into Eq.~\eqref{SIIIe16}, the expression for the orbital phase evolution takes the form  
\begin{equation}
\label{SIIIe20}
\phi(\nu)=\phi^\text{0PN}_\text{GR}(\nu)-\frac{5}{24\eta}(2m\pi\nu)^{-1}\ell+\mathcal{O}\left[\ell^2,(2m\pi\nu)^{-1/3}\right],
\end{equation}
where
\begin{equation}
\label{SIIIe21}
\phi^\text{0PN}_\text{GR}(\nu)=-\frac{1}{32\eta}(2m\pi\nu)^{-5/3}.
\end{equation}
Now, to compute the correction to the Fourier phase of the GW, we assume that its rate of change is much more rapid than the change of rate of the GW amplitude. This is known as the stationary phase approximation~\cite{Maggiore:2007ulw}. Hence, the Fourier phase can be written as 
\begin{equation}
\label{SIIIe22}
\Psi_\text{GW}(f)=2\phi(t_0)-2\pi f t_0,
\end{equation}
where $t_0$ is the stationary time such that $\nu(t_0)=f/2$ and $f$ is the Fourier frequency. At the leading order in the parameter $\ell$, we obtain the following expression 
\begin{equation}
\label{SIIIe23}
\Psi_\text{GW}(f)=\Psi^\text{GR}_\text{GW}-\frac{5}{24}\ell u^{-1}\eta^{-2/5}+\mathcal{O}(\ell^2,u^{-1/3}),
\end{equation}
with $u=\eta^{3/5}\pi m f$. In the last expression, $\Psi^\text{GR}_\text{GW}$ corresponds to the Fourier GW phase of GR; i.e.,
\begin{equation}
\label{SIIIe24}
\Psi^\text{GR,0PN}_\text{GW}(f)=-\frac{3}{128}u^{-5/3}.
\end{equation}
Finally, as mentioned above, we want to map our result to the PPE framework. In this framework, the parametrization is given by~\cite{Yunes:2009ke}
\begin{equation}
\label{SIIIe25}
\Psi_\text{GW}=\Psi^\text{GR}_\text{GW}+\beta u^b.
\end{equation} 
Therefore, after comparing with Eq.~\eqref{SIIIe23}, we found that $b=-1$ at 1PN and 
\begin{equation}
\label{SIIIe26}
\beta=-\frac{5}{24}\eta^{-2/5}\ell.
\end{equation} 
On the other hand, in the PPE parametrization used by LVC~\cite{Yunes:2016jcc} (see their Eq.~(28)), the expession for $\beta$ is given by 
\begin{equation}
\label{SIIIe27}
\beta=\frac{3}{128}\varphi_2\delta\varphi_2\eta^{-2/5},
\end{equation}
where (see the appendix B of Ref.~\cite{Khan:2015jqa}, Eq.~(B8))
\begin{equation}
\label{SIIIe28}
\varphi_2=\frac{3715}{756}+\frac{55}{9}\eta.
\end{equation}
Therefore, after comparing Eqs.~\eqref{SIIIe26} and \eqref{SIIIe27}, we obtain that 
\begin{equation}
\label{SIIIe29}
\ell=-\frac{9}{80}\varphi_2\delta\varphi_2.
\end{equation}

	Finally, we fit the publicly available posterior samples released by the LVK collaboration to obtain the constraints on the parameter $\ell$. We use the data released for GWTC-1~\cite{LVC:GWTC-1}, GWTC-2~\cite{LVC:GWTC-2} and GWTC-3~\cite{LVKC:GWTC-3} and follow the name convention from the corresponding released paper; i.e., Refs.~\cite{LIGOScientific:2019fpa}, \cite{LIGOScientific:2020tif}, and \cite{LIGOScientific:2021sio}, for GWTC-1, 2 and 3, respectively. 

\begin{figure*}[t]
\centering
\includegraphics[scale=0.47]{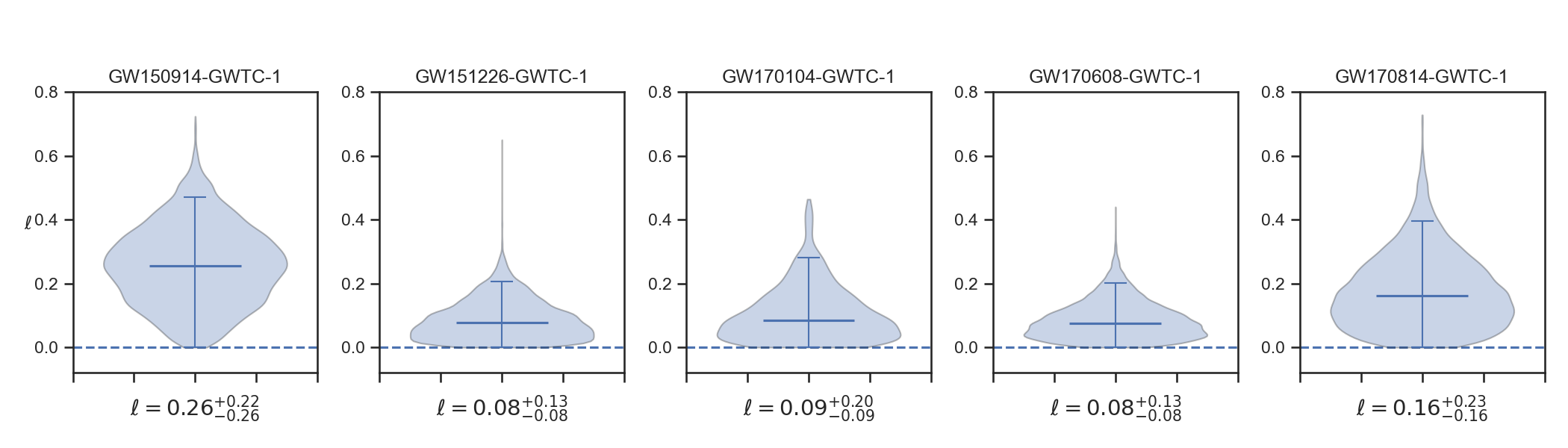}\\
\includegraphics[scale=0.47]{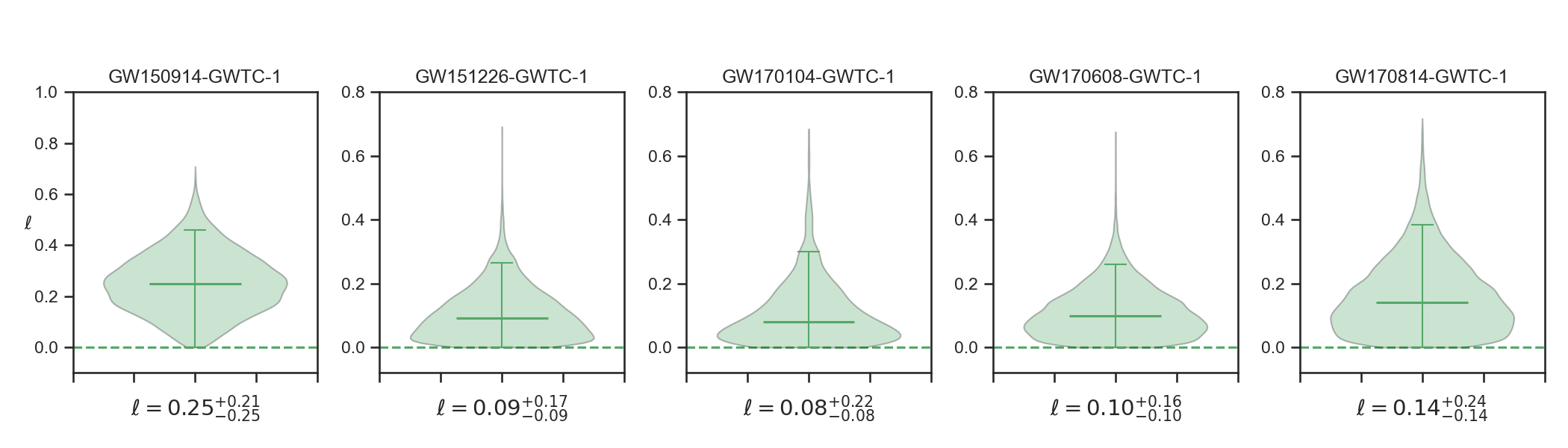}\\
\includegraphics[scale=0.47]{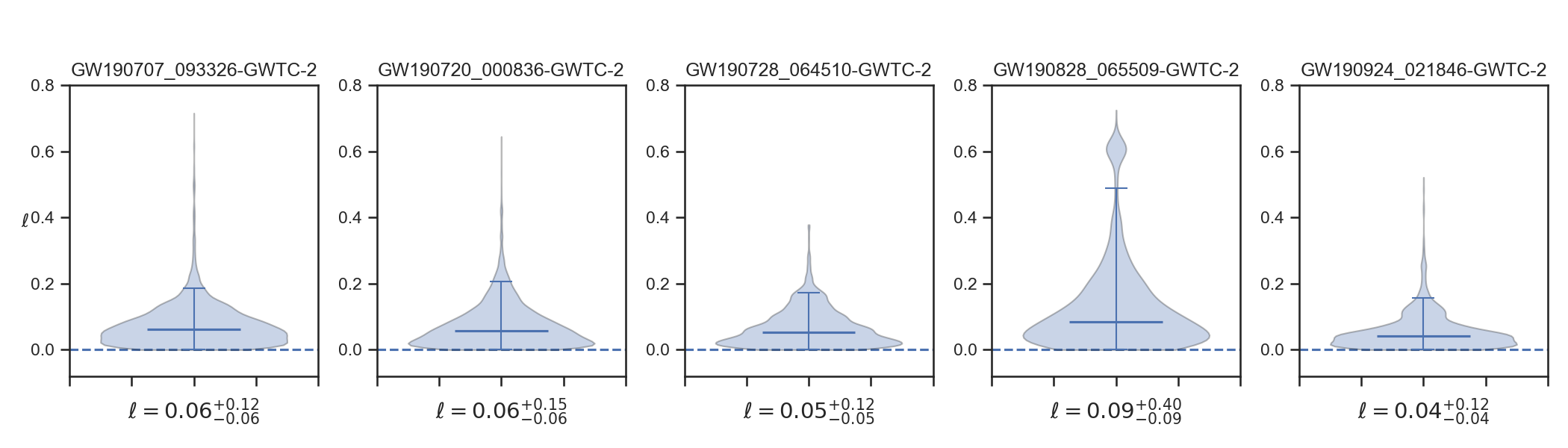}\\
\includegraphics[scale=0.47]{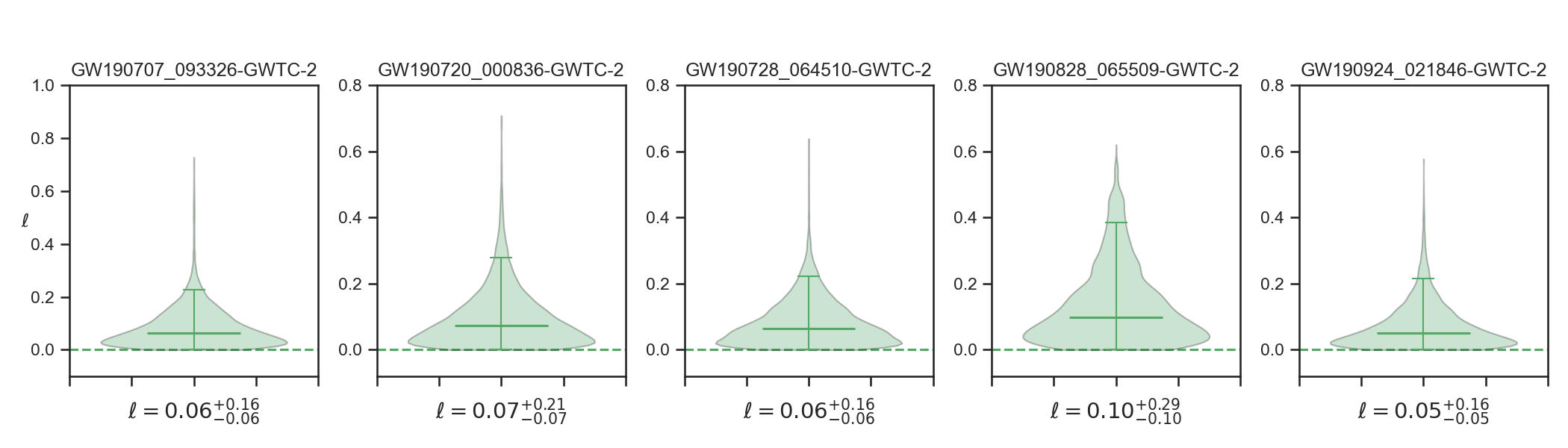}\\
\includegraphics[scale=0.47]{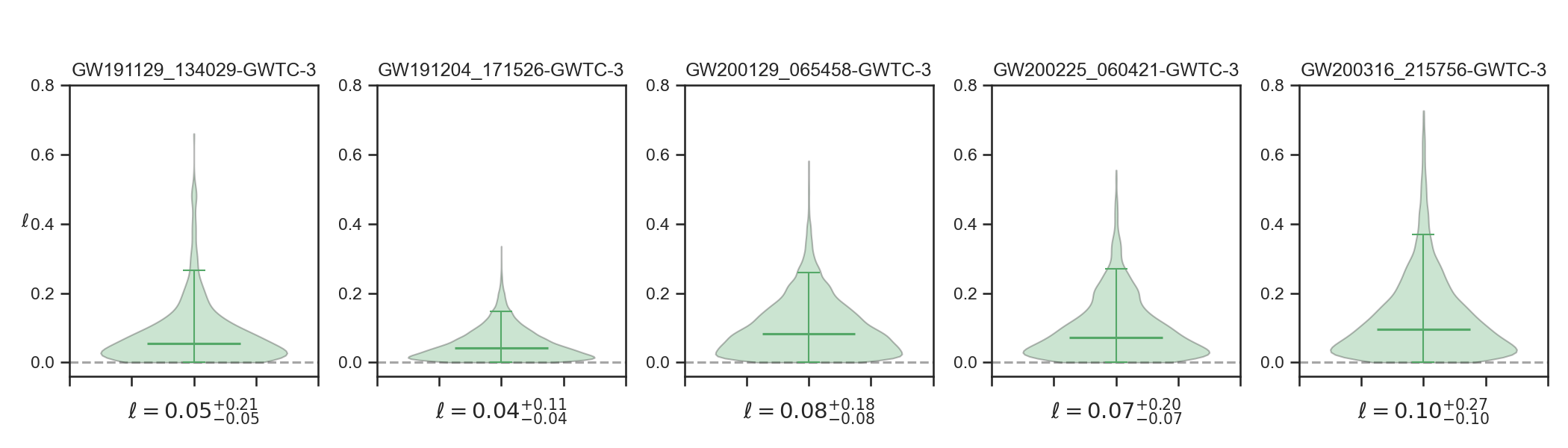}
\caption{Violin plots for $\ell$ using the GWTC-1, 2 and 3 data for SEOB (green) and IMRPhenom (blue) models. The horizontal lines indicate the $90\%$ credible interval around the indicated mean.}
\label{Fig3}
\end{figure*}

	The procedure to obtain constraints on the parameter $\ell$ from GW observations is simple, thanks to the analysis performed above. In our case, according to Eq.~\eqref{SIIIe29}, the parameter $\ell$ depends on $\varphi_2$ and $\delta\varphi_2$. The latter can be obtained directly from the data since LIGO, Virgo and KAGRA analyze the GW events to constrain all non-GR parameters. The former, on the other hand, can be computed using Eq.~\eqref{SIIIe28}, where $\eta$, the symmetric mass ratio, depends on the BH masses $m_1$ and $m_2$, also included in the GWTC data. It is important to remark that the EOS space-time assumes positive values of $\ell$ and, according to Fig.~\ref{Fig2}, the range of this parameter depends on the extremal case, which depends on the BH spin $a$. Since we consider $a=0$ in our analysis, the extremal value of $\ell\approx0.73576$. In this sense, while fitting $\ell$, we rule out all the negative values and consider only those for which $0\leq\ell<\ell_\text{extremal}$.  In Fig.~\ref{Fig3}, we show the violin plots for the free parameter $\ell$ usign the GWTC 1, 2 and 3 observations. We consider the SEOB~\cite{Bohe:2016gbl, Cotesta:2018fcv} and IMRP models~\cite{Husa:2015iqa, Hannam:2013oca}. In Table~\ref{table1}, we show the resutls with $90\%$ of confidence. We left the discussion of our results to Sec.~\ref{SecVII}.

\begin{table*}
\caption{\label{table1} Constraints on the free parameter $\ell$ for the BBH events in GWTC-1 and GWTC-2 with the IMRPhenomPv2 and SEOBNRv4P waveform models. The reported uncertainties correspond to the $90$\% confidence limit.}
\begin{ruledtabular}
\begin{tabular}{cccccc}
&\multicolumn{2}{c}{$\ell$}&&\\
GWTC-1&&&& \\
\hline
Event&IMRPhenomPv2&SEOBNRv4P&&&\\
\hline\\
GW150914 &$0.255^{+0.216}_{-0.255}$&$0.249^{+0.210}_{-0.249}$&&&\\\\
GW151226*&$0.077^{+0.129}_{-0.077}$&$0.091^{+0.174}_{-0.091}$&&&\\\\
GW170104*&$0.086^{+0.196}_{-0.085}$&$0.080^{+0.220}_{-0.080}$&&&\\\\ 
GW170608*&$0.076^{+0.127}_{-0.076}$&$0.099^{+0.163}_{-0.099}$&&&\\\\
GW170814 &$0.161^{+0.234}_{-0.161}$&$0.141^{+0.243}_{-0.141}$&&& \\
&&&&\multicolumn{2}{c}{$\ell$}\\
GWTC-2&&&&& \\
\hline
Event&IMRPhenomPv2&SEOBNRv4P&Event&IMRPhenomPv2&SEOBNRv4P\\
\hline\\
GW190408-181802*  &$0.092^{+0.271}_{-0.092}$&$0.129^{+0.349}_{-0.129}$& GW190412               &$0.129^{+0.151}_{-0.129}$&$0.120^{+0.169}_{-0.120}$ 
\\\\
GW190503-185404  &$0.357^{+0.322}_{-0.357}$&$0.282^{+0.345}_{-0.282}$& GW190512-180714  &$0.392^{+0.263}_{-0.392}$&$0.354^{+0.270}_{-0.354}$ 
\\\\
GW190513-205428  &$0.125^{+0.284}_{-0.125}$&$0.135^{+0.335}_{-0.135}$& GW190517-055101  &$0.309^{+0.369}_{-0.309}$&$0.401^{+0.286}_{-0.401}$
\\\\
GW190521-074359  &$0.188^{+0.262}_{-0.188}$&$0.123^{+0.235}_{-0.123}$& GW190602-175927  &$0.365^{+0.323}_{-0.364}$&$0.356^{+0.333}_{-0.356}$ 
\\\\
GW190630-185205* &$0.086^{+0.189}_{-0.086}$&$0.086^{+0.194}_{-0.086}$& GW190707-093326* &$0.061^{+0.124}_{-0.061}$&$0.064^{+0.163}_{-0.064}$ 
\\\\
GW190708-232457* &$0.080^{+0.170}_{-0.080}$&$0.078^{+0.170}_{-0.078}$& GW190720-000836* &$0.057^{+0.150}_{-0.057}$&$0.073^{+0.207}_{-0.073}$ 
\\\\
GW190727-060333 &$0.297^{+0.371}_{-0.285}$&$0.323^{+0.368}_{-0.323}$& GW190728-064510*  &$0.054^{+0.118}_{-0.054}$&$0.065^{+0.157}_{-0.065}$ 
\\\\
GW190828-063405 &$0.218^{+0.279}_{-0.218}$&$0.218^{+0.290}_{-0.218}$& GW190828-065509*  &$0.085^{+0.404}_{-0.085}$&$0.099^{+0.287}_{-0.099}$ 
\\\\
GW190910-112807 &$0.196^{+0.360}_{-0.196}$&$0.169^{+0.301}_{-0.169}$& GW190924-021846*   &$0.042^{+0.116}_{-0.042}$&$0.050^{+0.165}_{-0.050}$ 
\\\\
GWTC-3&&&&& \\
\hline
Event&SEOBNRv4P&Event&SEOBNRv4P&Event&SEOBNRv4P\\
\hline\\
GW191129-134029*&$0.054^{+0.212}_{-0.054}$&GW191204-171526*&$0.041^{+0.106}_{-0.041}$&GW191216-213338*&$0.072^{+0.126}_{-0.072}$ 
\\\\
GW200129-065458*&$0.082^{+0.178}_{-0.082}$&GW200202-154313*&$0.071^{+0.213}_{-0.071}$&GW200225-060421*&$0.072^{+0.197}_{-0.072}$
\\\\
GW200311-115853  &$0.144^{+0.294}_{-0.144}$&GW200316-215756*&$0.096^{+0.274}_{-0.096}$&&
\\\\
\end{tabular}
\end{ruledtabular}
\end{table*}

\begin{figure*}[t]
\centering
\includegraphics[scale=0.5]{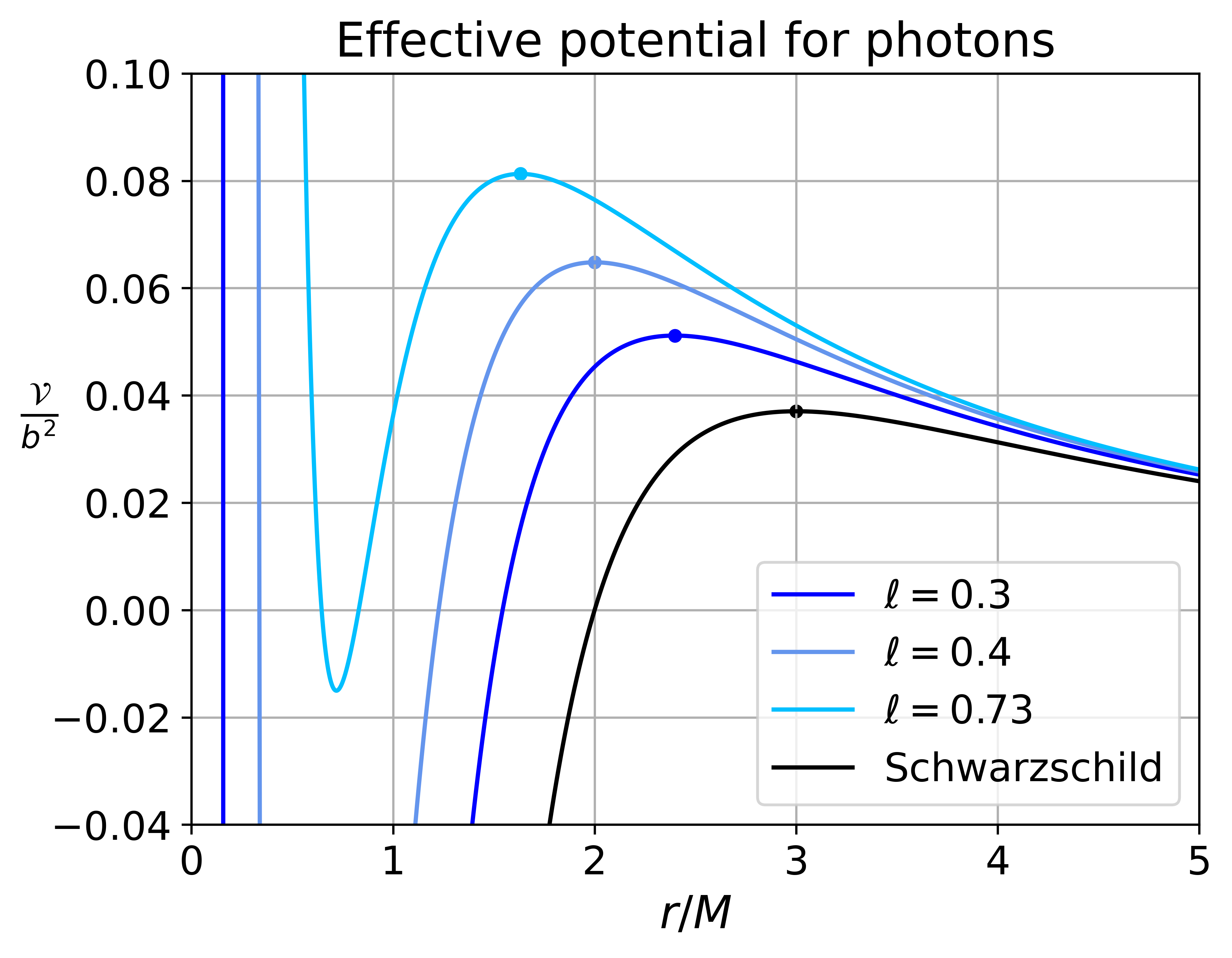}
\includegraphics[scale=0.5]{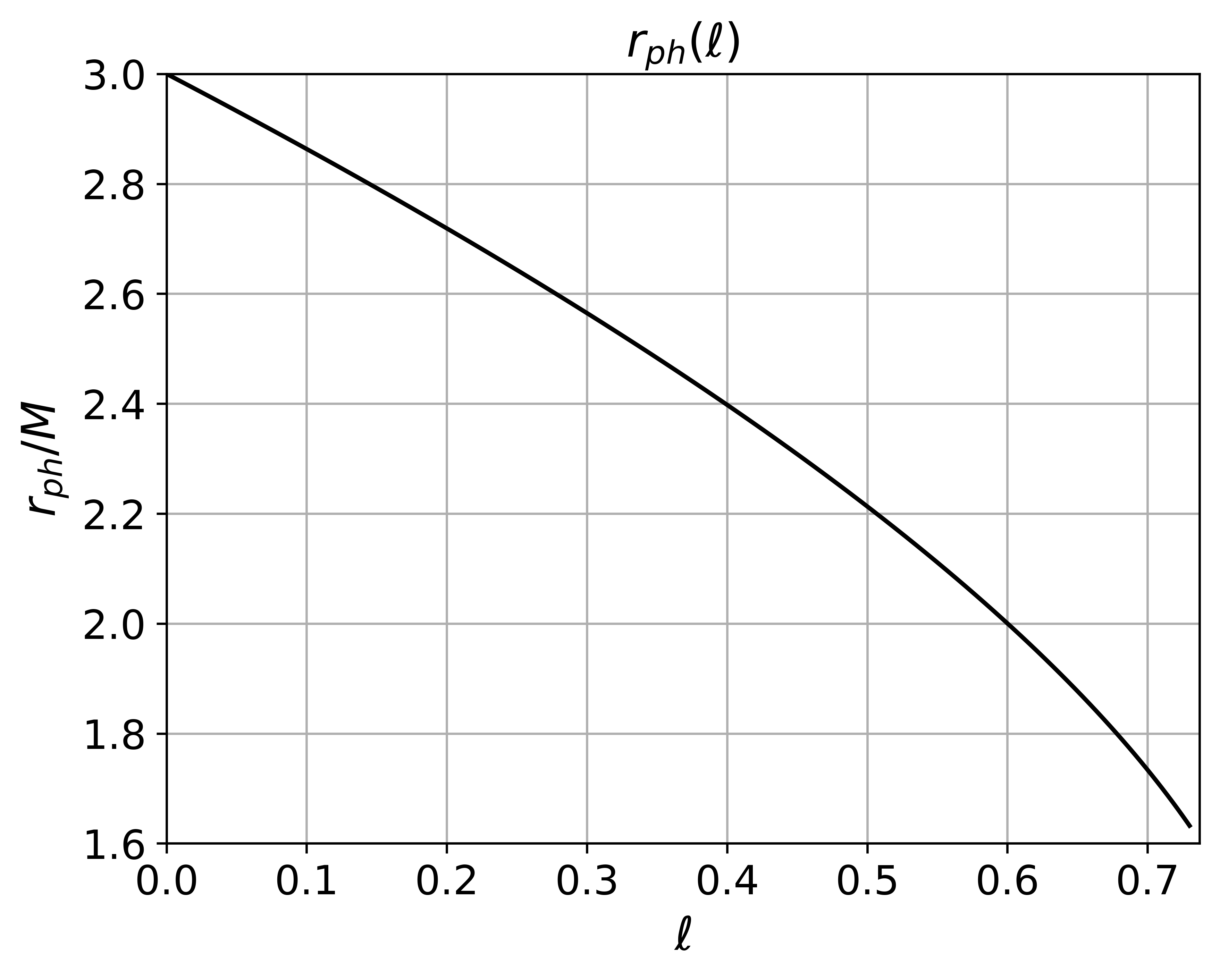}
\caption{Left panel: the effective potential for photons for different values of $\ell$. The dot markes show the location of $\mathcal{V}_\text{max}$. Right panle: $r_\text{ph}$ vs. $\ell$. In the plots, we use $M=1$.}
\label{Fig4}
\end{figure*}
	
\section{Constraining $\ell$ using EHT observations \label{SecV}}
	
	In this section we investigate the observational signatures of the EOS space-time using the EHT observations. To do so, we investigate the motion of photons using the system of equations obtained in Sec.~\ref{SecIII}. Without loss of generality, we constraint our analysis to the equatorial plane; i. e.,  $\dot{\theta} = 0$ and $\theta=\pi/2$. Under this condtions, the Carter's constant $\mathscr{Q}$ vanishes and $\mathcal{K}$ reduces to $L^2$, see Eq.~\eqref{SIIe7}, where we also assume $a=0$. Hence, in the case of photons ($\epsilon=0$), the radial equation \eqref{SIIe9c} takes the form
\begin{equation}
\label{SIVe1}
\left(\frac{dr}{ds}\right)^2=1-\frac{b^2}{r^2}\left(1-\frac{2Me^{-\ell M/r}}{r}\right)=1-\mathcal{V}_\text{eff}.
\end{equation}
Where, following Ref.~\cite{Gralla:2019xty}, we introduce a dimensionful parameter $s$ in such a way that $k^\mu=dx^\mu/d\lambda = Edx^\mu/ds$ is the four-momentum of the photon, with $E$ the conserved energy and, $b$, the impact parameter, defined as $L/E$.  On the other hand, the right-hand-side of Eq.\eqref{SIVe1} is the well-known effective potential, $\mathcal{V}_\text{eff}$, crucial for analyzing the motion of photons.  In Fig.~\ref{Fig4}, left panel, we plot the behavior of $\mathcal{V}_\text{eff}/b^2$ for different values of the parameter $\ell$. Note that for values of $\ell$ closer to the extremal case (for $a=0.0$, $\ell_\text{extremal}=0.73576$), it is possible to identify the maximun and minimun values of $\mathcal{V}_\text{eff}$. Nevertheless, as $\ell$ becomes closer to the Schwarzschild case ($\ell=0$), the minimun goes deeper.  The figure also shows the effect of $\ell$ on the radial location of $\mathcal{V}_\text{max}$, which decreses as $\ell$ increases. The exact location of $\mathcal{V}_\text{max}$,  known as the radio of the photon sphere, $r_\text{ph}$, can be computed by the condition 
 \begin{equation}
 \label{SIVe2}
 \frac{\partial \mathcal{V}_\text{eff}}{\partial r}=0.
 \end{equation} 
 In the case of the EOS space-time, the roots are computed numerically, due to  the mass distribution $M(r) = Me^{-\ell M/r}$. The right panel of Fig.~\eqref{Fig4} shows clearly the effect of the free parameter $\ell$ on $r_\text{ph}$. As expected, the photon sphere radius radius reduces to that of Schwarzschild ($r^S_\text{ph}= 3M$) when $\ell=0$.  

\subsection{Angular diameter of the shadow}

\begin{figure*}[t]
\centering
\includegraphics[scale=0.5]{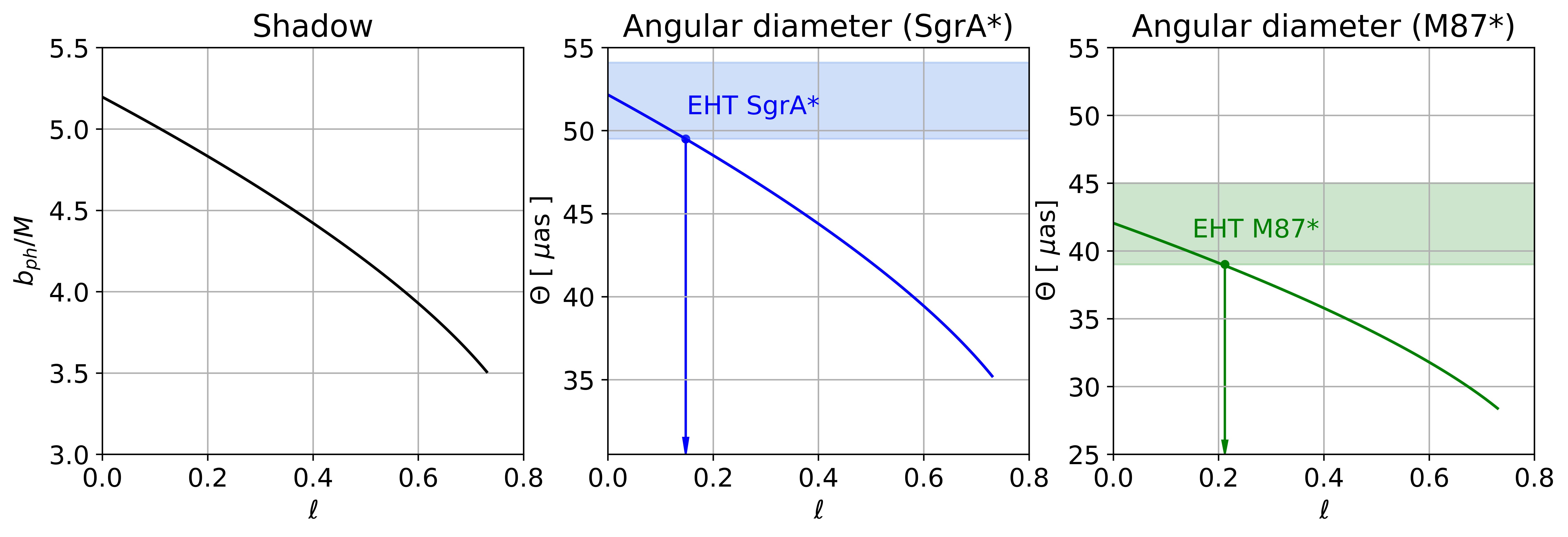}
\caption{Left panel: the apparent boundary (shadow) vs. $\ell$. Central panel: the angular diameter vs. $\ell$ for SgrA*. Right panle: the angular diameter vs. $\ell$ for M87*. In the figures, we show the angular diameter region given by the ETH observations for Sgr* and M87*, blue and green regions, respectively. We use $M=1$.}
\label{Fig5}
\end{figure*}

	From the physical point of view, the shadow of a BH involve a special curve on the image plane of the observer that Bardeen called the ``apparent boundary''~\cite{Bardeen:1973tla}. By definition, when traced backwards from its observation by a distant observer, a light ray from the apparent boundary will asymptotically approach a bound photon orbit (the photon sphere). Thus photons which are seen near the apparent boundary will have orbited the BH many times on their way to the observer. In the case of Schwarzschild, it is well-known that the bound orbit occurs at $r^S_\text{ph}=3M$, and that the apparent boundary is a circle of radius (impact parameter) $b^S_\text{ph}=3\sqrt{3}M$. To obtain the apparent boundary in the case of the EOS space-time, we consider first the photons moving on circular orbits around the BH. Hence, from the radial equation \eqref{SIVe1} with $dr/ds=0$ and evaluated at  $r=r_\text{ph}$, one obtains the shadow as 
\begin{equation}
\label{SIVe3}
b_\text{ph}=\frac{r_\text{ph}}{\sqrt{1-\frac{2M(r_\text{ph})}{r_\text{ph}}}}.
\end{equation}
In the left panel of Fig.~\ref{Fig5}, we show the behavior of $b_\text{ph}$ as a function of $\ell$. Note how the shadow shrinks from the Schwarzschild value ($b^S_\text{ph}\approx 5.19615$) as $\ell$ changes from zero to the extremal case $\ell_\text{extremal}=0.73576$. 

	Distant observers usually measure the angular diameter of the shadow, given by the relation~\cite{Perlick:2021aok}
\begin{equation}
\label{SIVe4}
\Theta = \frac{2b_\text{ph}}{D},
\end{equation}
where $D$ is the distance between the black hole and the observer. Giving $D$ in units of [Mpc] and introducing the mass of the black hole measured in solar masses, we obtain the angular diameter of the shadow in units of [$\mu$as] as
\begin{equation}
\label{SIVe5}
\frac{\Theta}{\text{ $\mu$as} }= \frac{6.191165 \times 10^{-8}}{\pi} \left(\frac{M}{M_\odot }\right) \left(\frac{D}{\text{Mpc}} \right)^{-1} \left(\frac{b_\text{ph}}{M}\right).
\end{equation}

    From the observational point of view, we know two BHs that could help to study the shadow size: Sgr A*, located at a distance $D=8.127 \text{ kpc}$ with a mass of $M = 4.14 \times 10^6 M_\odot$; and M87*, located at a distance $D=16.8 \text{ Mpc}$ with a mass of $M = 6.2 \times 10^9 M_\odot$. According to the observations from the Event Horizon Telescope (EHT) collaboration~\cite{EHT1, EHT2}, the angular diameter of the shadows reported for these two BHs are $51.8 \pm 2.3~\text{$\mu$as}$ and $42 \pm 3~\text{$\mu$as}$ with a $68\%$ of confidence, respectively. Therefore, it is possible to constrain the values of the free parameter $\ell$. In the central and left panels of  Fig.~\ref{Fig5}, we compare the predicted values of the angular size of the shadows for SgrA* and M87* with those reported by the EHT observations. From the figures, it is possible to see that  $0\leq \ell \leq 0.148$ and $0\leq \ell \leq 0.212$ for SgrA* and M87*, respectively.

	Now, we focus our attention on the appearence of the EOS BH. Our goal is to obtain the image cast by this RBH for an observer located at infinity. We begin by investigating the shadow and photon rings for different accretions flows. To do so, it is necessary to investigate how photons move in the sourroundings of the RBH.
	
\begin{figure*}[t]
\centering
\includegraphics[scale=0.5]{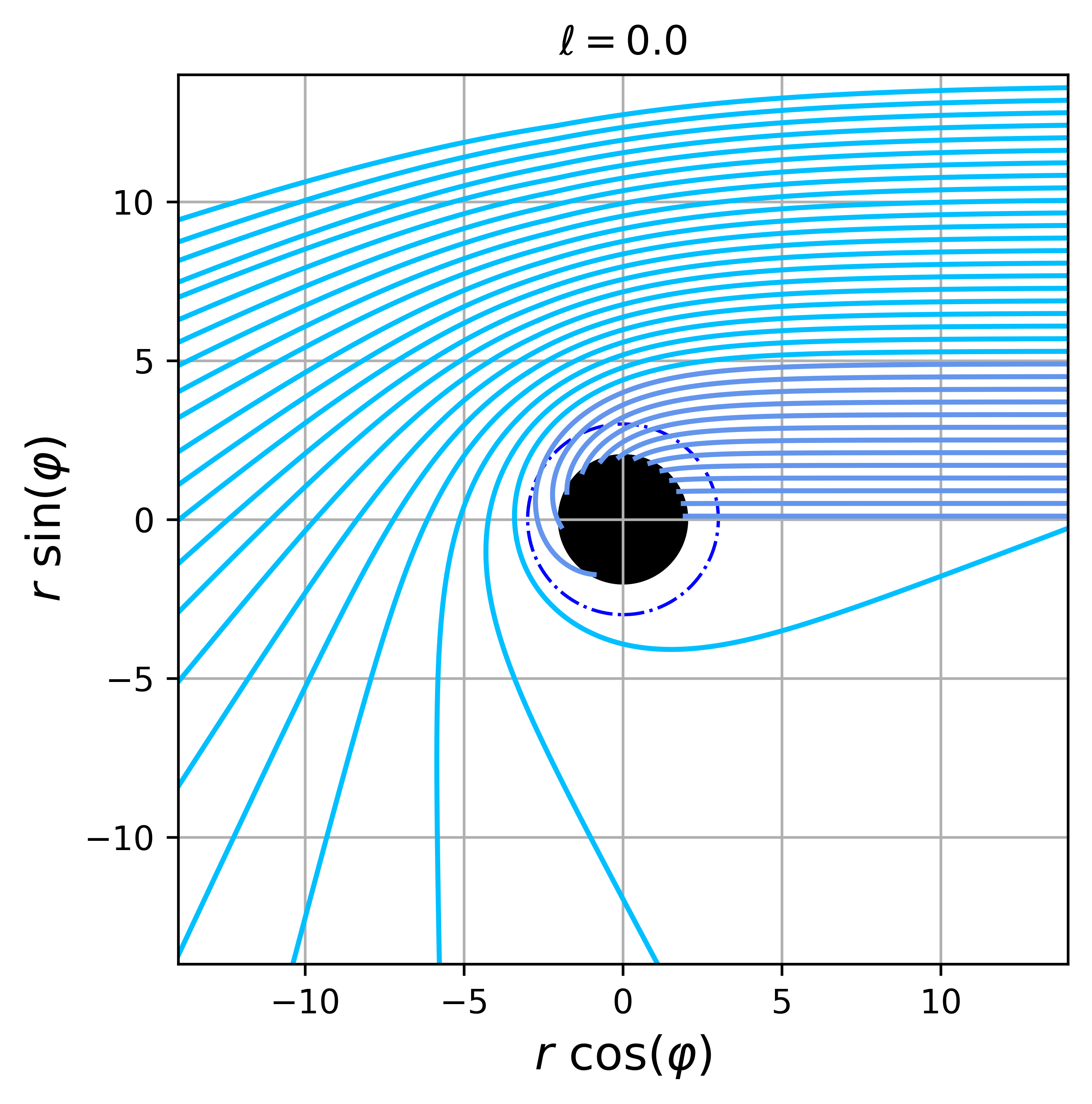}
\includegraphics[scale=0.5]{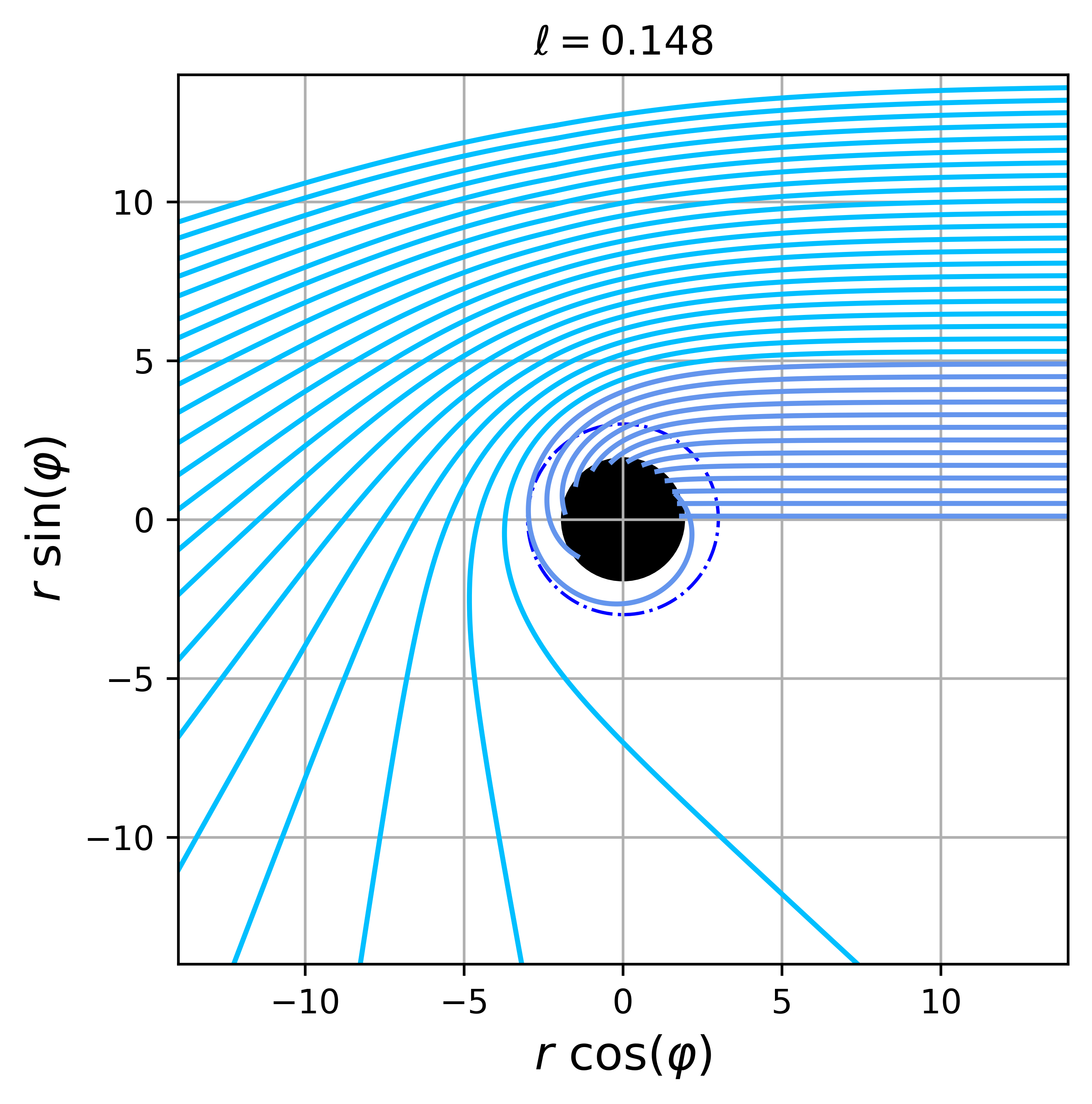}
\includegraphics[scale=0.5]{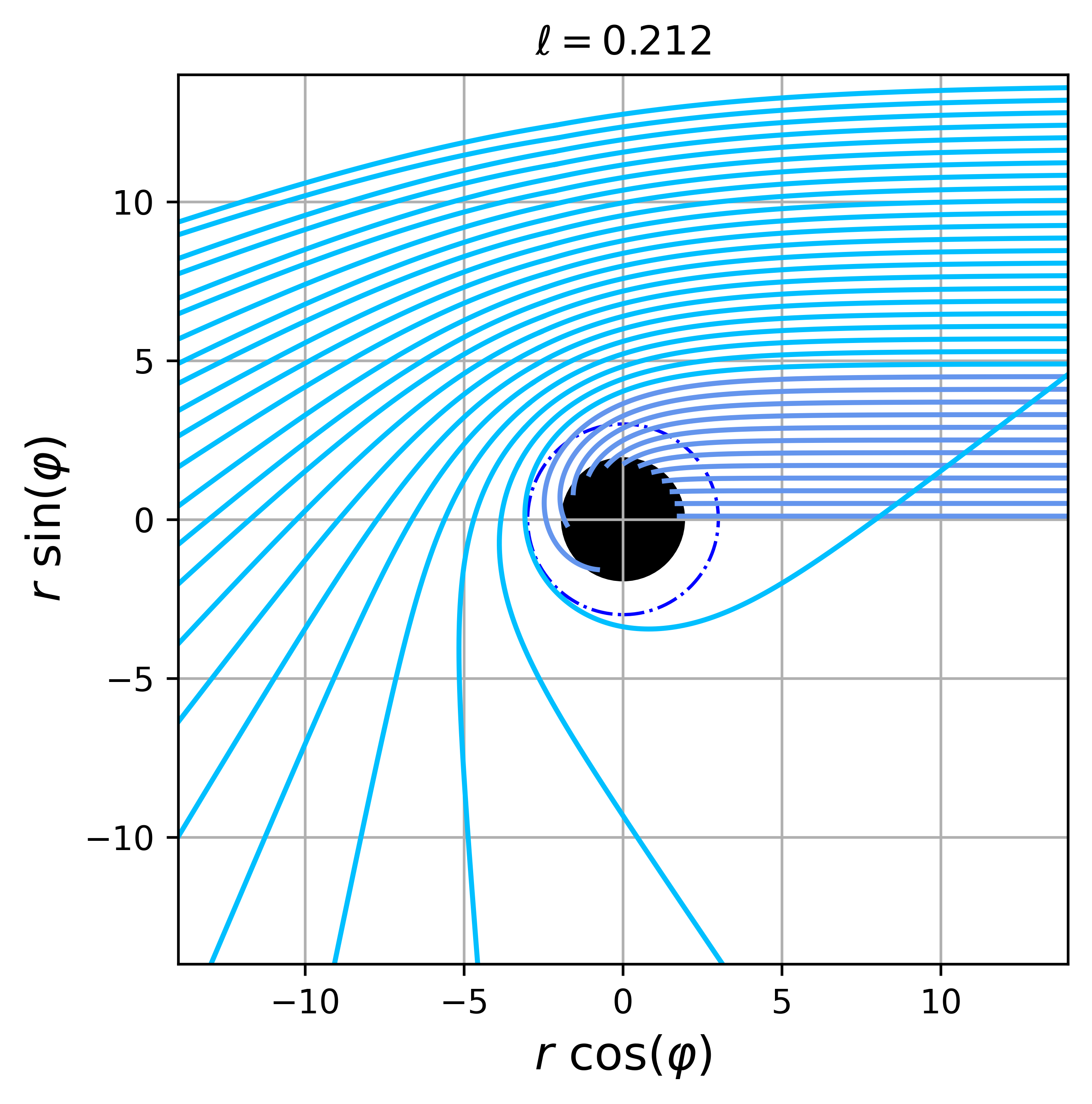}\\
\includegraphics[scale=0.5]{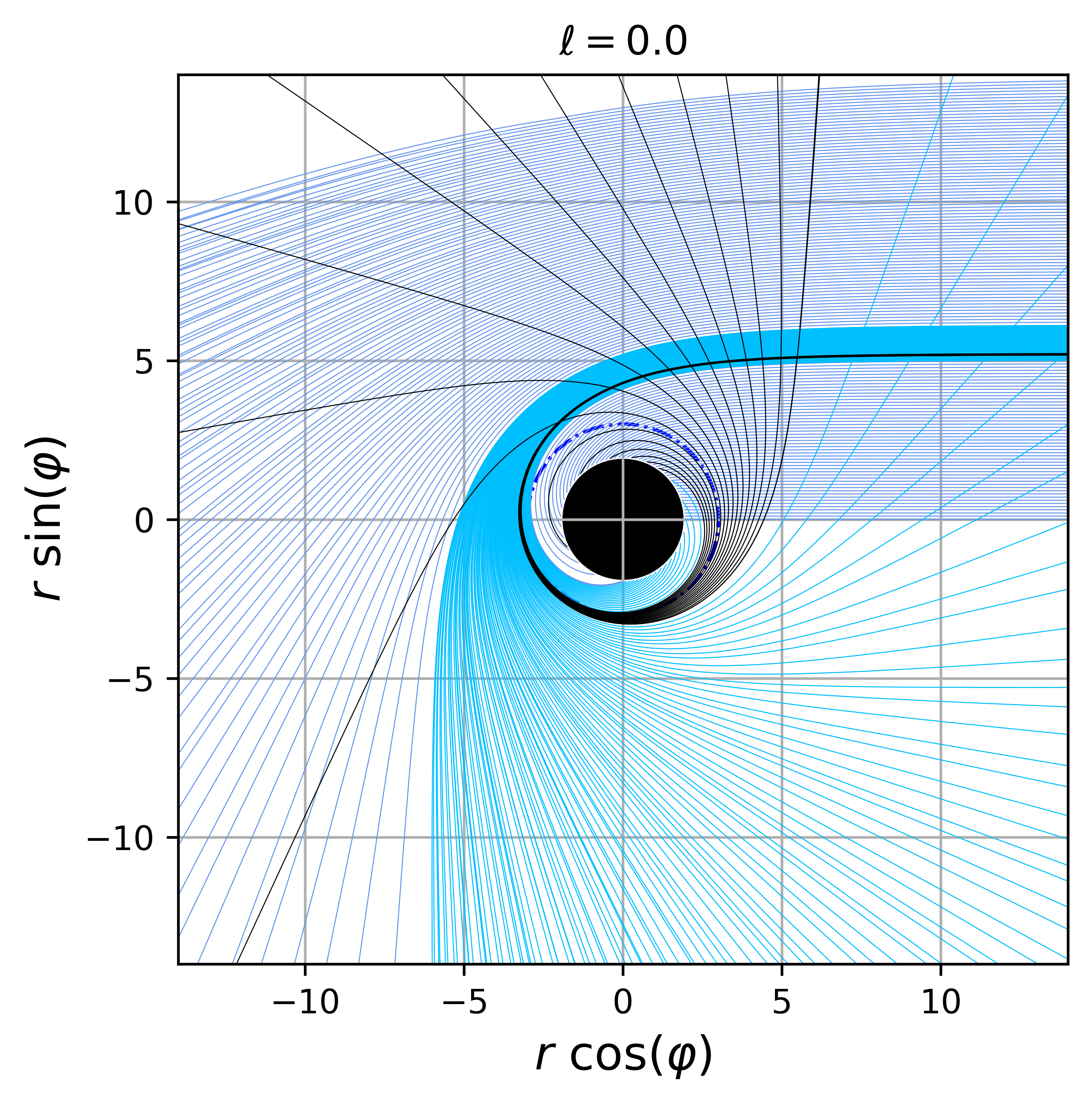}
\includegraphics[scale=0.5]{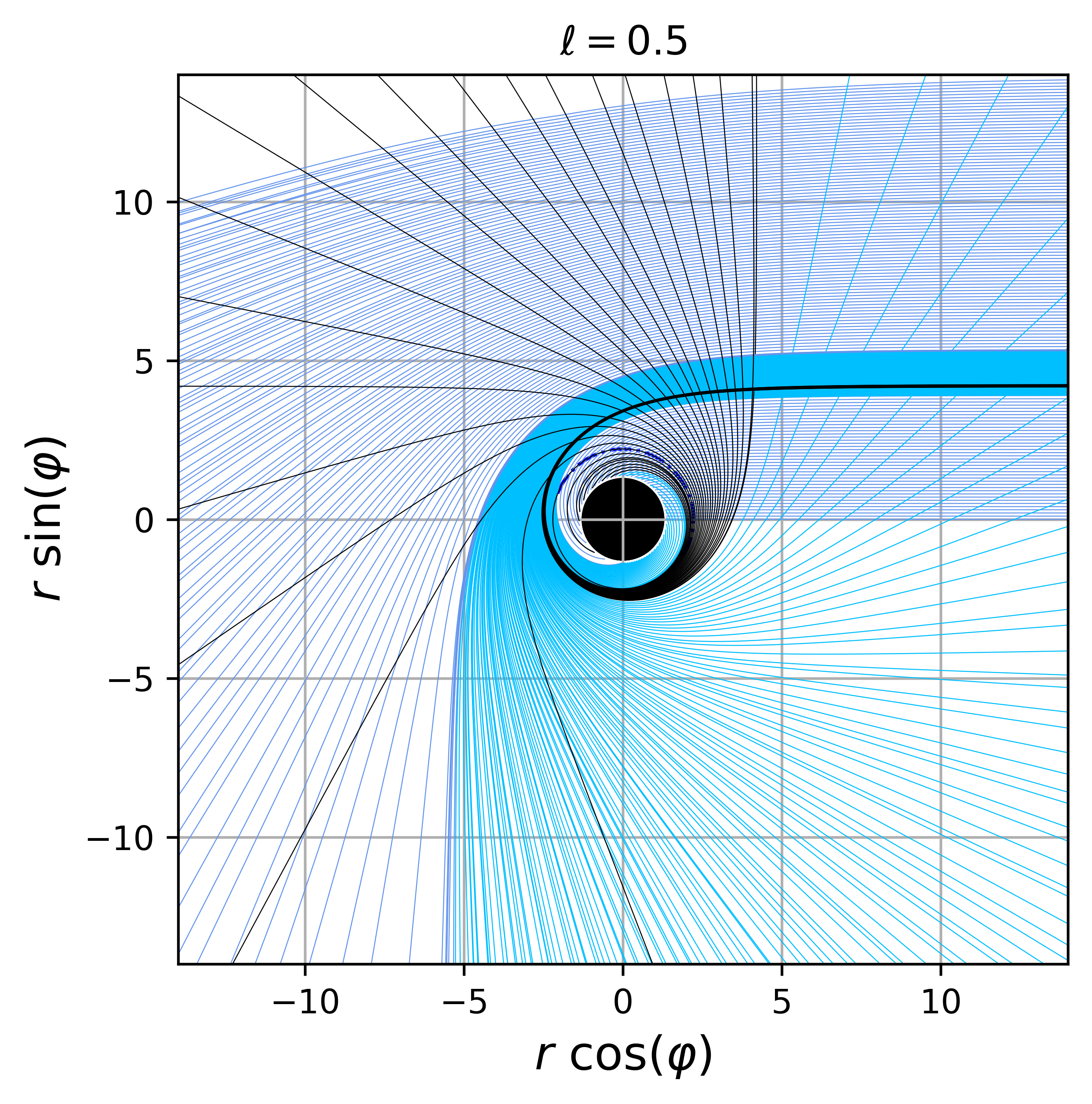}
\includegraphics[scale=0.5]{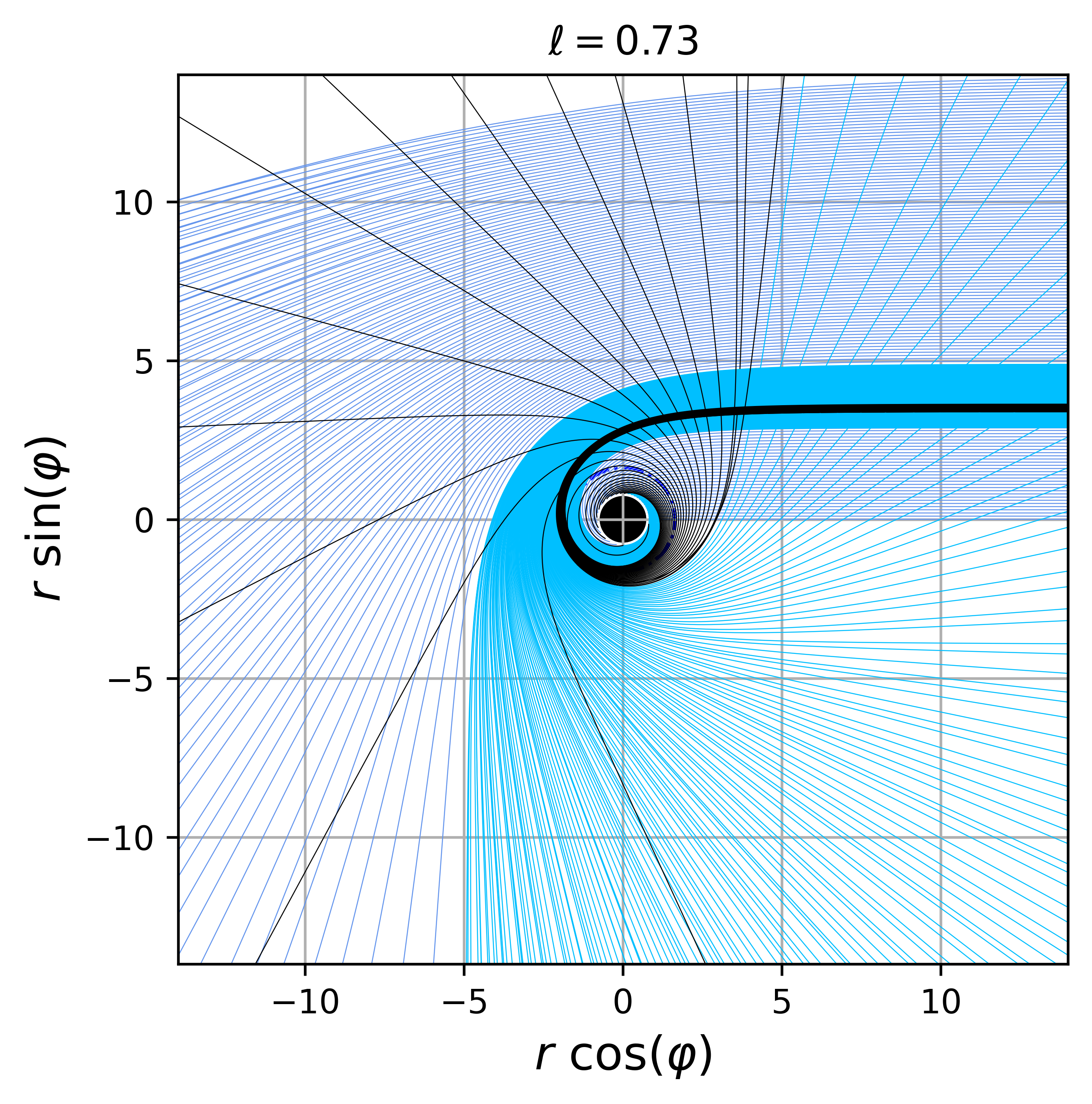}
\caption{First row: Examples of photon trajectories for different values of $\ell$. Second row: behavior of photon rays accoring to the direct (cornflowerblue), lensed ring (deepskyblue) and photon ring (black) emissions for different values of $\ell$. In the figures the BH is depicted by a black dot and the photon sphere is represented by a blue dash-dotted line. We use $M=1$.}
\label{Fig6}
\end{figure*}

	From Eq.~\eqref{SIIe9d}, with $\dot{\theta}=0$ and $\theta=\pi/2$, we obtain that 
\begin{equation}
\label{SIVe5}
\frac{d\varphi}{ds}=\frac{b}{r^2},
\end{equation}
where we also consider the dimensionful parameter $s$. Hence, with the help of Eq.~\eqref{SIVe1} and changing the radial coordinate to $u=1/r$, we obtain the differential equation:
\begin{equation}
\label{SIVe6}
\left(\frac{du}{d\varphi}\right)^2 = 2M(u)u^3-u^2+\frac{1}{b^2},
\end{equation}
with $M(u)=Me^{-\ell M u}$. After integration, we obtain the following expression:
\begin{equation}
\label{SIVe7}
\Delta\varphi=\pm\int^{u_\text{s}}_0\frac{du}{\sqrt{2M(u)u^3-u^2+\frac{1}{b^2}}}.
\end{equation}
Here, the $\pm$ signs determine the photon's direction of motion; i.e., ``minus'' when photons travel in a clockwise direction ($L<0$) and ``plus'' when traveling in a counterclockwise direction ($L>0$).  Note that the integral \eqref{SIVe7} traces back the photon trajectory form the observer located at infinty ($u=0$) to the source, $u_\text{s}$, which depends on the impact parameter $b$. In the case of photons with $b<b_\text{ph}$, we already know they will be captured by the BH. In this sense, $u_\text{s}$ must be closer to the photon sphere radius, $u_\text{ph}$. On the other hand, photons trajectories with $b>b_\text{ph}$ have a turning point, $u_t$,  which satisfies the following condition:
\begin{equation}
\label{SIVe8}
2M(u_t)u^3_t-u^2_t+\frac{1}{b^2}=0.
\end{equation}  
In this sense, for $b>b_\text{ph}$, the integral in Eq.~\eqref{SIVe7} gives us the first half of the motion by tracing  back the photon's trajectory from the observer at infinity to the turning point, $u_t$. Then, by symmetry, one obtains the second half of the motion. In Fig.~\ref{Fig6},  we show the motion of photons near the RBH for different impact parameters $b$ and different values of $\ell$. 

\subsection{Shadow and photon rings}

	From an astrophysical perspective, matter in the form of gas or plasma surrounds BHs, forming what is known as an accretion disk. The light rays radiated by the disk illuminate the BH, enabling the formation of an image for a distant observer. In this subsection, we examine the shadow and photon rings of the EOS BH. Initially, we consider static and infalling spherical accretions, and then we move on to the more realistic scenario in which we consider a thin disk accretion.    

\subsubsection{Static spherical accretion}
	The observed specific intensity $I(\nu_0)$ is defined as the integral of the specific emissivity $j(\nu_e)$ along the photon trajectory $\gamma$~\cite{Zeng:2020dco, Jaroszynski:1997bw, Bambi:2013nla}
\begin{equation}
\label{SIVe9}
I(\nu_0)  = \int_\gamma g^3 j(\nu_e) d\ell_\text{prop},
 \end{equation}
where $\nu_0$ and $\nu_e$ are the observed and emitted photon frequencies; respectively. The redshift factor, $g$, is the ratio between these frequencies and is written in terms of the metric tensor as
\begin{equation}
\label{SIVe10}
g=\frac{\nu_0}{\nu_e} = \sqrt{-g_{tt}}.
\end{equation}
The proper length, $d\ell_\text{prop}$, can be computed from the line element \eqref{SIIIe1} taking $dt=d\theta =0$ and $\theta = \frac{\pi}{2}$,
\begin{equation}
\label{SIVe11}
d\ell_\text{prop} = \sqrt{g_{rr} + r^2 \left( \frac{d\varphi}{dr} \right)^2} dr,
\end{equation}
where the derivative inside the square root comes from the equations of motion (with $a=0$) as
\begin{equation}
\label{SIVe12}
\left(\frac{d\varphi}{dr}\right)^2 = -\frac{g_{rr}}{r^2} \left( \frac{b^2 g_{tt}}{r^2 + b^2 g_{tt}}\right). 
\end{equation}
Usually,  the emissivity per unit volume (measured in the static frame) for the static spherically symmetric accretion flow is modeled as $j(\nu_e) \propto 1/r^2$ for monochromatic radiation emitted with frequency $\nu_e$. Hence, after the change of variable $u=1/r$, the specific intensity takes the form
\begin{equation}
\label{SIVe13}
I_o =I(\nu_0) = \int^{u_s}_0\left[\frac{(1-2M e^{-\ell M u} u)^2}{\left[b^2(2M e^{-\ell M u} u^3-u^2)+1\right]}\right]^{1/2}du,
\end{equation}
where the observer is at infity ($u=0$). 

\begin{figure}
\centering
\includegraphics[scale=0.5]{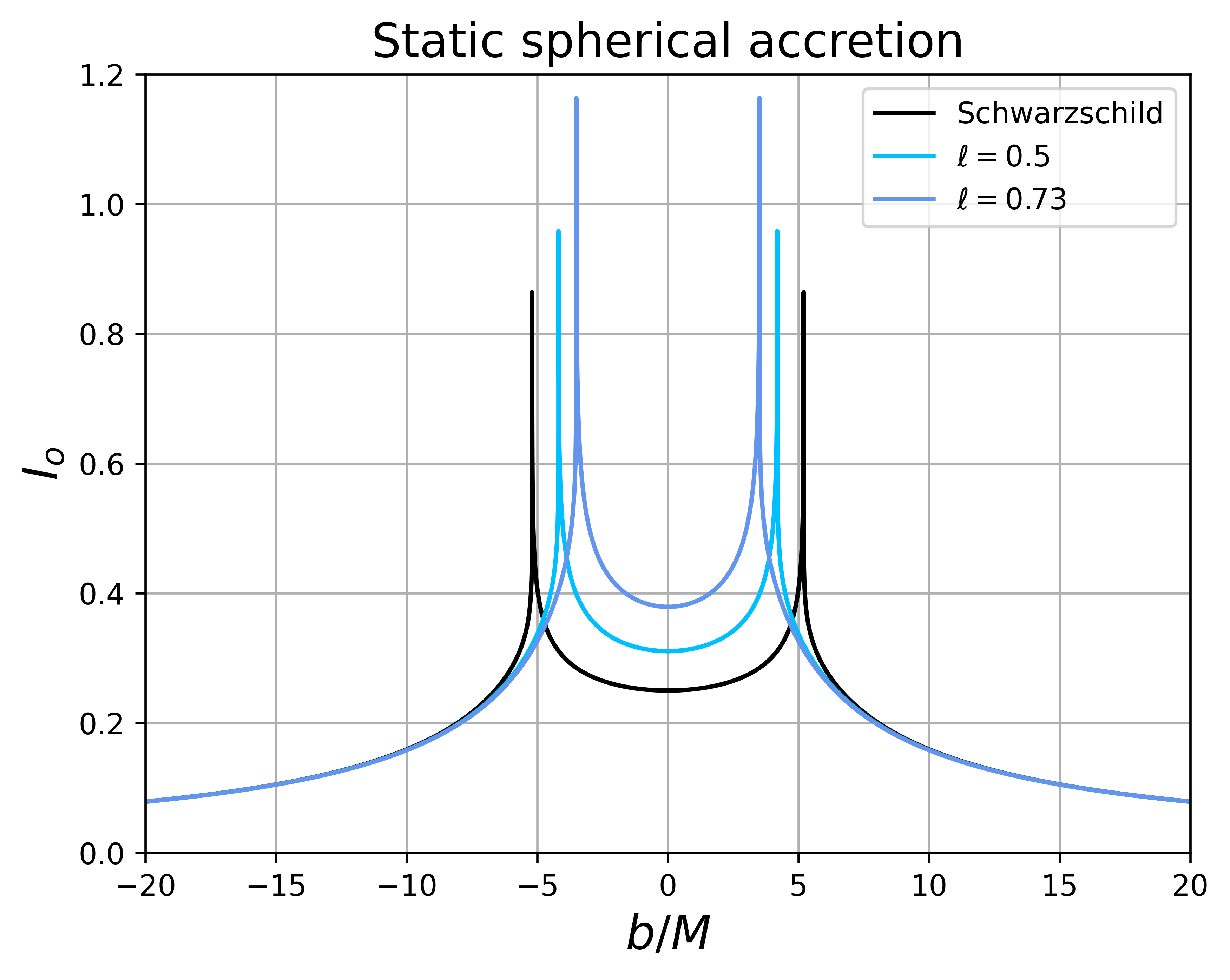}
\caption{Dependence of the observer specific intensity $I_o$ on the impact parameter for different values of $\ell$ in the case of the static spherical accretion. We use $M=1$.}
\label{Fig7a}
\end{figure}

	The observed specific intensity depends on the impact parameter $b$ and the free parameter $\ell$. In Fig.~\ref{Fig7a}, we show the behavior of $I_o$ radiated by a static spherical accretion flow surronding the EOS BH (with $a=0$). From the figure, it is possible to see how the specific intensity increases as $|b| < b_\text{ph}$. At $b=b_\text{ph}$, the intensity diverges since the photon is on the photon sphere. Then, for $|b|>b_\text{ph}$, the intensity decreases. This behavior does not depend on the free parameter $\ell$. Nevertheless, the free parameter $\ell$ does affect the magnitude of $I_o$, which increases as $\ell$ goes from $0$ (Schwarzschild) to $0.73$.  Note how the shadows shrinks as the free parameter increases its value in agreement with the first panel of Fig.~\ref{Fig5} .  

\subsubsection{Infalling sphercial accretion}
Since matter, such as plasma can be trapped by the BH with its initial velocity, the infalling spherical accretion is a more realistic scenario to model the observer's specific intensity. Similar to the static spherical accretion, $I_o$ can be calculated using Eq.~\eqref{SIVe9}.  Nevertheless, the red-shift factor, $g$, takes the form~\cite{Bambi:2013nla}
\begin{equation}
\label{SIVe14}
g=\frac{k_\mu u^\mu_0}{k_\nu u^\nu_e},
\end{equation}   
where $u^\mu_0$ and $u^0_e$ are the four velocity of the distant oberver and accretion matter, respectively. Following, Refs.~\cite{Bambi:2013nla} and \cite{Hu:2022lek}, we assume a stationary distance observer; i. e., $u^\mu_0=(1,0,0,0)$. 

	On the other hand, in the EOS BH (with $a=0$), the components of the four-velocity for the accretion matters are given by
\begin{equation}
\label{SIVe14}
\begin{aligned}
u^t_e&=\left(1-\frac{2M(r)}{r}\right)^{-1},\\\\
u^r_e&=-\sqrt{\frac{2M(r)}{r}},\\\\
u^\theta_e&=u^\varphi_e=0.
\end{aligned}
\end{equation}
Hence, the redshift factor reduces to
\begin{equation}
\label{SIVe15}
g=\frac{1}{u^t_e+\frac{k_r}{k_t}u^r_e}.
\end{equation}
The relationship between $k_r$ and $k_t$ can be inferred from the null condition $k_\alpha k^\alpha=0$, from which
\begin{equation}
\label{SIVe16}
k_r=\pm \sqrt{\left(1-\frac{2M(r)}{r}\right)^{-1}\left[\left(1-\frac{2M(r)}{r}\right)^{-1}-\frac{b^2}{r^2}\right]}.
\end{equation}
Here the ``plus'' (``minus'') sing corresponds to photons approaching to (going away from) the BH. 

\begin{figure}
\centering
\includegraphics[scale=0.5]{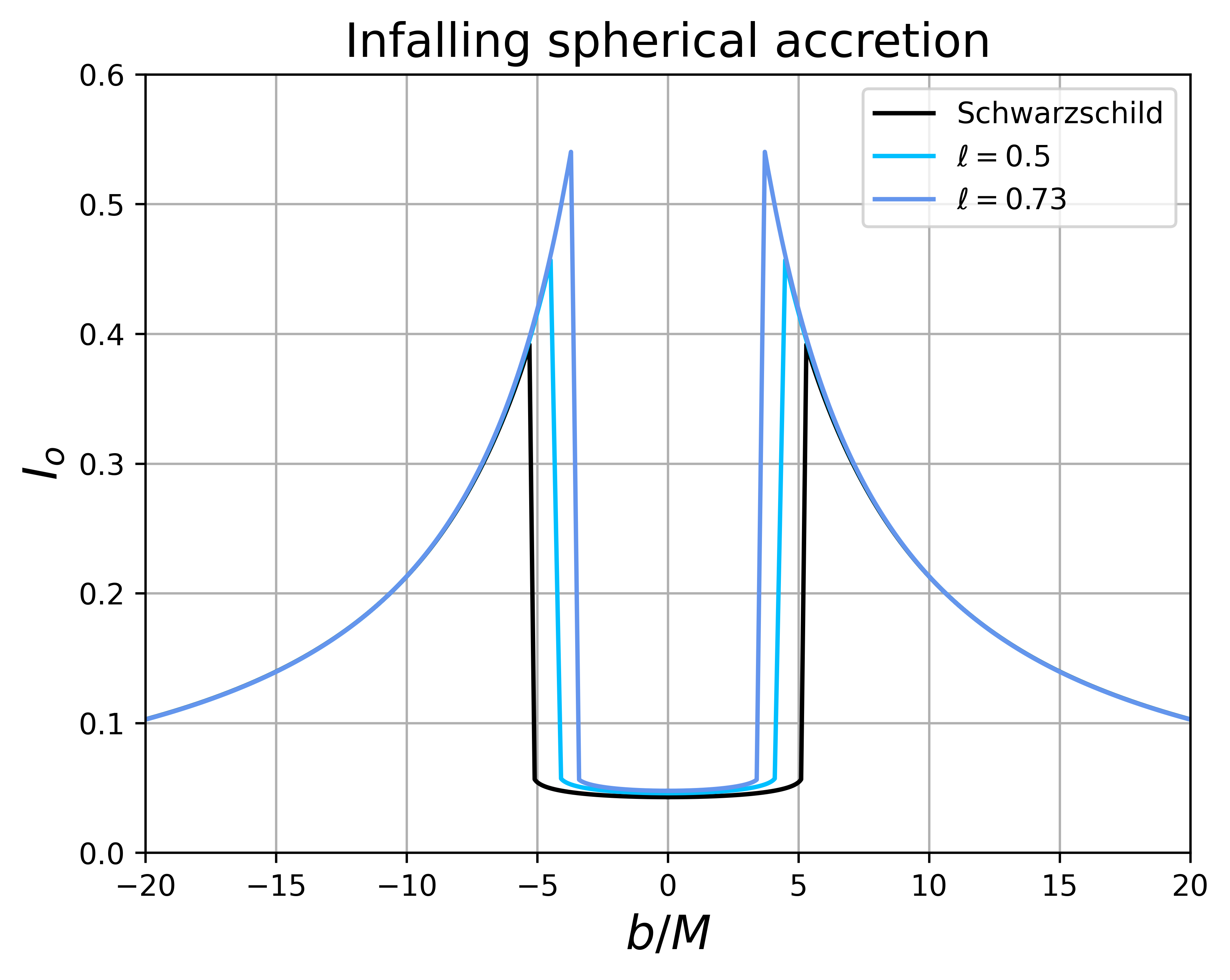}
\caption{Dependence of the observer specific intensity $I_o$ on the impact parameter for different values of $\ell$ in the case of the infalling spherical accretion. We use $M=1$.}
\label{Fig7b}
\end{figure}

\begin{figure*}[t]
\centering
\includegraphics[scale=0.41]{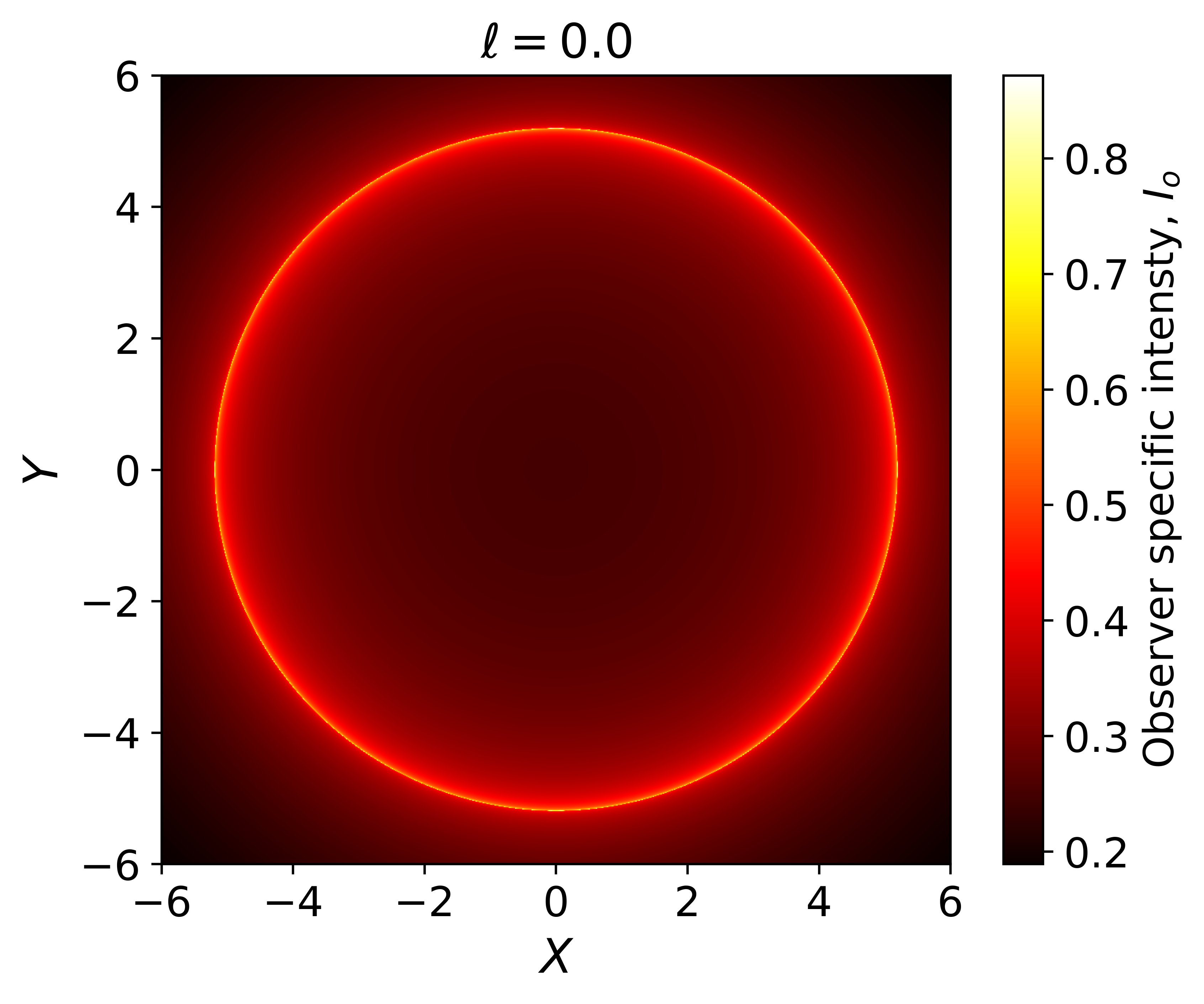}
\includegraphics[scale=0.41]{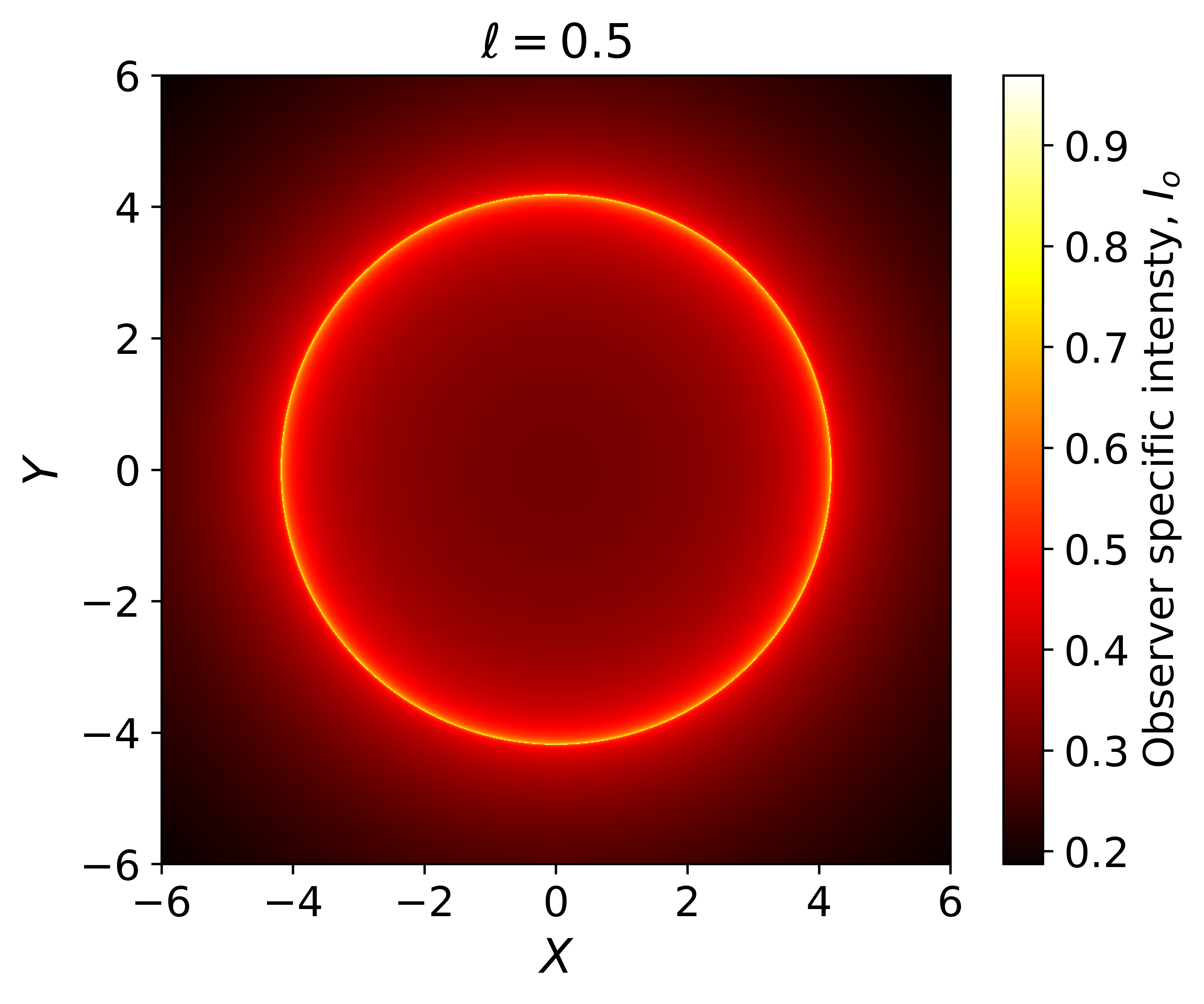}
\includegraphics[scale=0.41]{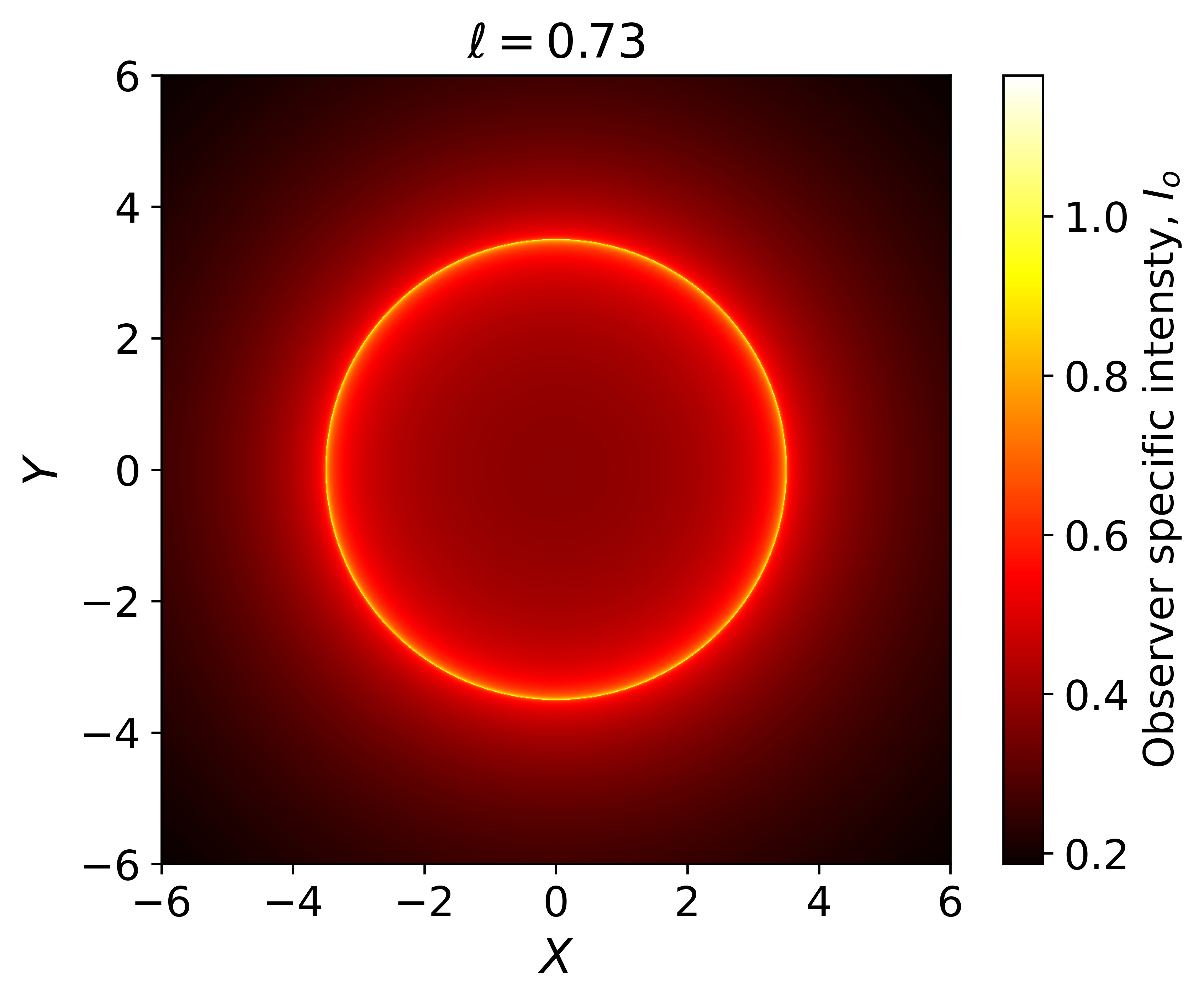}\\
\includegraphics[scale=0.4]{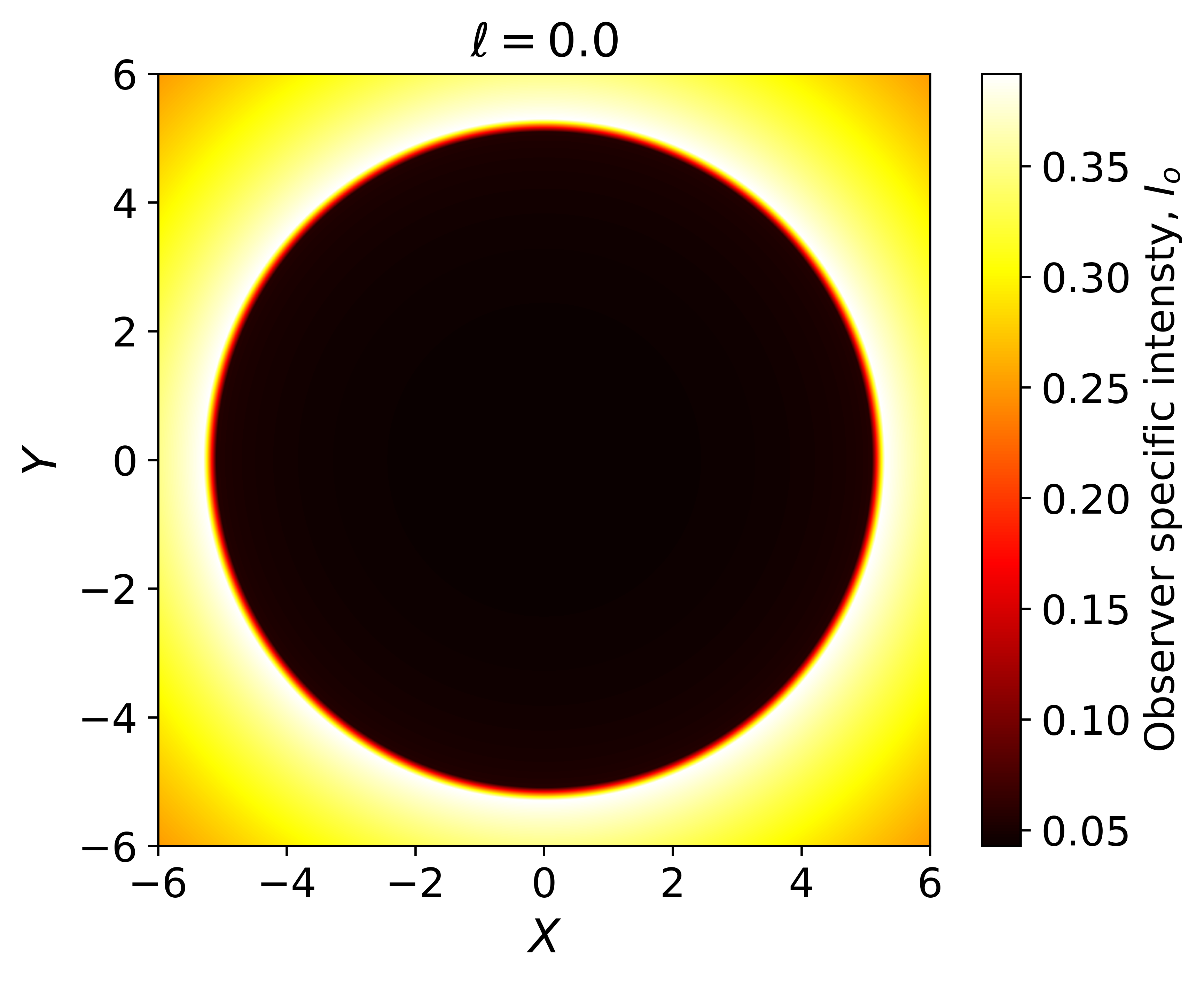}
\includegraphics[scale=0.4]{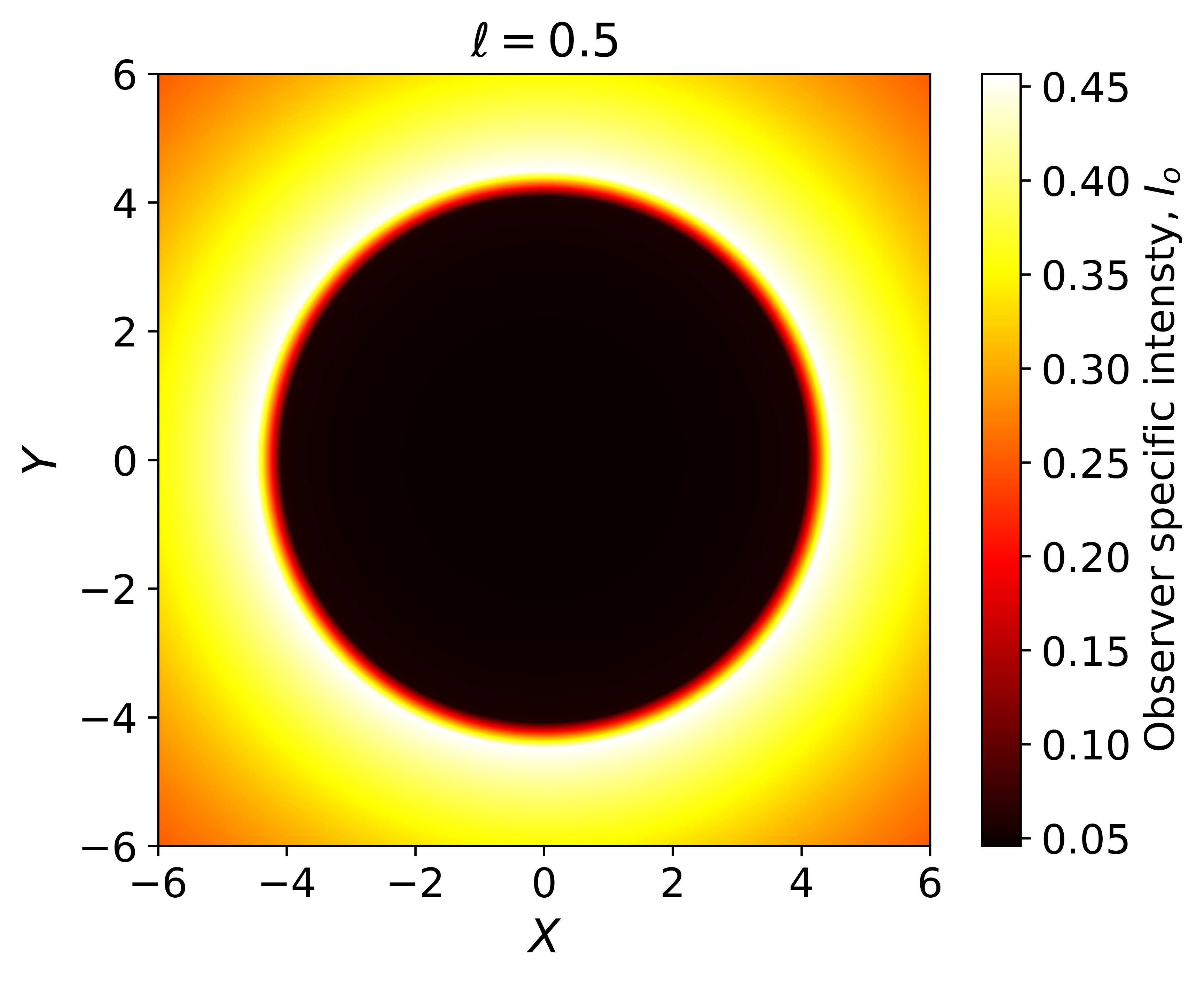}
\includegraphics[scale=0.4]{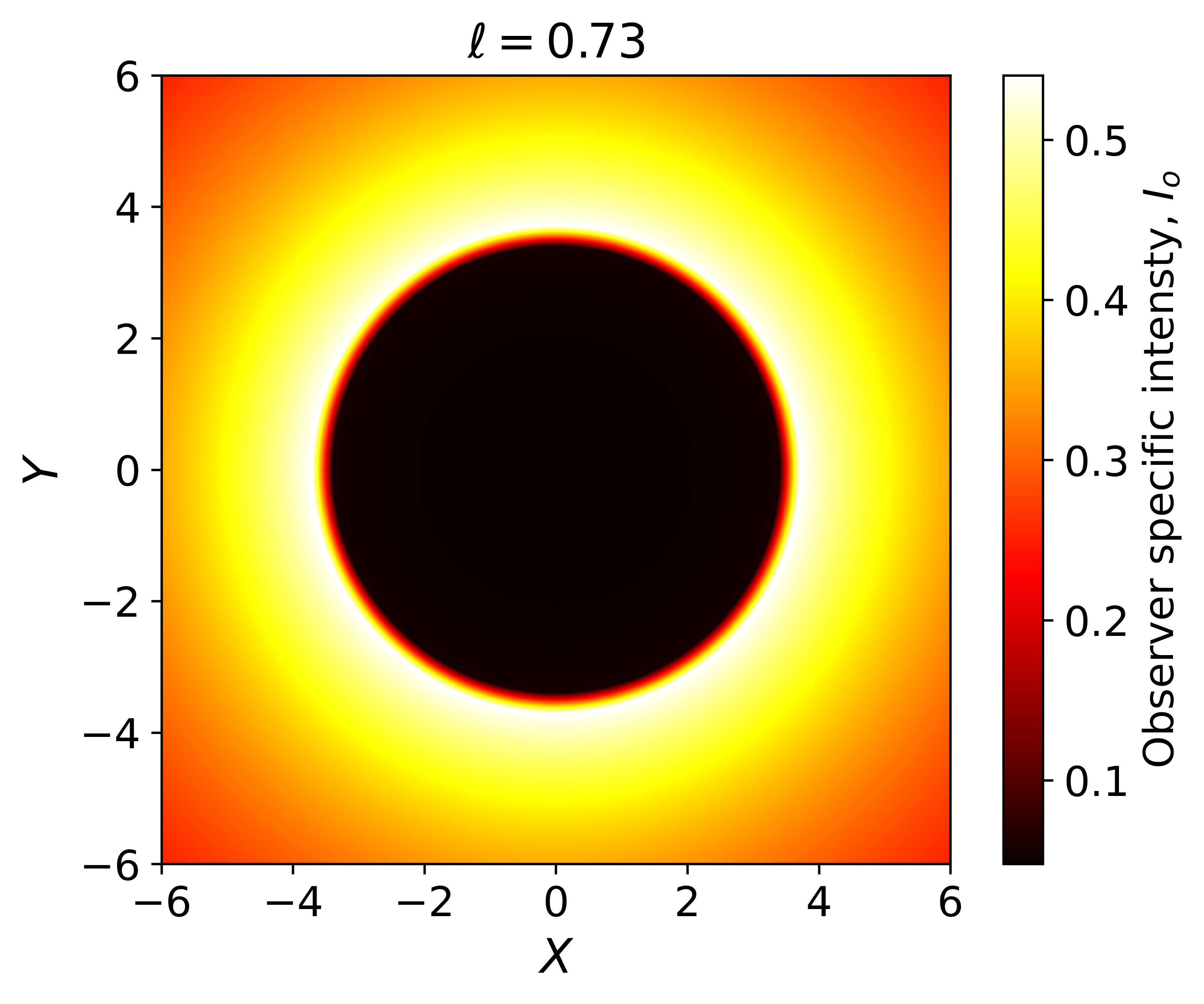}
\caption{First row: The observed intensity, $I_o$, for the static spherical accretion. Second row: The observed intensity, $I_o$, for the infalling spherical accretion. We consider different values of the free parameter $\ell$. We use $M=1$.}
\label{Fig8}
\end{figure*}

The infinitesimal proper length is given by~\cite{Bambi:2013nla}
\begin{equation}
\label{SIVe16}
dl_\text{prop}=\frac{k_t}{|k_r|}dr.
\end{equation}
Therefore, the observed specific intensity for the infalling spherical accretion takes the form
\begin{equation}
\label{SIVe17}
I_0=\int_\gamma\frac{g^3 k_t}{r^2|k_r|}dr=\int^{u_s}_0\frac{\left[1-2M(u)u\right]^4}{\Delta^{1/2} \left[1+\sqrt{2M(u)u}\Delta^{1/2}\right]^3}du,
\end{equation}
where
\begin{equation}
\label{SIVe18}
\Delta = b^2(2M e^{-\ell M u} u^3-u^2)+1.
\end{equation}
In Fig.~\ref{Fig7b}, we show the behavior of $I_o$ radiated by an infalling spherical accretion for the EOS BH (with $a=0$). Similar to the static spherical accretion, the figure shows how the specific intensity increases as $|b| < b_\text{ph}$, while it decreases for $|b|>b_\text{ph}$. At the photon sphere, the intensity diverges. Again, this behavior is independent of the free parameter $\ell$. However, the magnitude of $I_o$ increases when we consider different values of the free parameter $\ell$.

	Using the condition $b^2=X^2+Y^2$, we obtain the distribution of specific intensity $I_o$ on the observers' plane $(X, Y)$ for the static and infalling spherical accretions; see the first and second row of Fig.~\ref{Fig8}, respectively. In the center of the figures, a faint luminosity region corresponding to the BH shadow is visible. Moreover, the figure clearly shows the previous results, i.e., the shadow shrinks while the magnitude of the specific intensity increases as the free parameter $\ell$ changes from $0$ to $0.73$. 

	In the case of spherical static accretion, the figure shows the brightest ring surrounding the shadow: the photon ring; note that the photon ring for $\ell=0.73$ is brighter than the other two cases; this has to do with the fact that photons move with higher energy ($b=L/E$) closer to the BH's center. 
	
	On the other hand, in the case of spherical infalling accretion, the luminosity of the central part is much lower than the spherical static accretion but significantly higher outside the shadow region; this is due to the Doppler effect caused by the initial velocity of the matter in the spherical infalling accretion case~\cite{Hu:2022lek}. 
	
\begin{figure*}[t]
\centering
\includegraphics[scale=0.4]{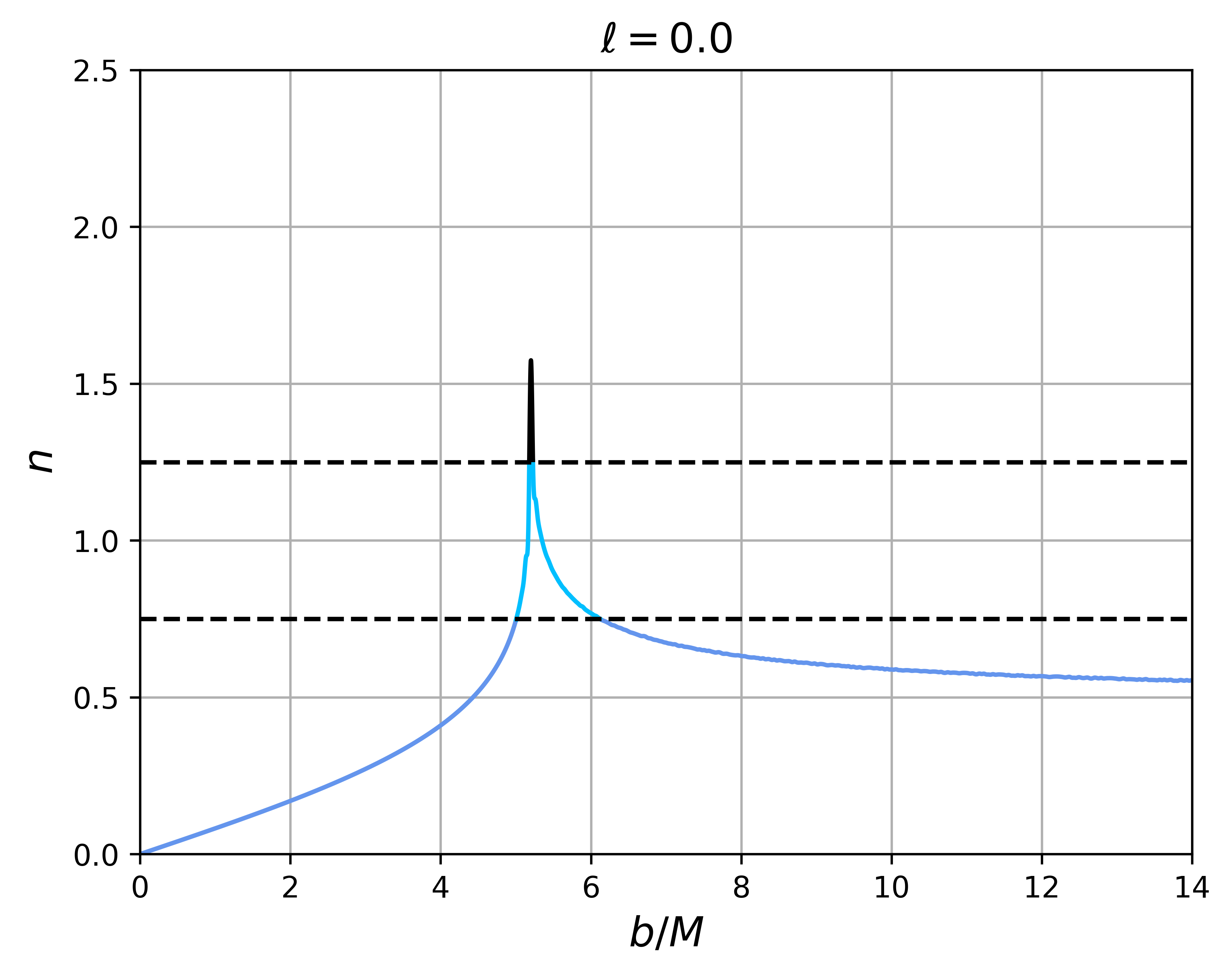}
\includegraphics[scale=0.4]{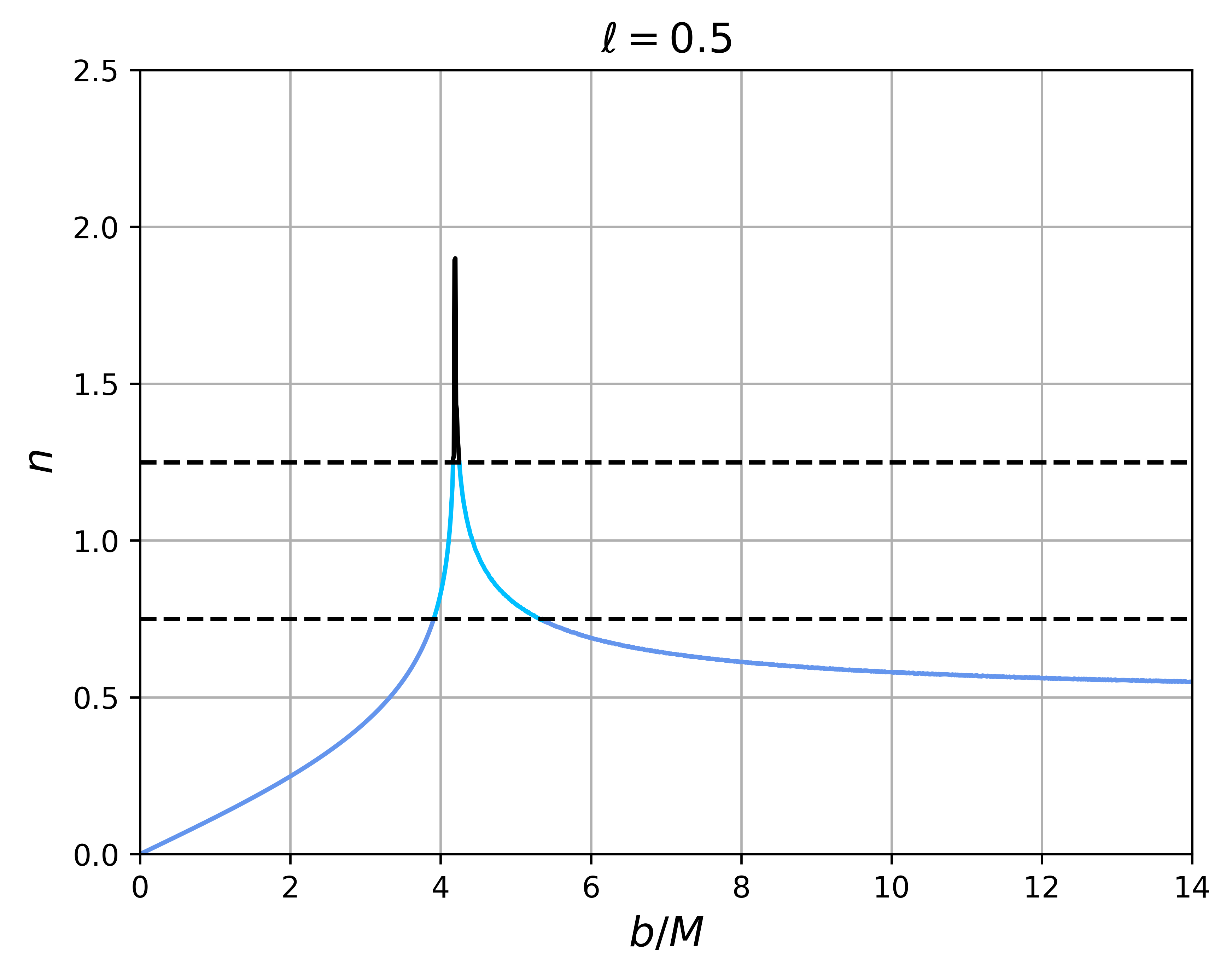}
\includegraphics[scale=0.4]{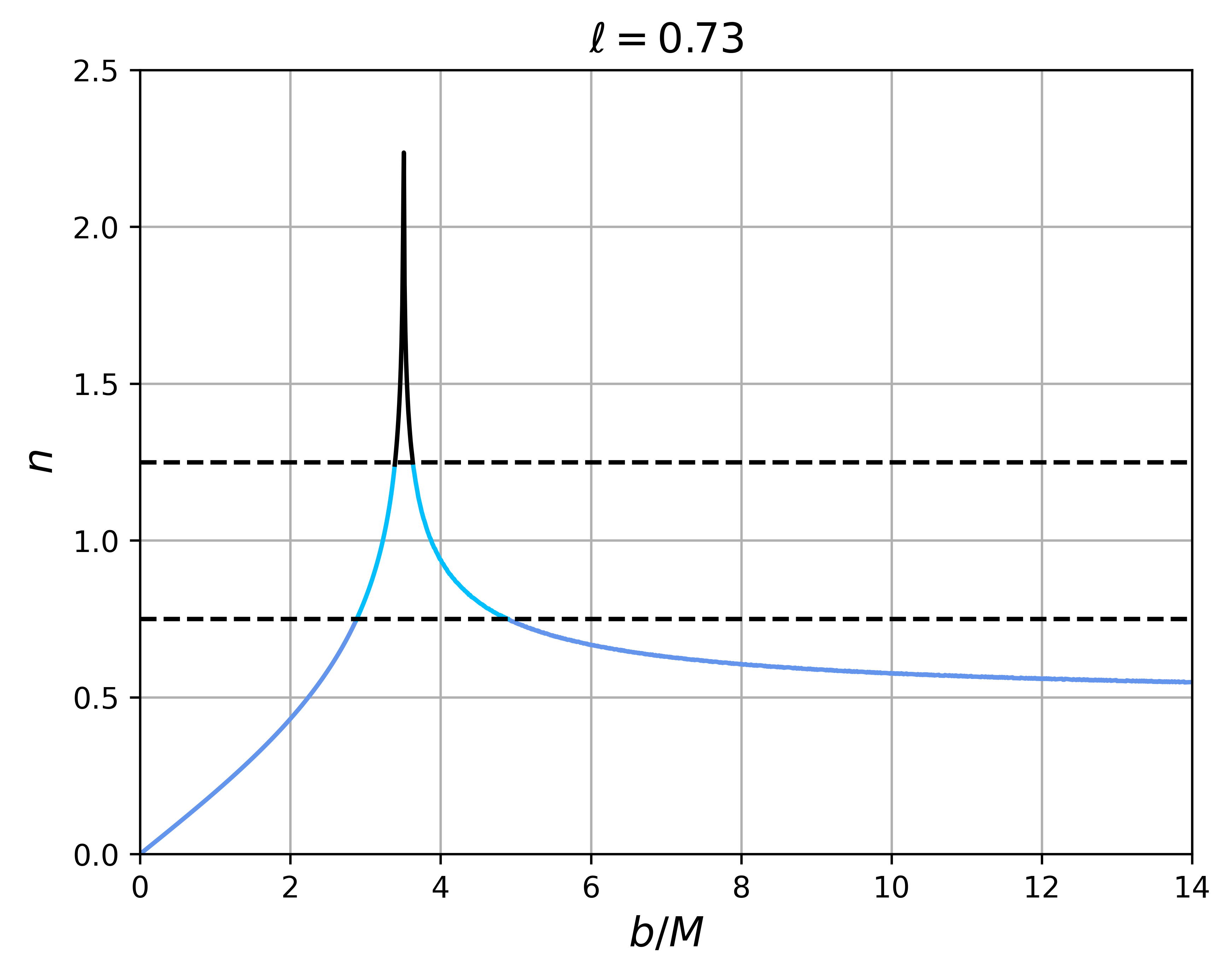}\\
\includegraphics[scale=0.47]{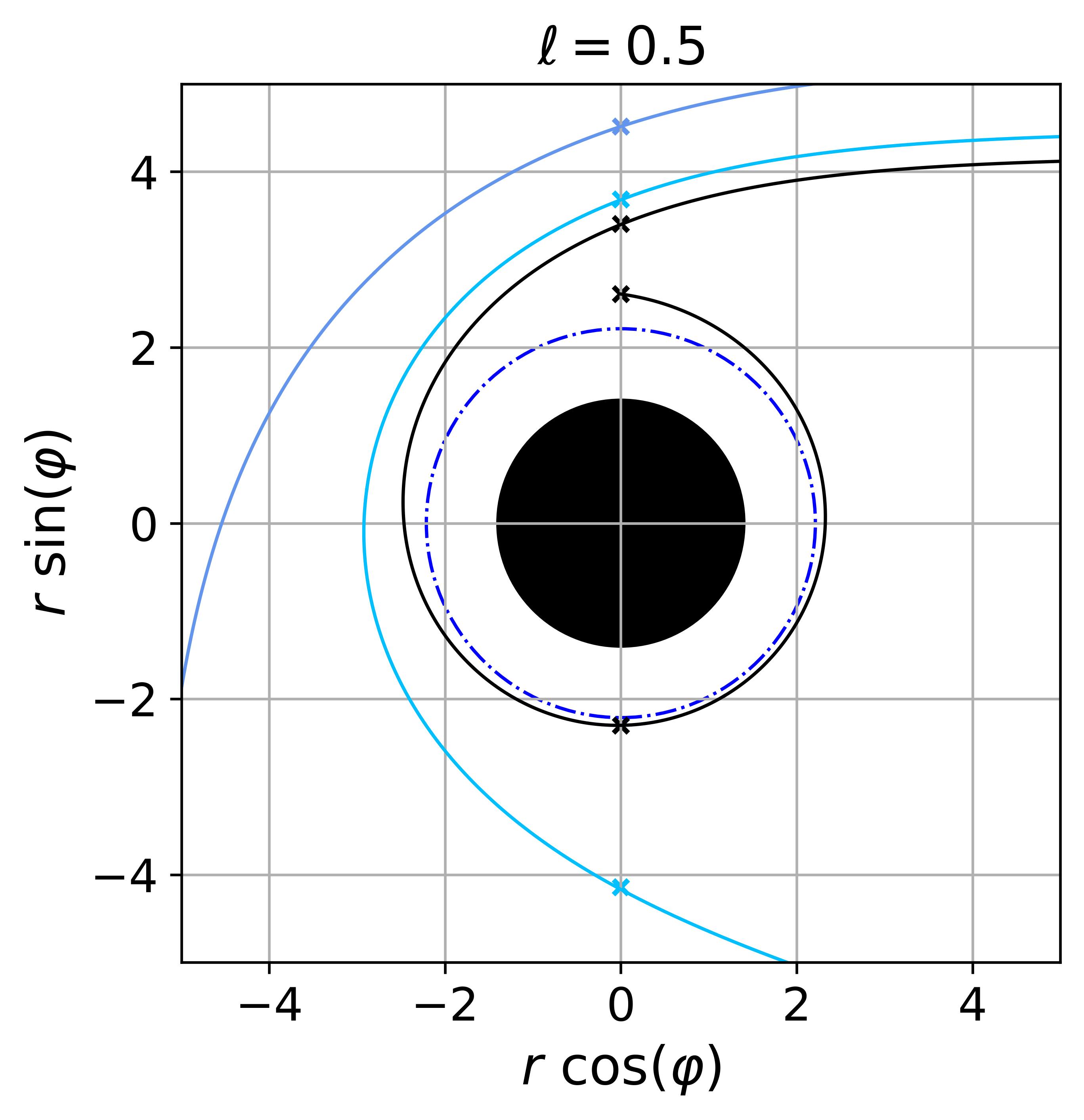}
\includegraphics[scale=0.47]{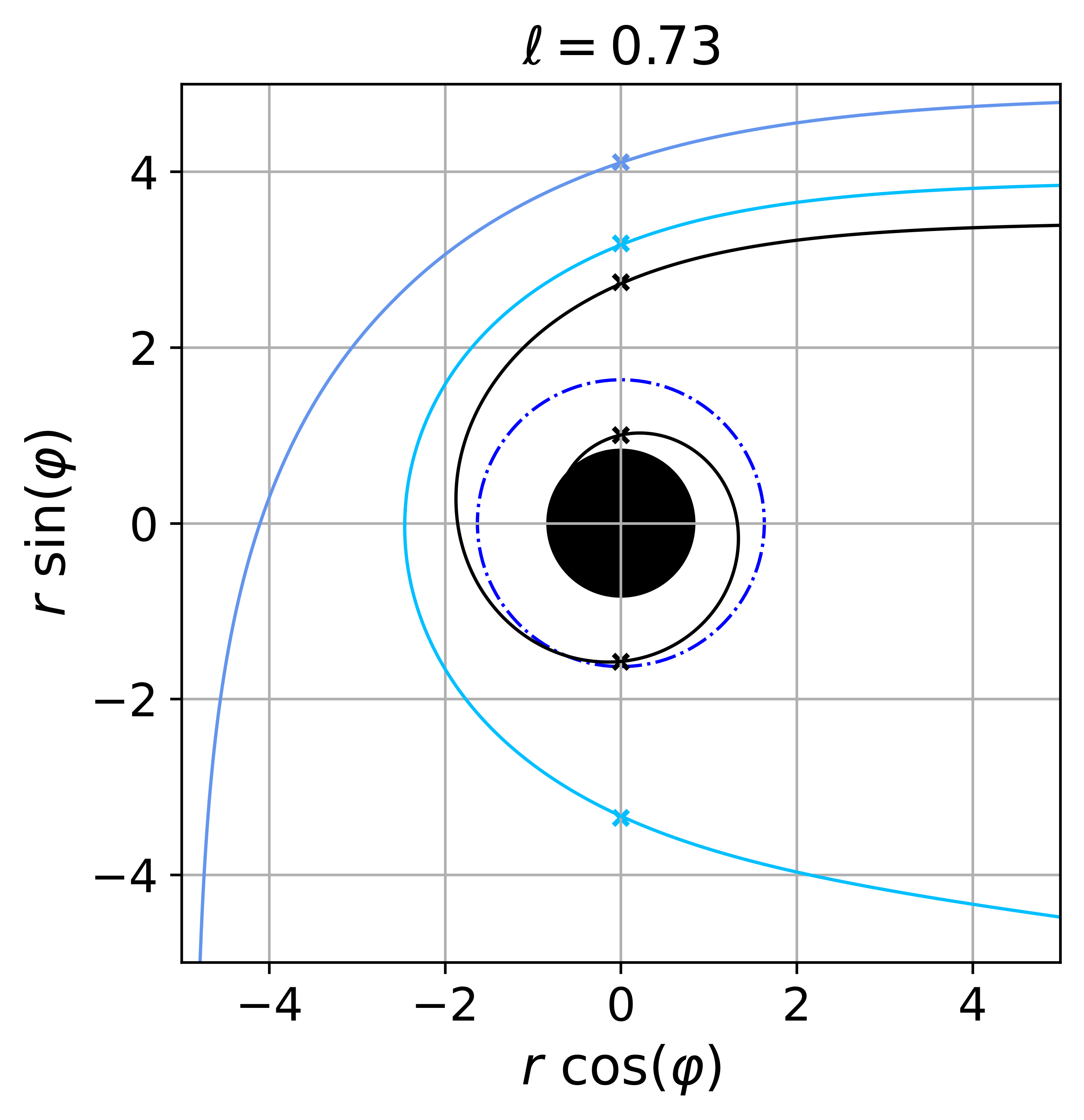}
\includegraphics[scale=0.47]{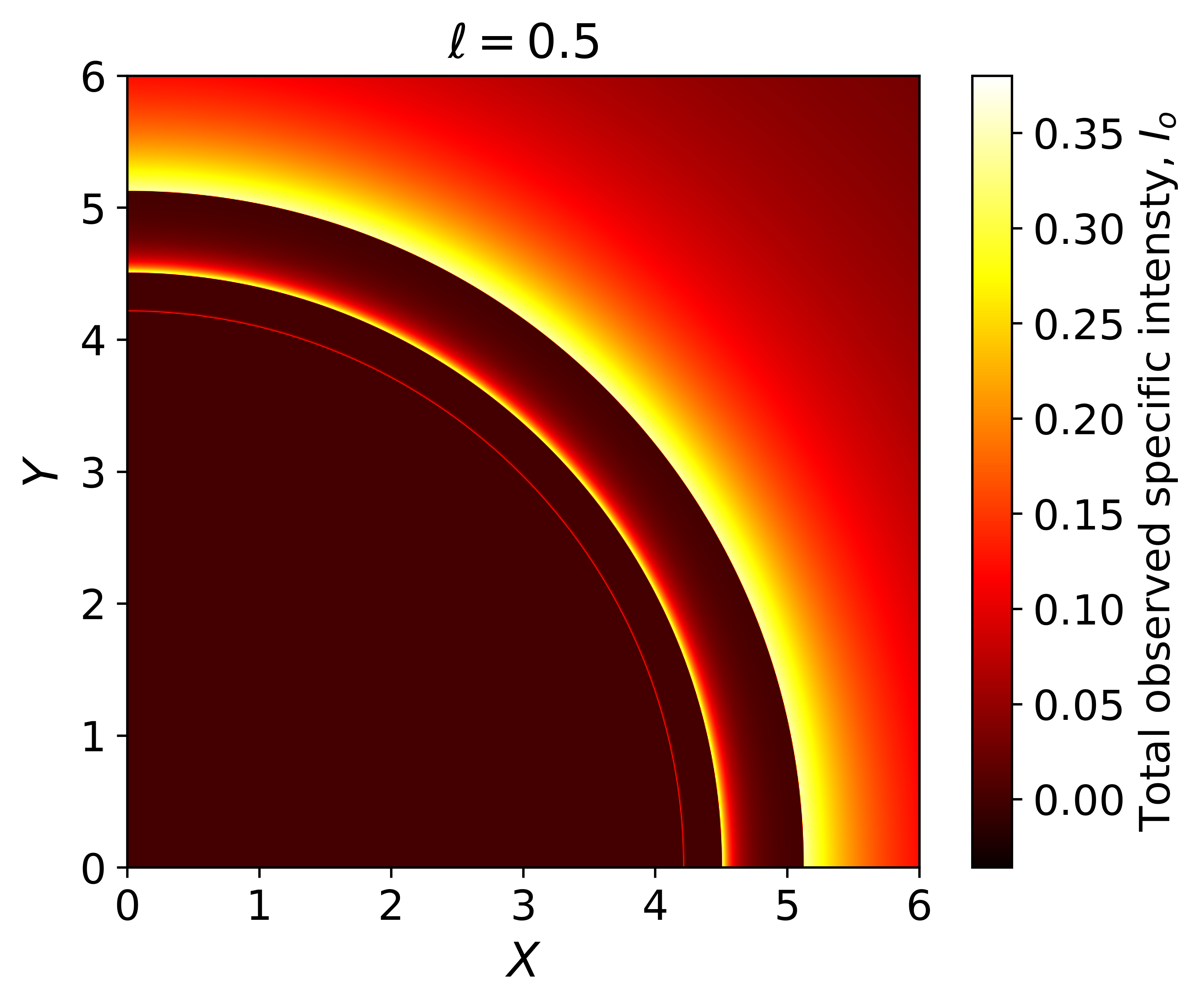}
\caption{First row: the total number of photons orbits, $n$, as a function of the impact parameter, $b$, for different values of the free parameter $\ell$. Second row: left and central panels: photon trajectories showing the first three transfer functions, $r_q$, in the case of $\ell=0.5$ and $\ell=0.73$.  The location of $r_q$ is depicted using a cross marker. Right panel: zoom of the first row and right panel of Fig.~\ref{Fig11}. The direct, lensed ring, and photon ring emissions are represented using cornflowerblue, deepskyblue, and black colors, respectively. We use $M=1$.}
\label{Fig9}
\end{figure*}

\begin{table*}
\caption{\label{table2} Intervals for the impact parameter $b$ corresponding to the direct, lensed ring and photon ring emissions of the EOS BH (with $a=0$) for different values of the free parameter $\ell$.}
\begin{ruledtabular}
\begin{tabular}{cccccc}
\hline
$\ell$&Direct emission&Direct emission&Lensed ring emission&Lensed ring emission&Photon ring emission\\
\hline\\
0.0  &$b<5.00479$&$b>6.11929$&$5.00479<b<5.17999$&$5.22829<b<6.11929$&$5.17999<b<5.22829$\\\\
0.5  &$b<3.90999$&$b>5.34999$&$3.90999<b<4.16599$&$4.24779<b<5.34999$&$4.16599<b<4.24779$\\\\
0.73&$b<2.87946$&$b>4.89012$&$2.87946<b<3.39298$&$3.62879<b<4.89012$&$3.39298<b<3.62879$\\\\
\end{tabular}
\end{ruledtabular}
\end{table*}

\begin{figure*}[t]
\centering
\includegraphics[scale=0.38]{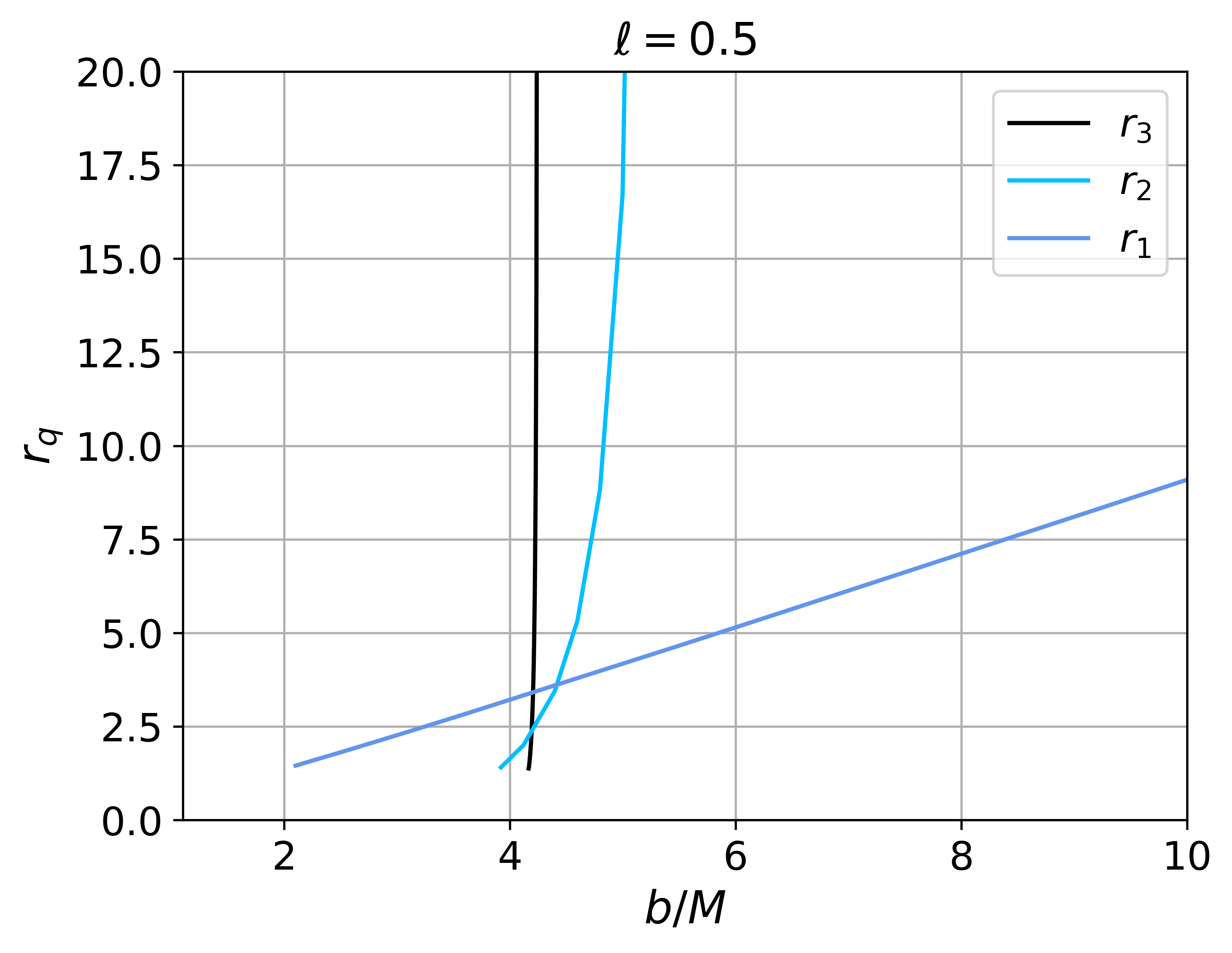}
\includegraphics[scale=0.38]{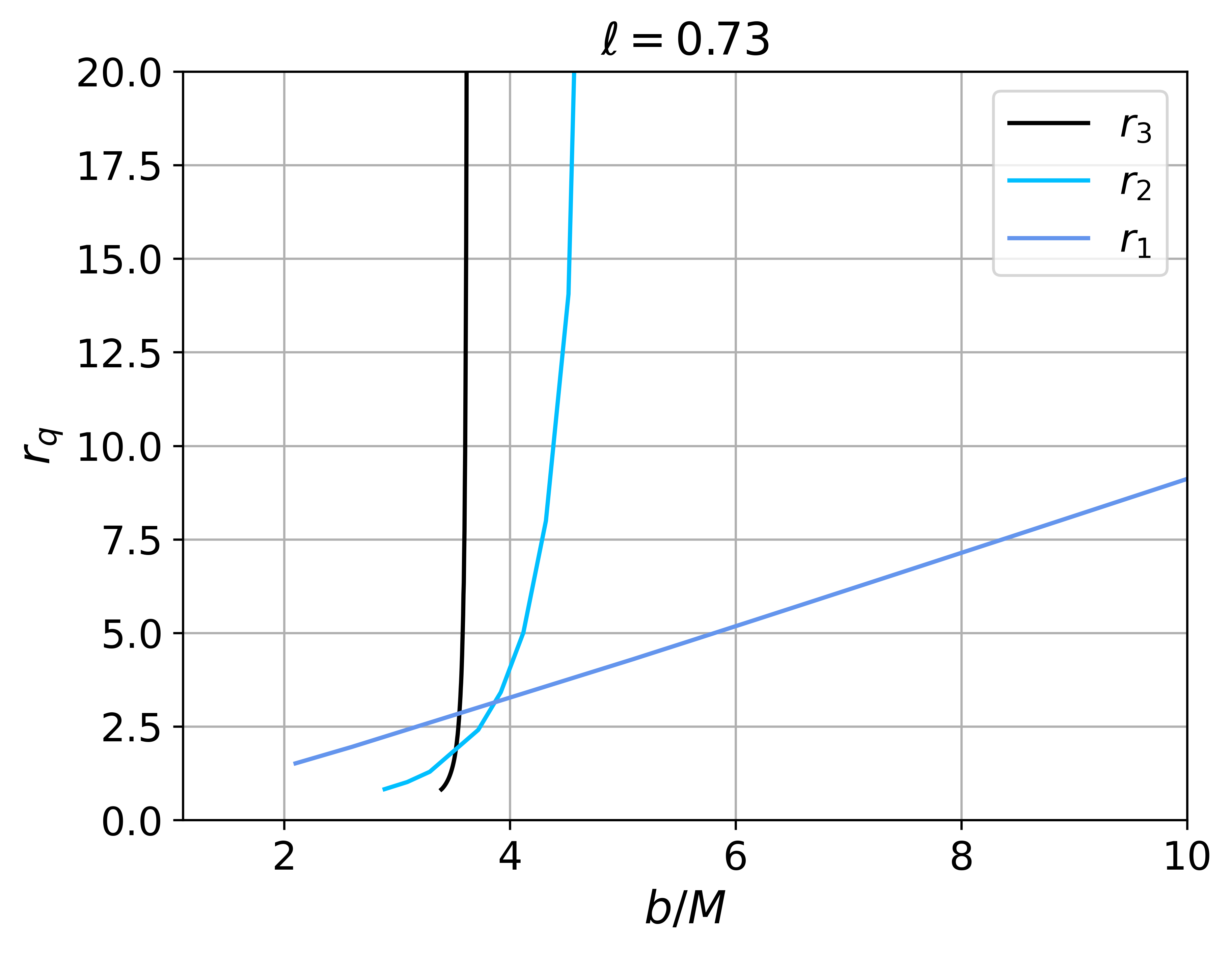}
\includegraphics[scale=0.38]{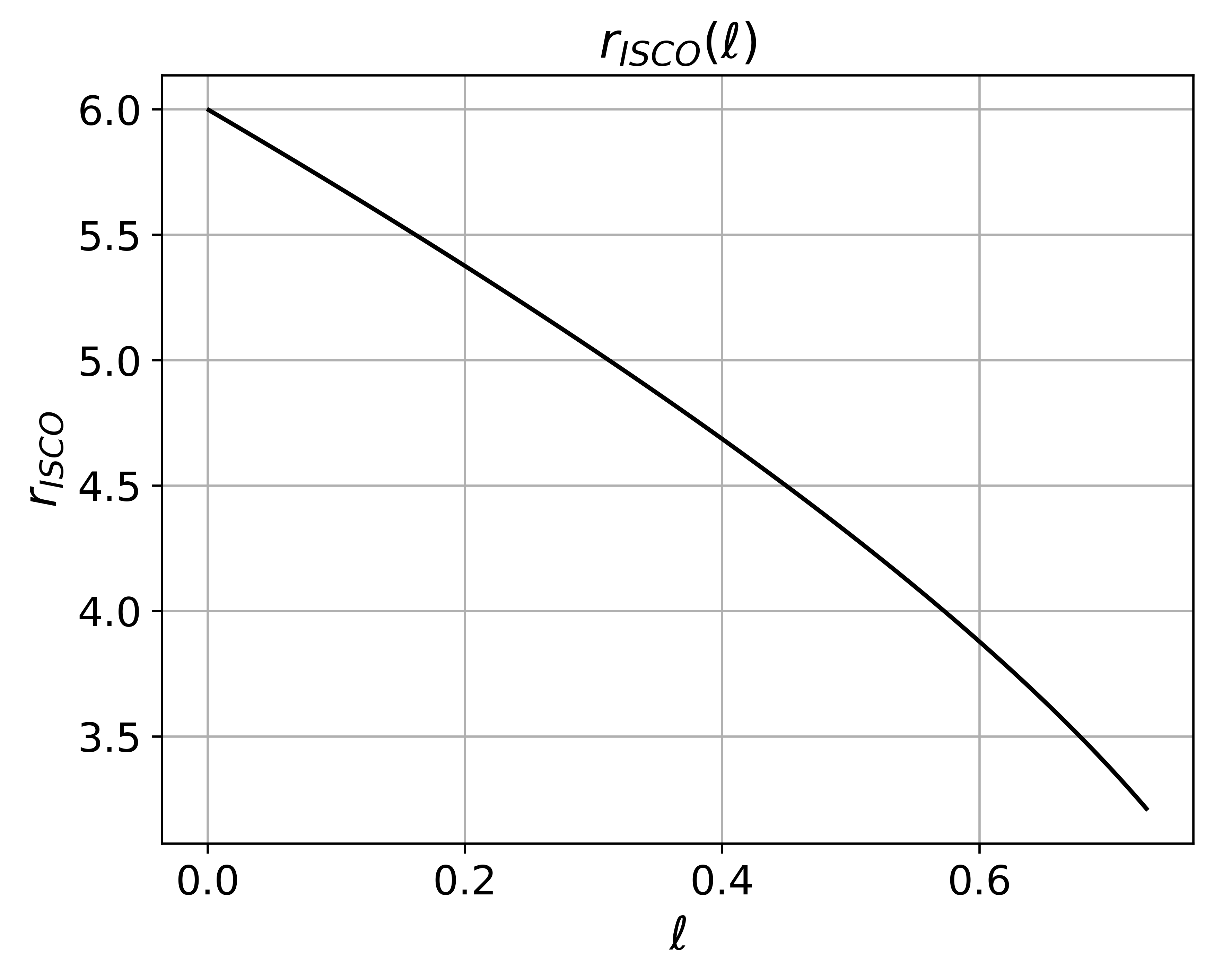}
\caption{Left and central panels: Observational appearance of a thin, optically thin disk of emission near the EOS BH  (with $a=0$) , viewed from a face-on-orietation for $\ell=0.5$ and $\ell=0.73$. Right panel: The radius of the ISCO as a function of $\ell$. We use $M=1$.}
\label{Fig10}
\end{figure*}

\subsubsection{Shadows and rings with thin disk accretions}

	Now, we consider an optically and geometrically thin accretion disk located at the equatorial plane, assuming that the emission from the accretion disk is isotropic and that the distant observer is in the north pole direction (face-on orientation); therefore, the light rays emitted from the disk-shaped accretion flow transmit their luminosities to the remote observer in the z-direction.
	
	Light rays can intersect the accretion disk several times before reaching the distant observer. Hence, according to Ref.~\cite{Gralla:2019xty}, the deflection angle of the light for $q^\text{th}$ intersection is given by.
\begin{equation}
\label{SIVe19}
\Delta \varphi= \frac{\pi}{2}+(q-1)\pi.
\end{equation} 
Therefore, the total number of photons orbits, $n$, is defined by the relation~\cite{Gralla:2019xty,Hu:2022lek}
\begin{equation}
\label{SIVe20}
n=\frac{\Delta \varphi}{2\pi}=\frac{q}{2}-\frac{1}{4}.
\end{equation}
In this sense, the light around the BH can be divided in three different categories depending on the value of $n$. These categories are:
\begin{enumerate}
\item \textbf{Direct emission:} light rays hitting the accretion disk only once; i.e.,  $0<n<0.75$.
\item \textbf{Lensed ring emission:} light rays hitting the accretion disk twice; i.e., $0.75<n<1.25$.
\item \textbf{Photon ring emission:} light rays hitting the accretion disk at least three times; i.e., $n>1.25$. 
\end{enumerate}
In the second row of Fig.~\ref{Fig6}, the light rays of the direct, lensed ring, and photon ring emissions are represented by cornflower blue, deep sky blue, and black lines, respectively. The relationship between the total number of photon orbits, $n$, and the impact parameter, $b$, are shown in the first row of Fig.~\ref{Fig9} for different values of $\ell$. The figure shows how $n$ increases as $b$ reaches the critical value $b_\text{ph}$, where the photon reaches the photon sphere radius, orbiting the BH many times. Then, for $b>b_\text{ph}$, the total number of photon orbits decreases asymptotically to a fixed value of the impact parameter. Note how the critical value $b_\text{ph}$ changes its location as the free parameter $\ell$ increases its value, getting closer to the BH's center. Furthermore, the figures show how the intervals for each kind of emission become broader as the free parameter $\ell$ goes from $0$ to $0.73$. For example, in the case of $\ell=0.5$, direct emission corresponds to the intervals $b<3.90999$ and $b>5.34999$ ($\Delta b=1.44$), while for $\ell=0.73$, the intervals for direct emission are $b<2.8843$ and $b>4.882845$ ($\Delta b=1.99855$). See Table~\ref{table2} for details. In this sense, the intervals for lensed and photon ring emissions get larger while direct emissions shrink as the free parameter $\ell$ increases. The second row of Fig.~\ref{Fig6} shows this behavior clearly, where the region for lensed and photon ring emissions become thicker as $\ell$ increases.

According to Liouville's theorem, the ratio between the specific intensity and the frequency of the emitted light $I_e(r)/\nu_e^3$ is invariant along the photon's trajectory, satisfying the equation~\cite{Bromley:1996wb}
\begin{equation}
\label{SIVe21}
\frac{I_e(r)}{\nu^3_e}=\frac{I_\text{ray}(r)}{\nu^3_0},
\end{equation}
where $I_\text{ray}(r)$ and $\nu_0$ are the specific intensity and frequency of the observed light. Thus, we have the following relation~\cite{Hu:2022lek}
\begin{equation}
\label{SIVe22}
I_\text{ray}(r)=\left(1-\frac{2Me^{-\ell M/r}}{r}\right)^{3/2}I_e(r).
\end{equation}
After integrating over the whole range of the observed photon frequencies $v_0$, the total observed specific intensity is given by 
\begin{equation}
\label{SIVe23}
I_o=\int_\gamma I_\text{ray}(r)dv_0=\left(1-\frac{2Me^{-\ell M/r}}{r}\right)^2I_\text{emit}(r).
\end{equation}
Here, the total emission of the accretion disk is defiined by~\cite{Hughes:2001gg}. 
\begin{equation}
\label{SIVe24}
I_\text{emit}(r)=\int_\gamma I_e(r)dv_e.
\end{equation} 
As mentioned above, there are three categories for photon trajectories depending on the number of photon orbits, $n$. In this sense, light from the emission becomes brighter each time the photon trajectory intersects the accretion disk. Consequently, the total observed specific intensity, $I_o$, should be redefined as:
\begin{equation}
\label{SIVe25}
I_\text{o}=\sum_q \left(1-\frac{2Me^{-\ell M/r}}{r}\right)^2I_\text{emit}(r)|_{r=r_q(b)}.
\end{equation}
where $r_q(b)$ is defined as the transfer function~\cite{Gralla:2019xty,Hu:2022lek}, which is the correlation between the impact parameter of the photon in the distant observer sky and the radial coordinate of the $q_\text{th}$ intersection of the light ray and the accretion disk. In the second row of Fig.~\ref{Fig9}, we show the location of the first three transfer functions for $\ell=0.5$ and $\ell=0.73$ while in Fig.~\ref{Fig10}, the behavior of $r_q$ as a function of the impact parameter $b$. In Fig.~\ref{Fig10}, the first transfer function, $r_1(b)$, is represented by the cornflower blue line while the second and third transfer functions, $r_2(b)$ and $r_3(b)$ are represented by deep sky blue and black lines, respectively. According to the figure, $r_1(b)$ increases proportionally as the impact parameter, $b$, increases. Note that the minimum value of the first transfer function is $r_1\approx r_h$; below that value, photon trajectories do not hit the accretions disk. In the case of the second and third transfer functions, on the other hand, the increment is not in direct proportion to the impact parameter $b$ since $r_1(b)$ and $r_2(b)$ increases drastically in a small interval of $b$ (in the vicinity of $b_\text{ph}$ for the third transfer function) Therefore, lensed/photon ring images are high/extremely demagnified~\cite{Hu:2022lek}. In this sense, the observed intensity of the photon ring is negligible. In contrast, direct emissions dominate the total observed intensity. This behavior is independent of $\ell$. 

\begin{figure*}[t]
\centering
\includegraphics[scale=0.38]{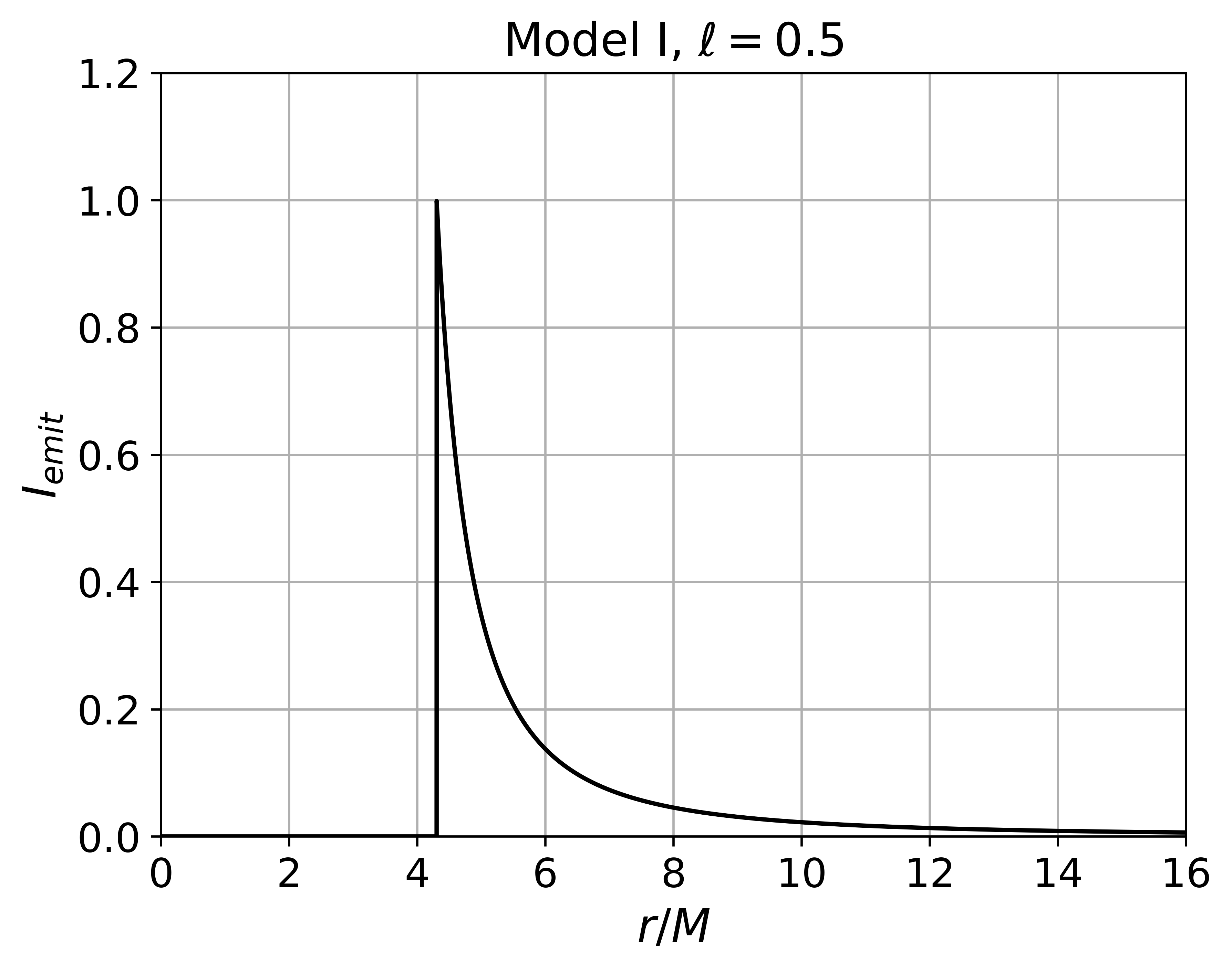}
\includegraphics[scale=0.38]{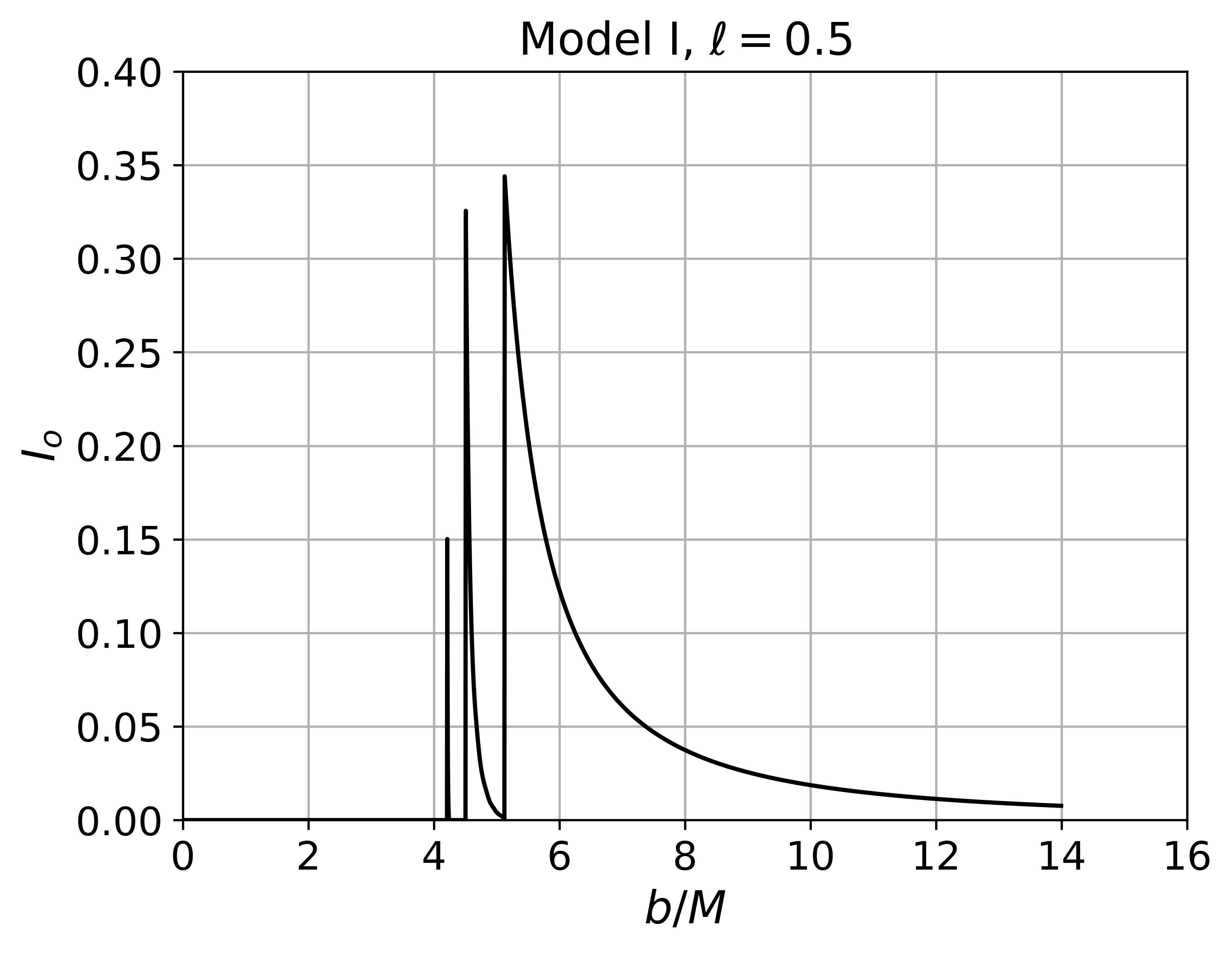}
\includegraphics[scale=0.38]{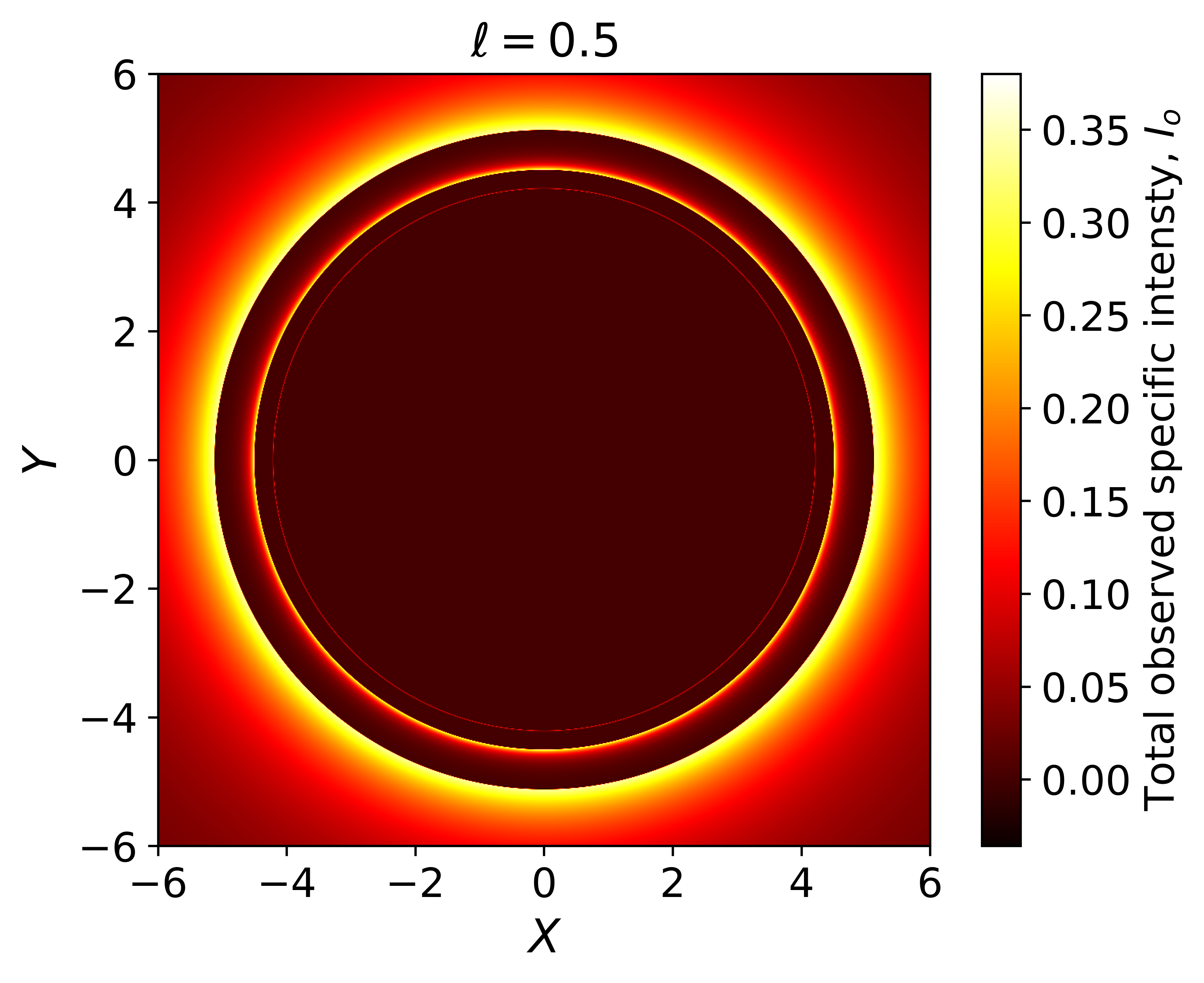}\\
\includegraphics[scale=0.38]{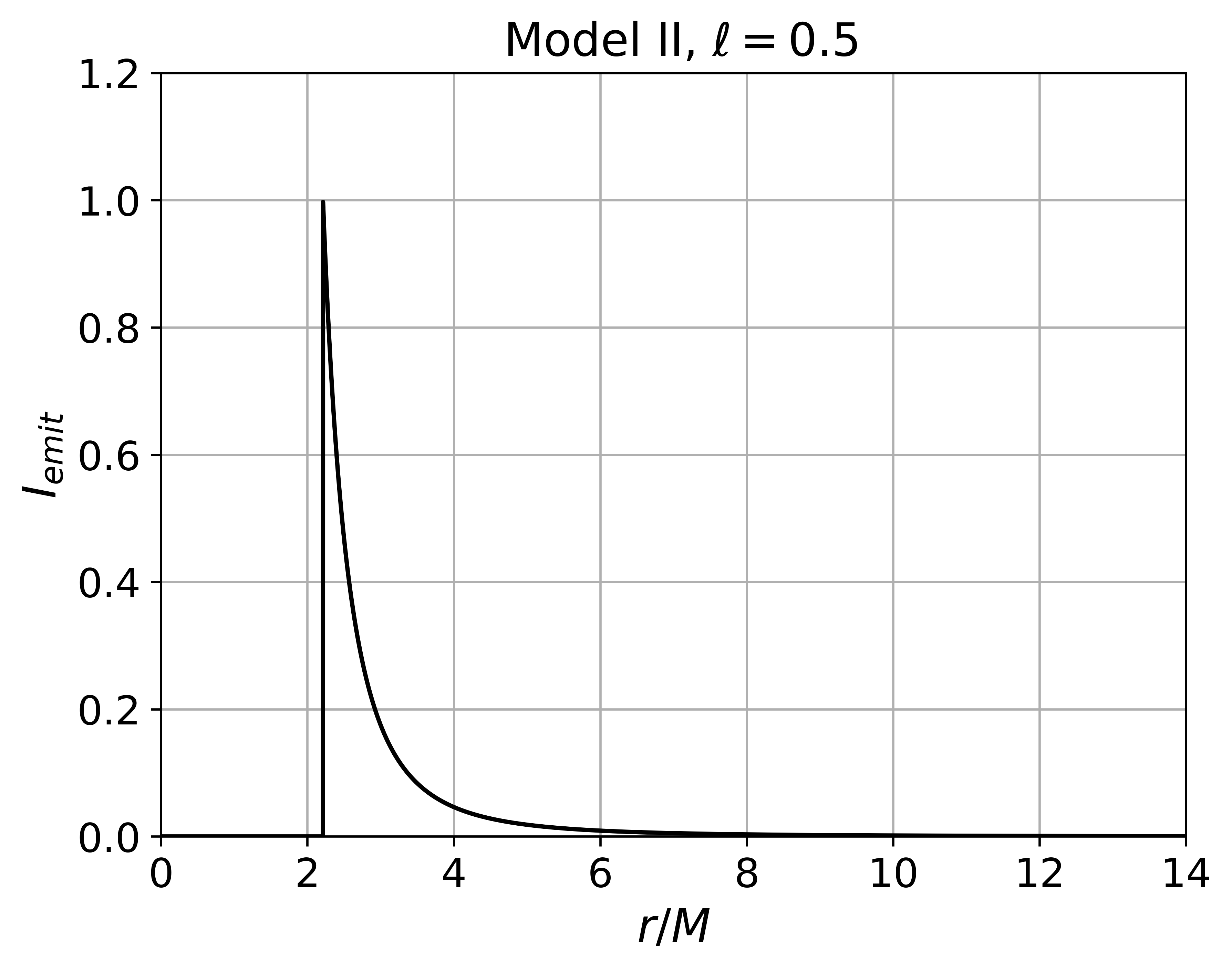}
\includegraphics[scale=0.38]{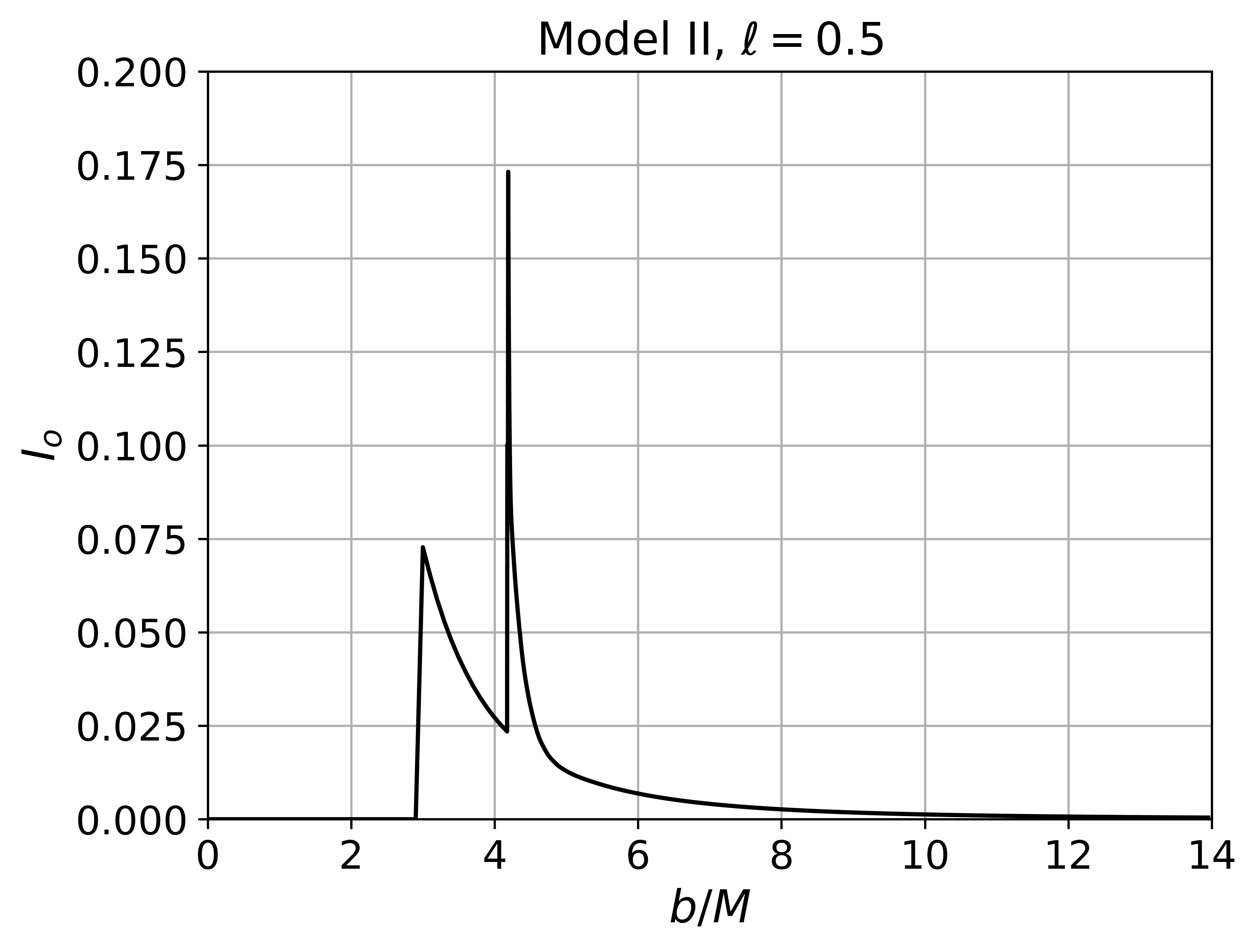}
\includegraphics[scale=0.38]{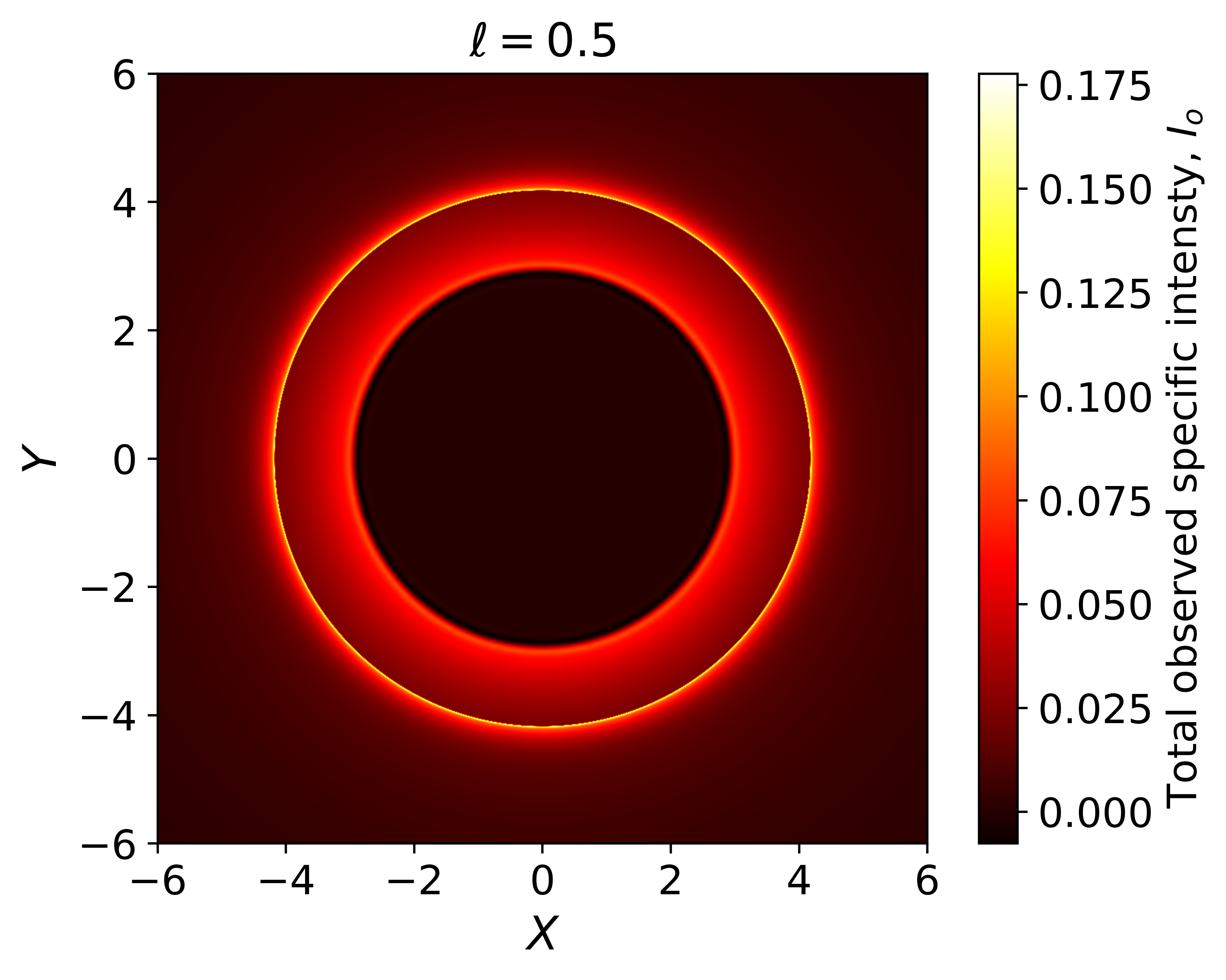}\\
\includegraphics[scale=0.38]{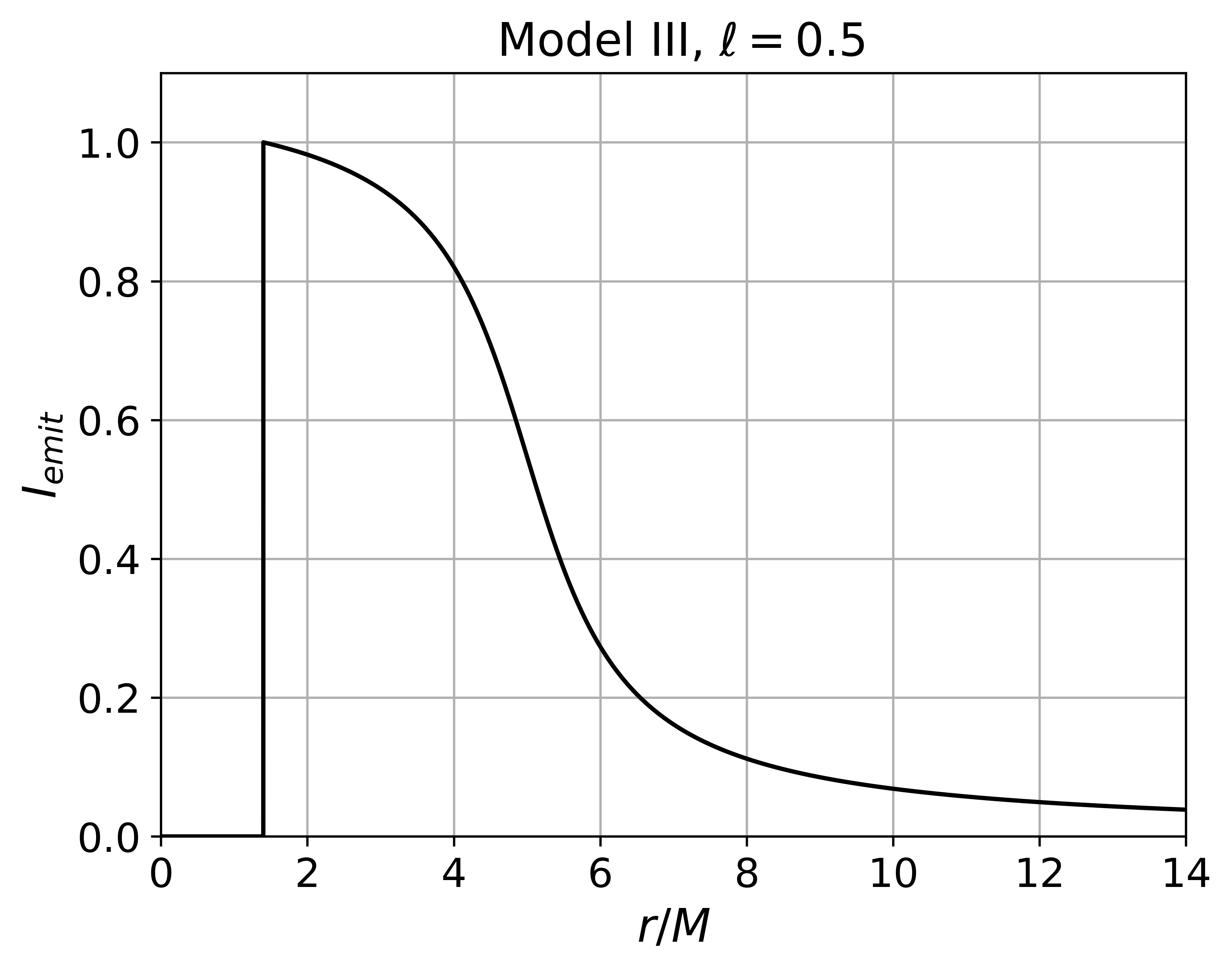}
\includegraphics[scale=0.38]{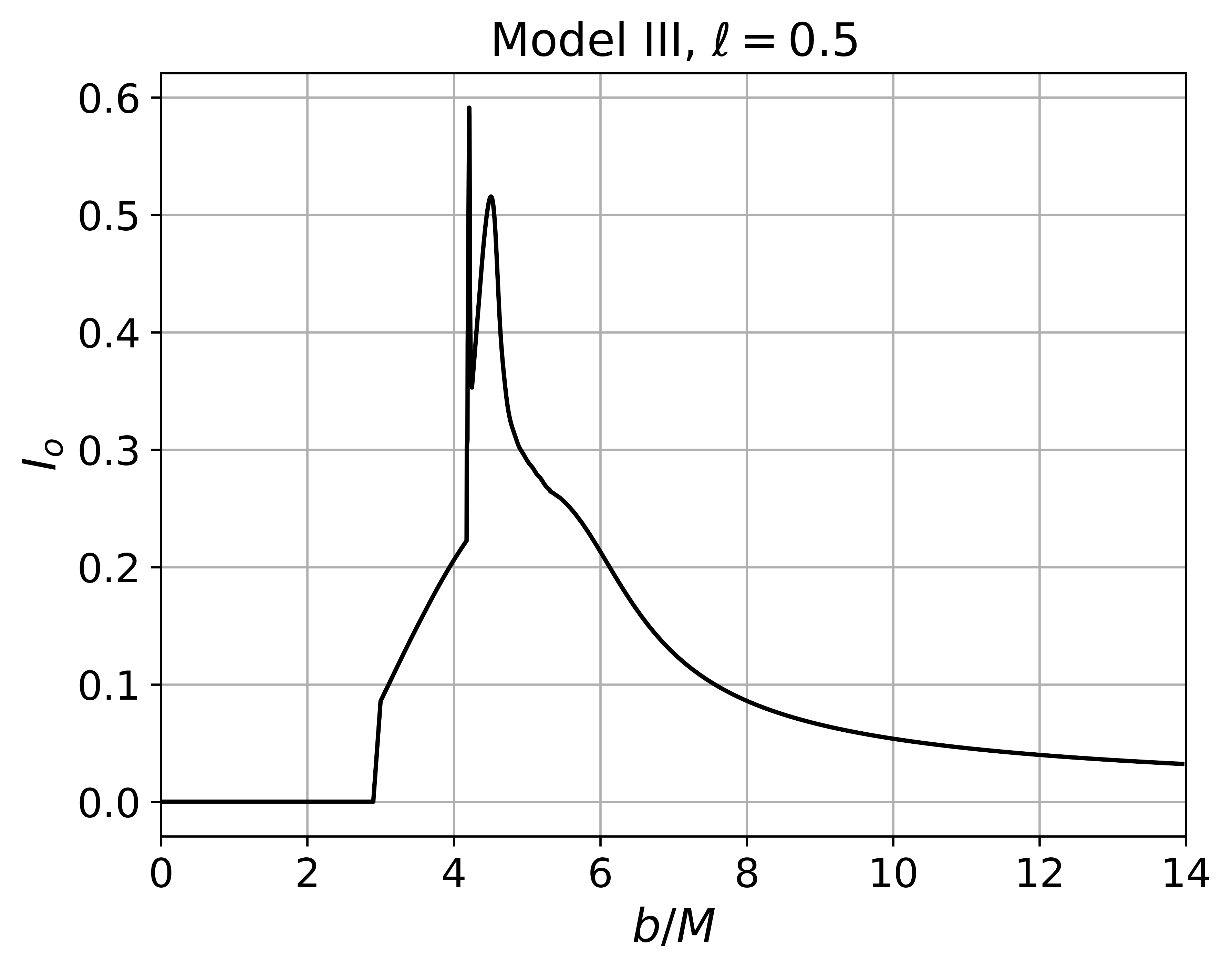}
\includegraphics[scale=0.38]{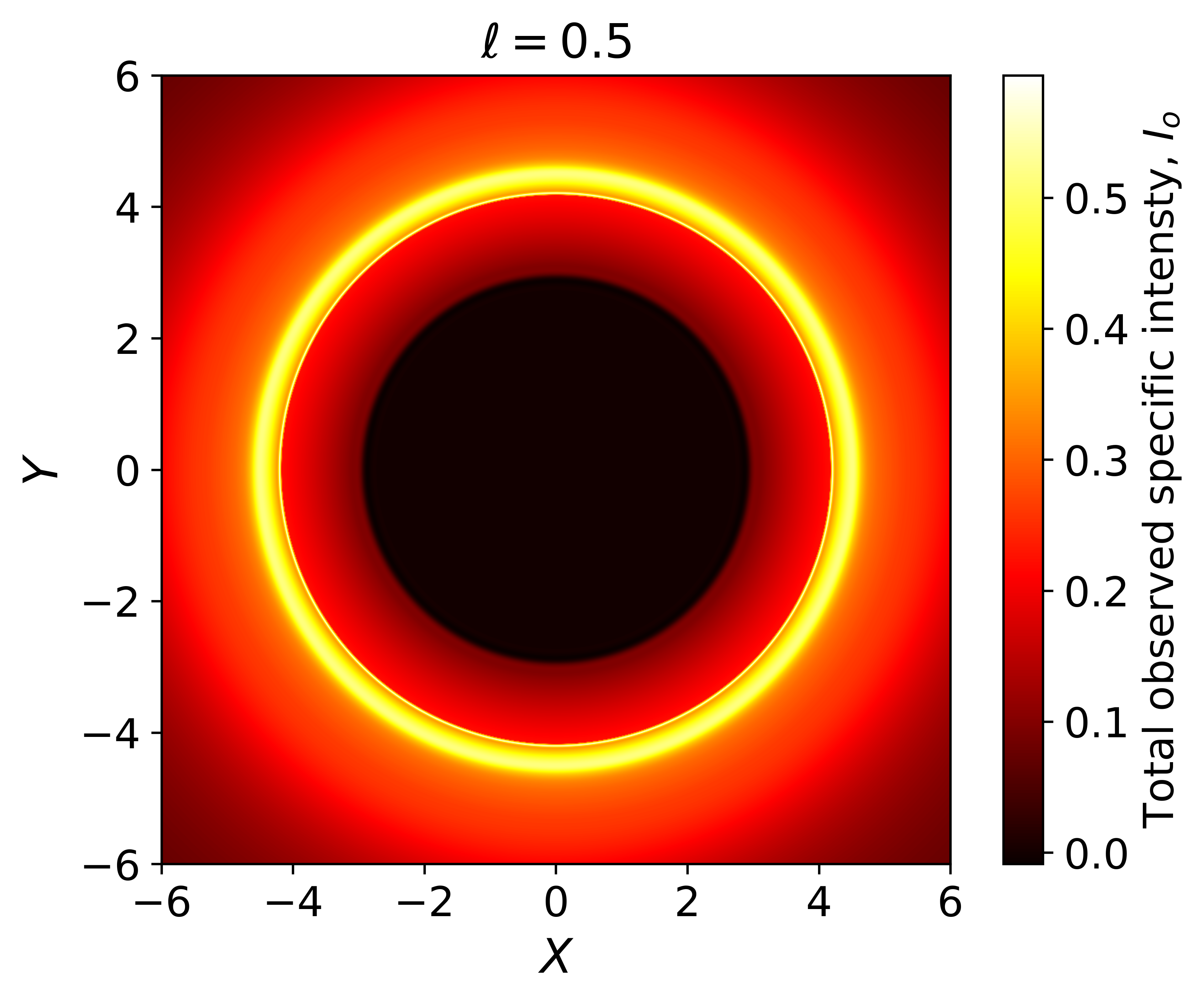}\\
\caption{Left column: total emission intensities $I_\text{emit}$ of the optical and geometrically thin accretion disk as a function of the radius $r$. Central column: the total observed specific intensities as a function of the impact parameter $b$. Right column: optical appearances of the EOS BH ($a=0$) with a thin accretion disk, viewed from a face-on-orietation. The emissions profiles, from top to botton, correspond to Models, I, II and III, respectively. We use $M=1$ and $\ell=0.5$.}
\label{Fig11}
\end{figure*}

\begin{figure*}[t]
\centering
\includegraphics[scale=0.38]{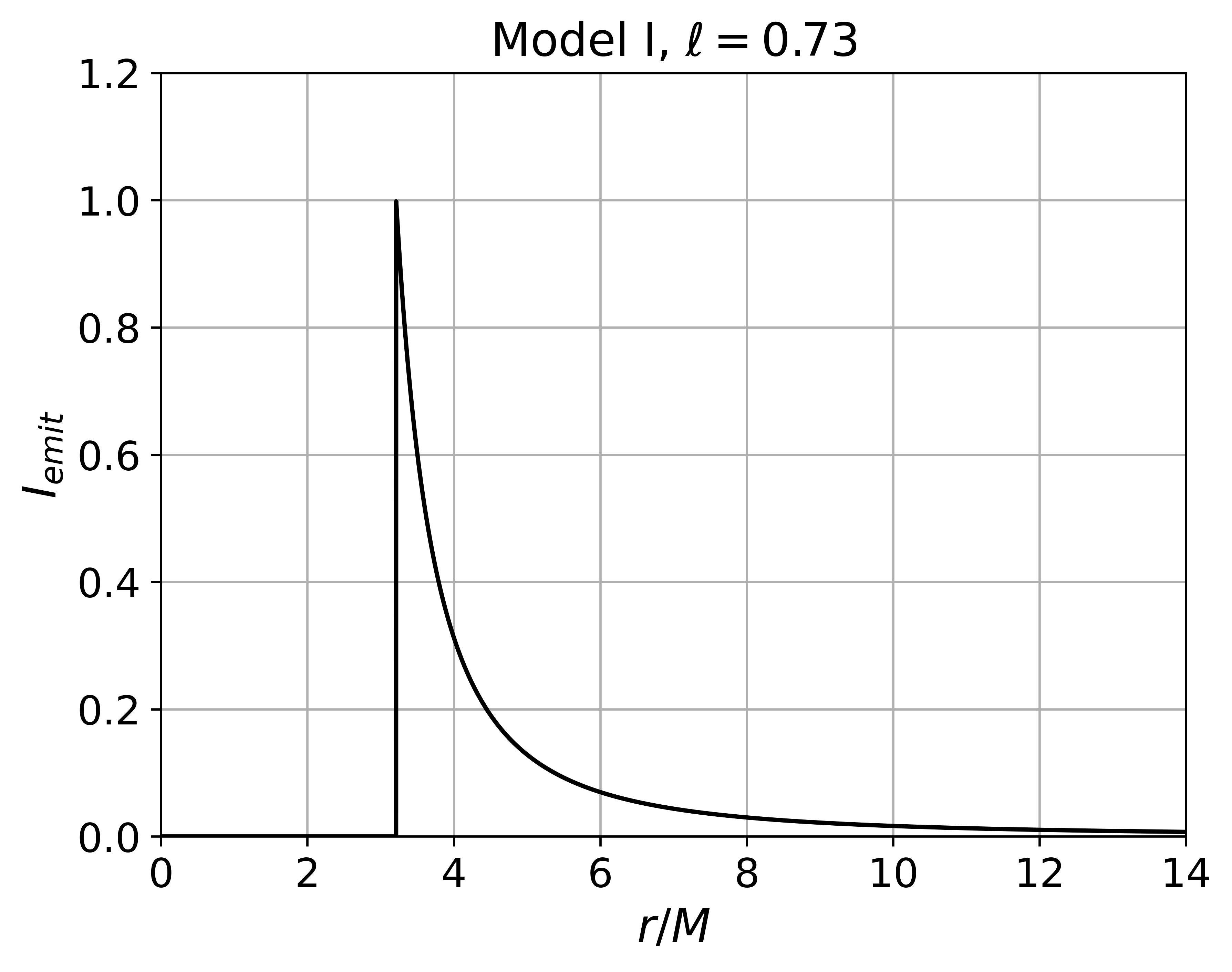}
\includegraphics[scale=0.38]{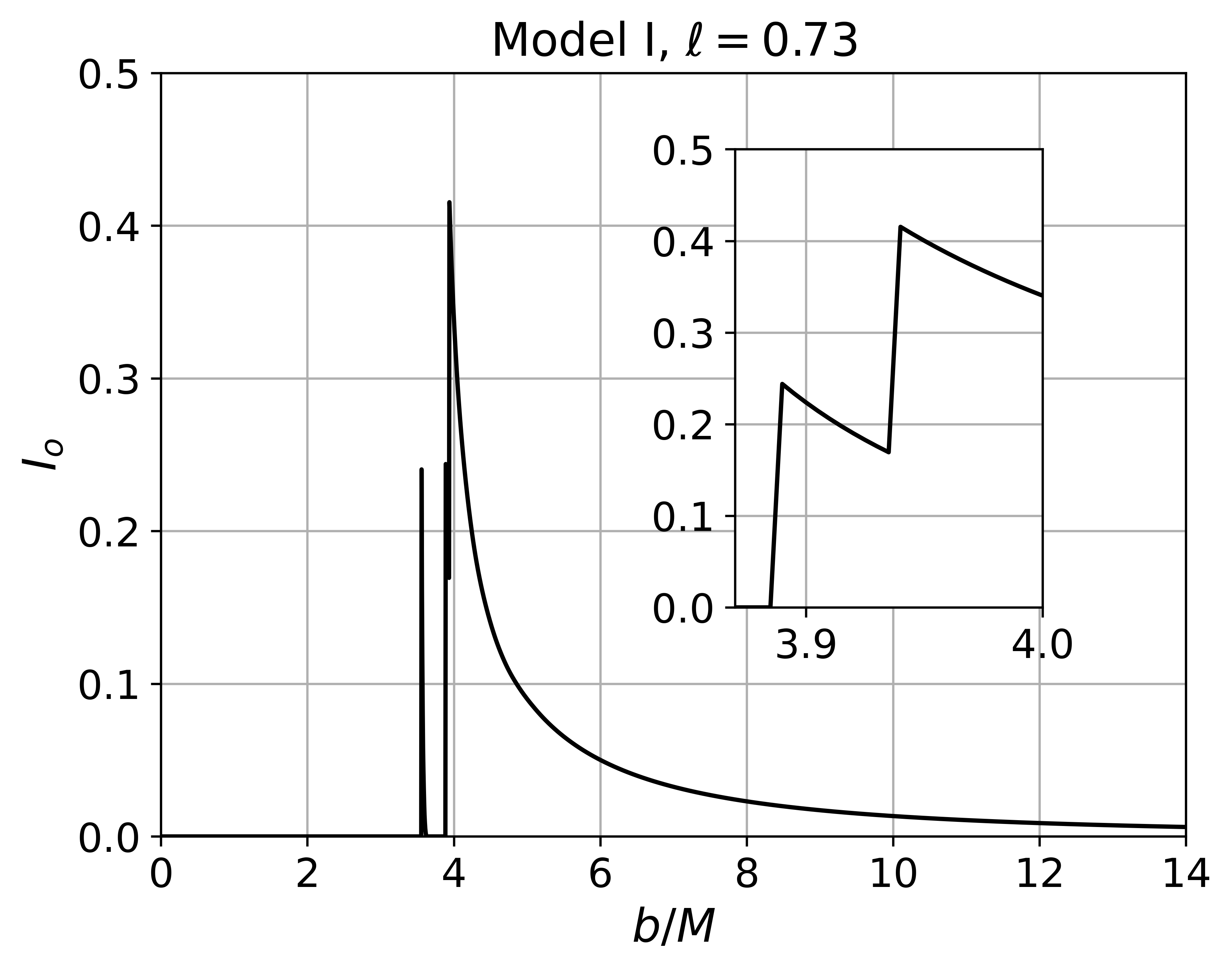}
\includegraphics[scale=0.38]{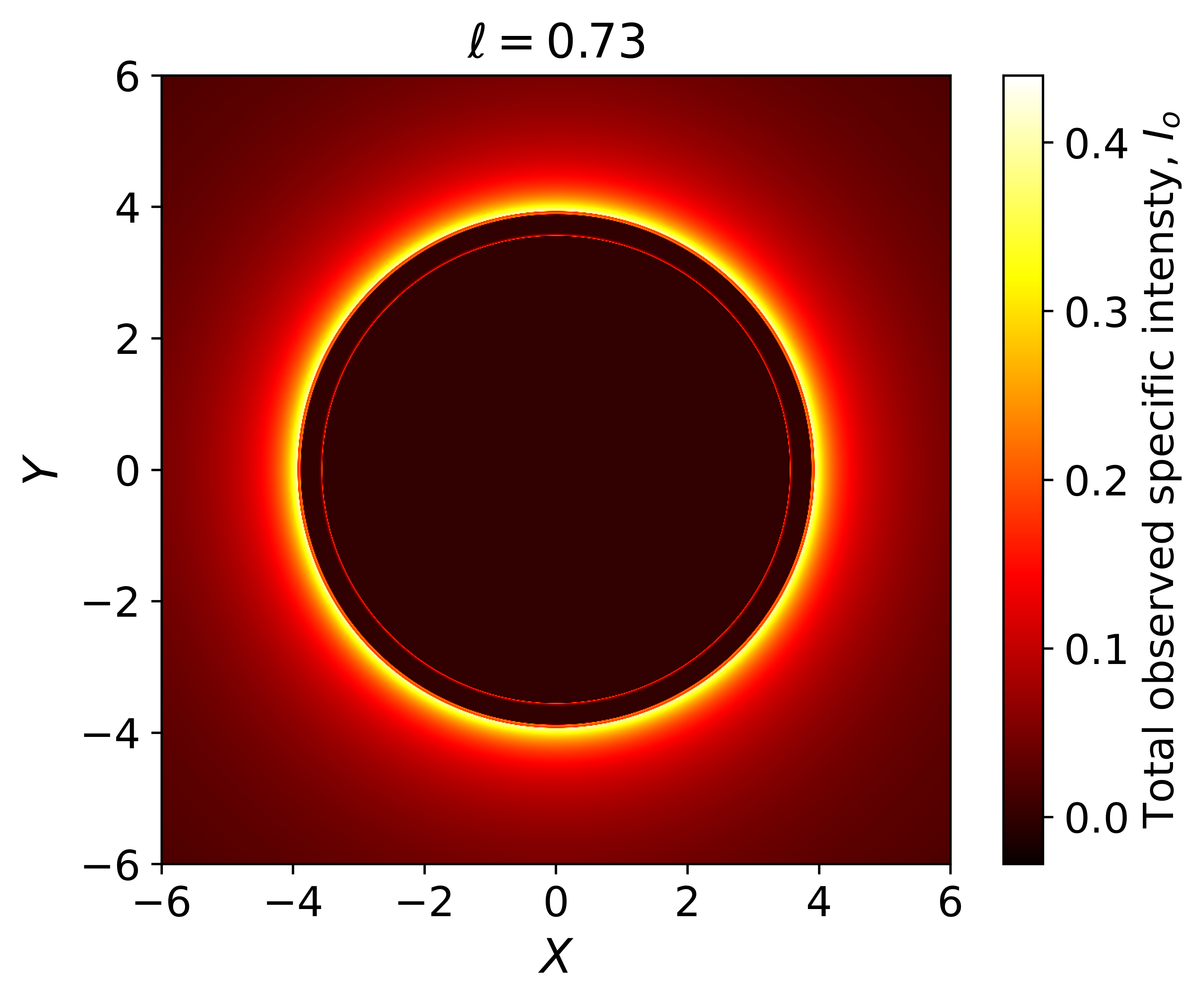}\\
\includegraphics[scale=0.38]{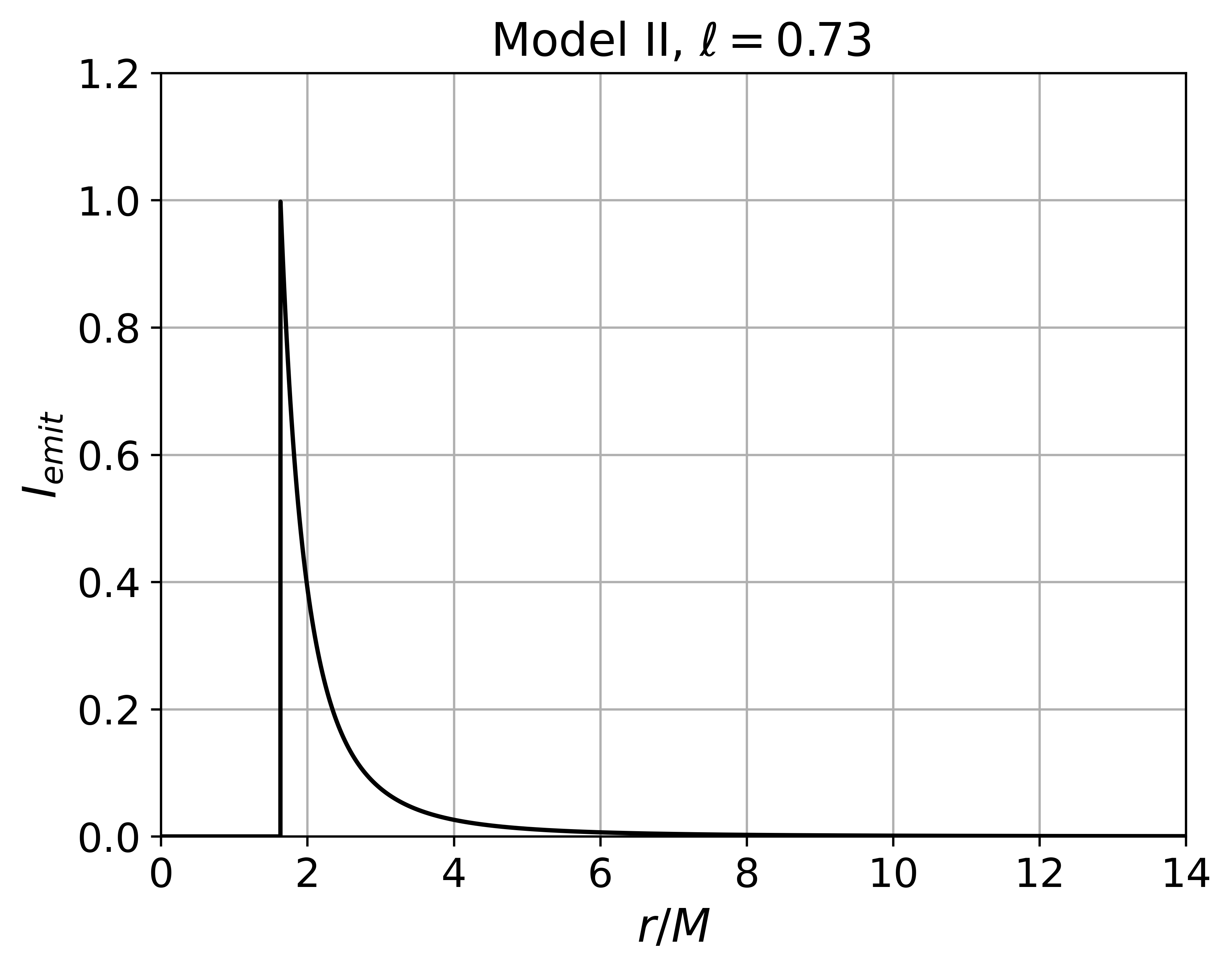}
\includegraphics[scale=0.38]{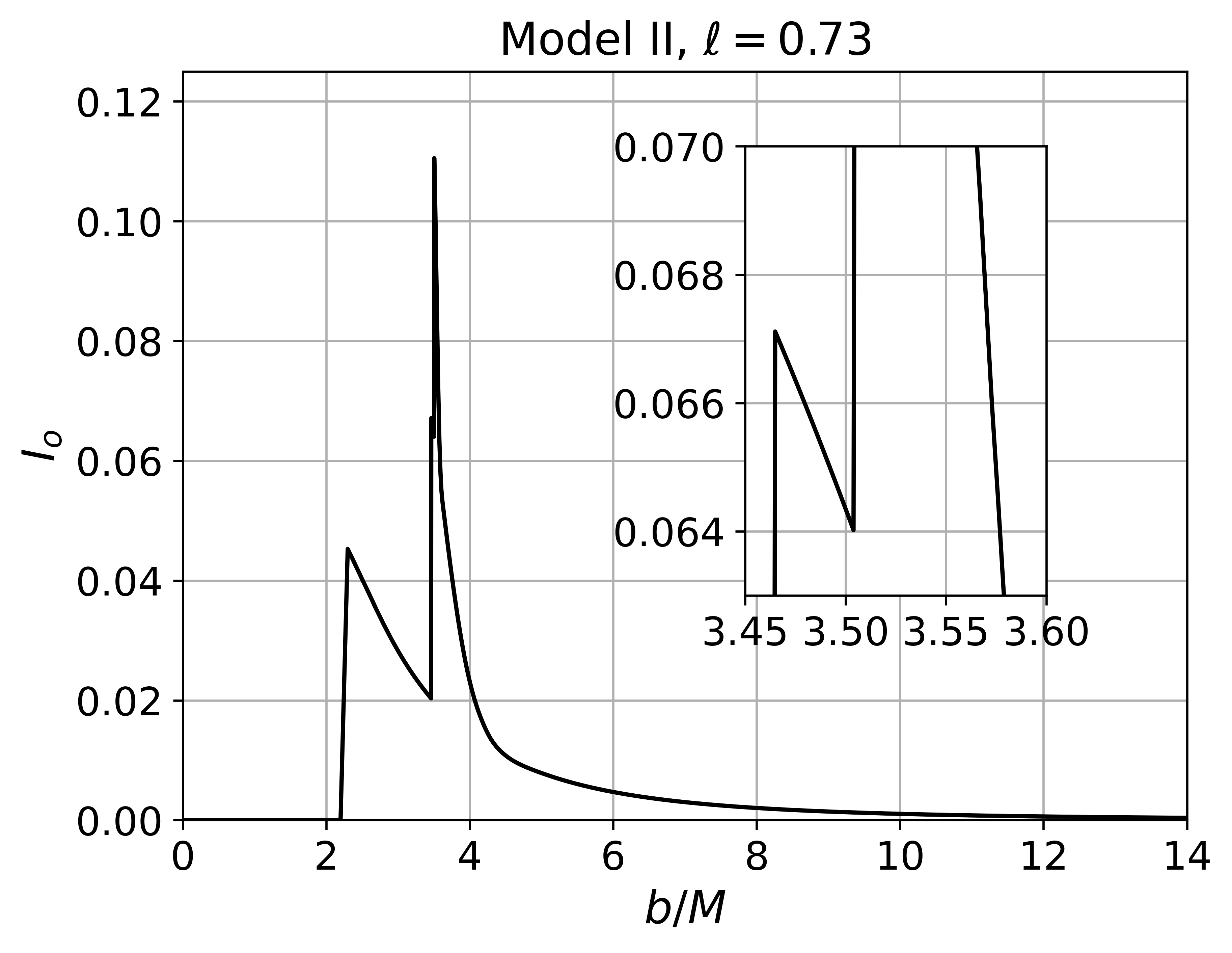}
\includegraphics[scale=0.38]{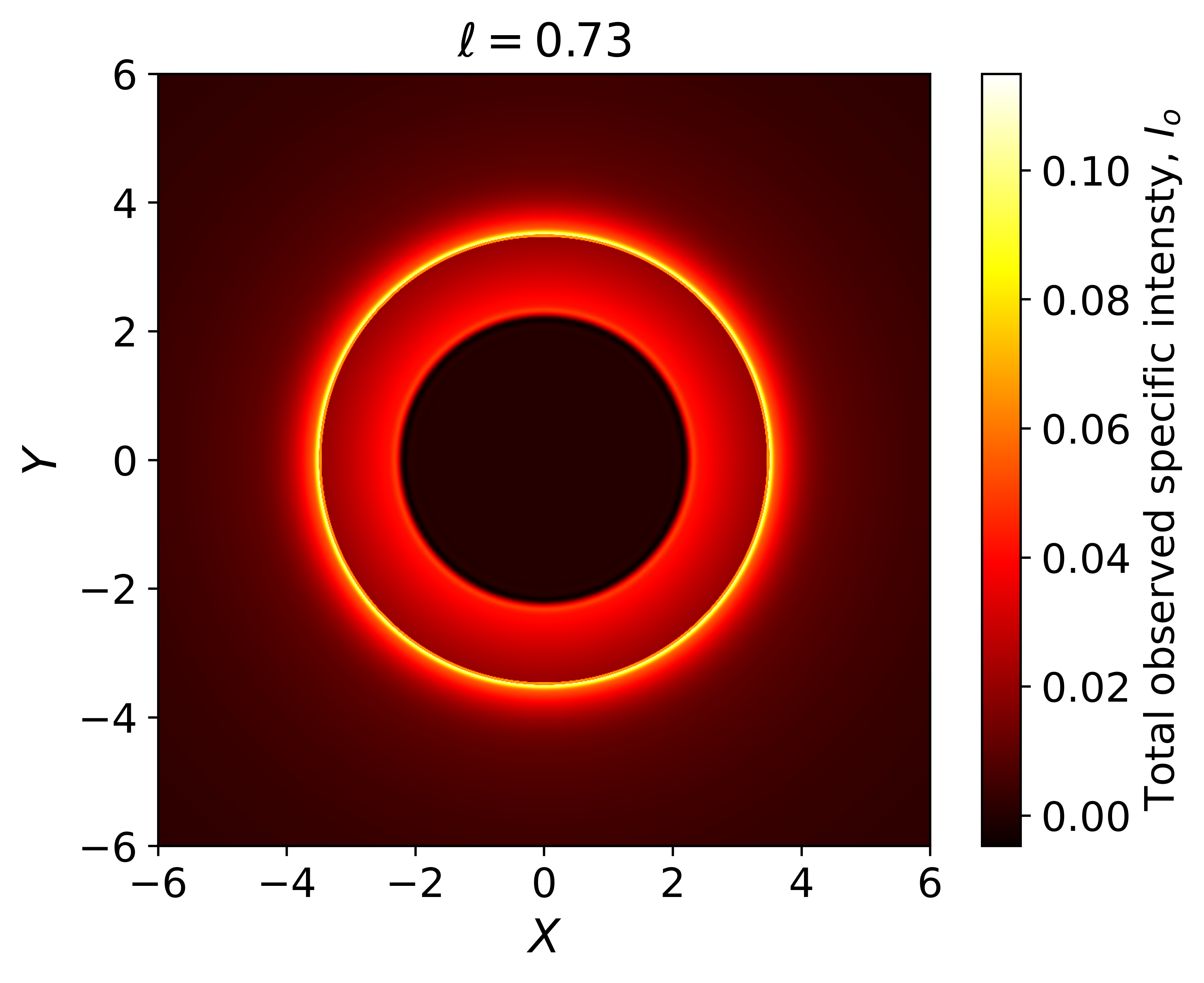}\\
\includegraphics[scale=0.38]{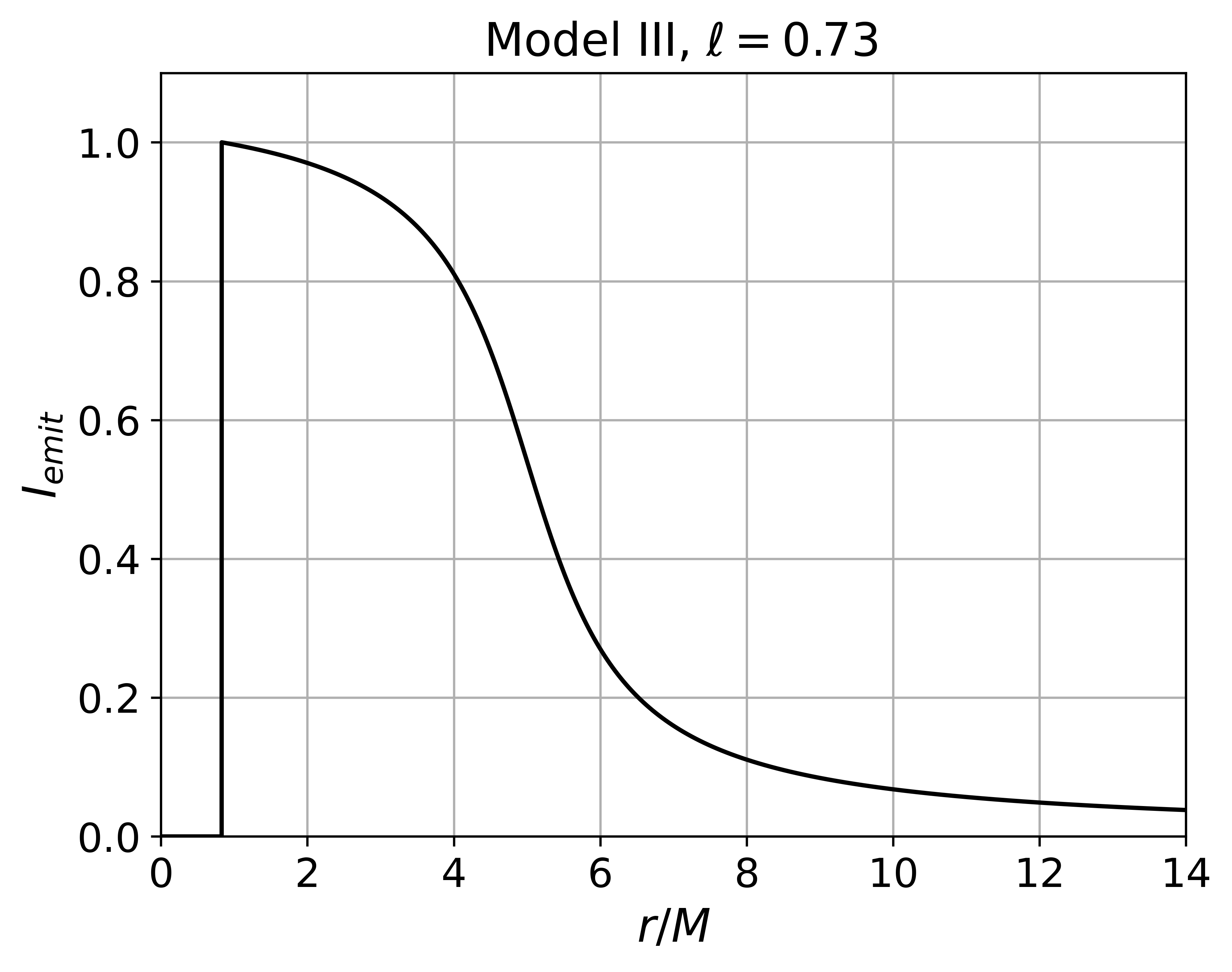}
\includegraphics[scale=0.38]{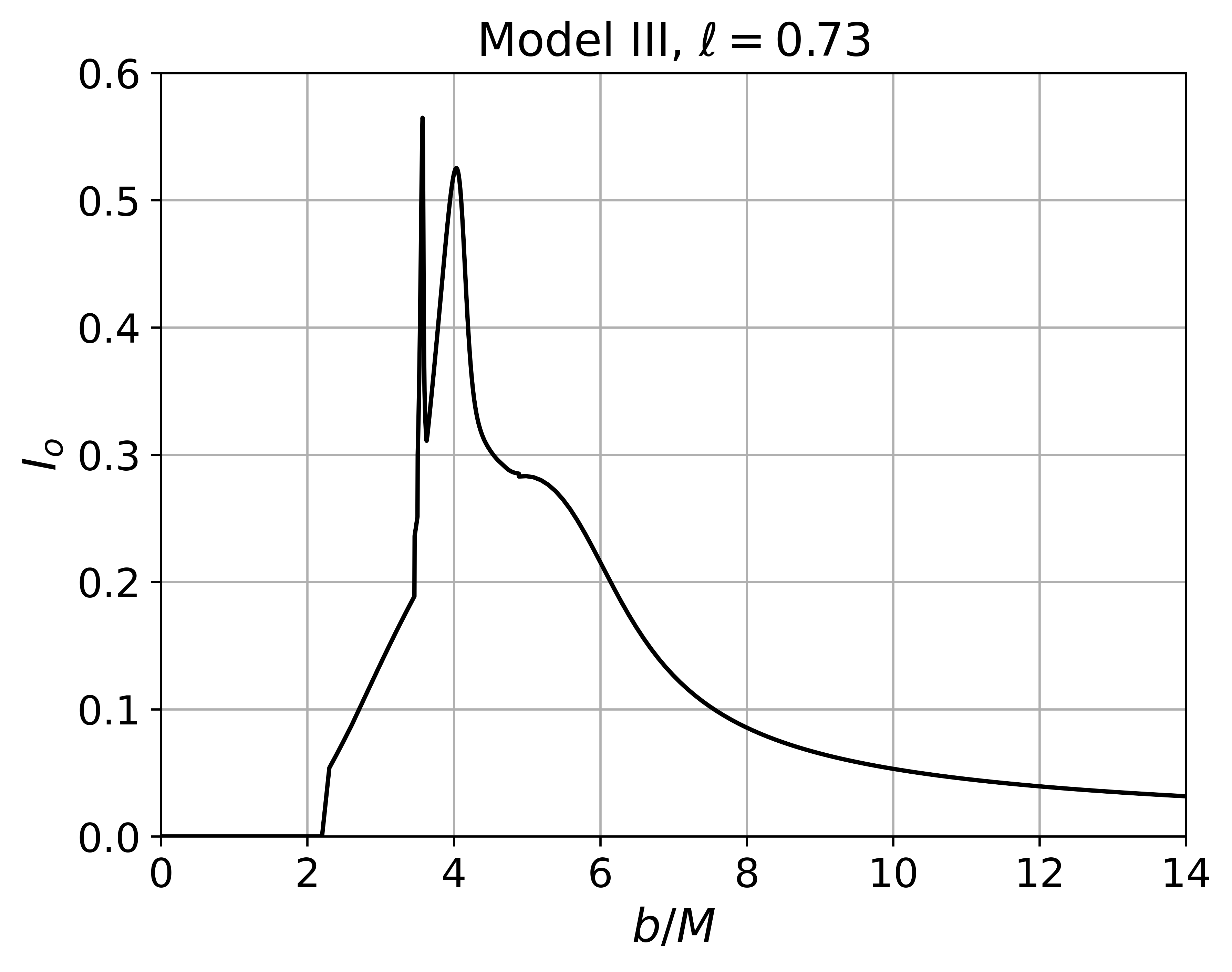}
\includegraphics[scale=0.38]{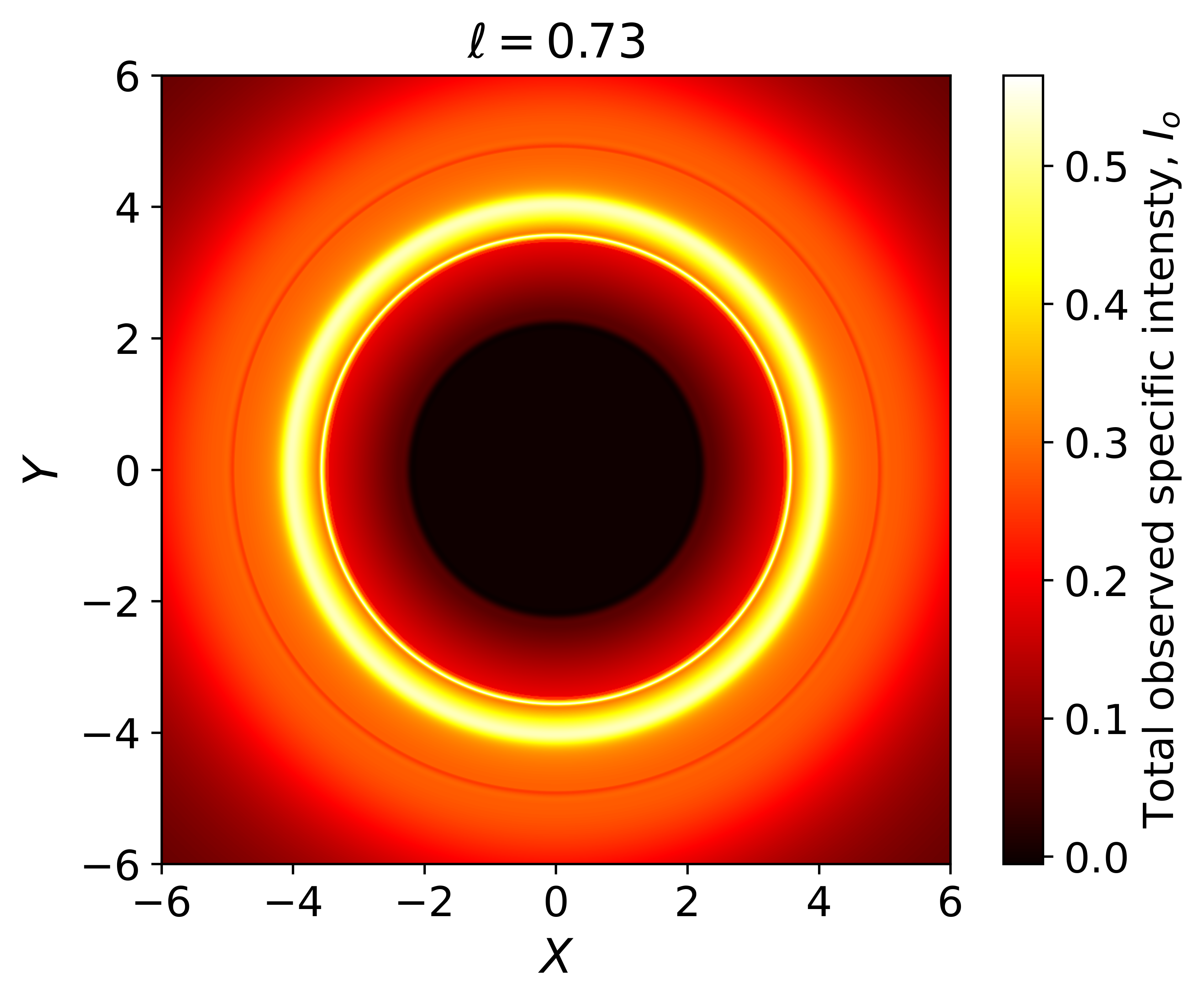}\\
\caption{Left column: total emission intensities $I_\text{emit}$ of the optical and geometrically thin accretion disk as a function of the radius $r$. Central column: the total observed specific intensities as a function of the impact parameter $b$. Right column: optical appearances of the EOS BH ($a=0$) with a thin accretion disk, viewed from a face-on-orietation. The emissions profiles, from top to botton, correspond to Models, I, II and III, respectively. We use $M=1$ and $\ell=0.73$.}
\label{Fig12}
\end{figure*}

	To investigate the observational signatures of the EOS BH, we consider three different models for the emission of the accretion disk. In the first model, the emission begins from the radius of the ISCO (the behavior of the ISCO as a functio of the free parameter $\ell$ is shown in In Fig.~\ref{Fig10}), and the total emission intensity $I_\text{emit}$ has the form~\cite{Hu:2022lek}
\begin{equation}
\label{SIVe26a}
I_\text{emit}(r)= \left\{ 
\begin{array}{lc} 
\frac{1}{\left[r-(r_\text{ISCO}-1)\right]^2} &  r > r_\text{ISCO} \\\\
0  & r\leq r_\text{ISCO}.
\end{array}\right.
\end{equation}
Note that the disk emission is sharply peaked near $r_\text{ISCO}$, and it ends sudently at $r=r_\text{ISCO}$, well outside of the photon sphere.  

	In the second model, the emission begins at the photon sphere, $r_\text{ph}$, and it decays with a cubic power when $r>r_\text{ph}$; i.e.~\cite{Hu:2022lek},
\begin{equation}
\label{SIVe26b}
I_\text{emit}(r)= \left\{ 
\begin{array}{lc} 
\frac{1}{\left[r-(r_\text{ph}-1)\right]^3} &  r > r_\text{ph} \\\\
0  & r\leq r_\text{ph}.  
\end{array}\right.
\end{equation} 
Finally, we consider a third model in which the emission intensity decays in a more moderate way when compared with the previous two models, and assuming that the emission begins at the horizon. The profile is given by the following relation~\cite{Guerrero:2022msp}
\begin{equation}
\label{SIVe26c}
I_\text{emit}(r)= \left\{ 
\begin{array}{lc} 
\frac{\frac{\pi}{2}-\arctan(r-5)}{\frac{\pi}{2}-\arctan(r_h-5)} &  r \geq r_h\\\\
0  & r < r_h.  
\end{array}\right.
\end{equation} 
The profile of $I_\text{emit}(r)$ for each model is shown in the first column of Figs.~\ref{Fig11} and \ref{Fig12} for $\ell=0.5$ and $\ell=0.73$, respectively. The figures also show the total observed specific intensity $I_o$ (central column) and the optical appearances of the EOS BH viewed from a face-on orientation (right column). Note that the starting points of the emission profile change because $r_\text{h}$, $r_\text{ph}$ and $r_\text{ISCO}$ are functions of the free parameter $\ell$. 

	On the other hand, the total observed specific intensity $I_o$ shows different behaviors associated with a particular ring structure depending on the model used. In the model of emission I, for example, the results show three different rings: the first one (the closest to the BH center) corresponds to the photon ring emissions, and the other two are associated with the lensed and direct emissions, respectively. Note that $I_o$ is much smaller for the photon ring than for the other two. Hence, from the optical point of view, it is more difficult to see the photon ring emissions, and it is necessary to zoom the figure to identify it. 
	
	When $\ell=0.73$, the first model also produces a three-ring structure. Nevertheless, the second ring, associated with the lensed emissions, becomes much closer to the third ring, almost merging. Note that the observed intensity of the third ring increases while that of the second ring decreases when compared with the case in which $\ell=0.5$. In the case of the inner ring, the observed intensity also increases its value when contrasted with the $\ell=0.5$ case. See the first row and the second panel of Fig.~\ref{Fig12}.  At this point, it is necessary to point out that the value $\ell=0.73$ is close to the extremal case. As was mentioned above, the EOS space-time models a regular BH for $\ell<0.73576$ (when $a=0$)~\cite{Guerrero:2022msp}. Therefore, this behavior could be related to the transition of a black hole to a naked regular compact object.
		
	In the second model, we can identify two rings when $\ell=0.5$. The inner one, related to the direct emissions, decreases its total observed specific intensity as the impact parameter approaches $b_\text{ph}$, where $I_o$ increases abruptly along the photon ring emission region. Then, it reduces its value asymptotically to zero. Note that the observed intensity of the inner ring is lower than the outer one, which, in this case, corresponds to the photon ring emissions. In the $\ell=0.73$ case, on the other hand, we can see a similar behavior; however, in contrast to the previous case ($\ell=0.5$), a second ring emerges closer to the outer ring. Note that the magnitude of the observed intensity of the inner and outer rings decreases when compared to the $\ell=0.5$ case. See the second row and the second panel of Fig.~\ref{Fig12}.  
	
	Finally, in the last model, we also see two rings. The inner ring is narrower than the second one. Note that the observed intensity is higher for the inner ring when contrasted with the outer ring. In general, the behavior of the third model is similar for different values of the free parameter $\ell$; the only differences surface when we compare the location of the rings (closer to the BH center in the case of $\ell=0,73$) and the value of the observed intensity for the first ring, which decreases its value as we increase $\ell$. Note that the observed intensity in the case of the second ring does not change considerably.  
\section{Image cast by the EOS black hole \label{SecVI}}

	In this section, we show the image cast by the EOS BH  (with $a=0$). To do so, we use the Hamiltonian formalism. The reason behind this choice is mainly numerical as Eq.~\eqref{SIIe9c} presents terms with square roots that generate difficulties in numerical solutions due to the change of signs in the turning points~\cite{Hughes:2001gg}.
	
	From an astrophysical perspective, we examine models of BHs that involve accreting material, particularly in supermassive black holes believed to reside at the centers of active galactic nuclei. The accreting material will likely possess high angular momentum, leading to an axisymmetric rather than a spherical accretion process, forming a thin, flattened accretion disk~\cite{Luminet1979}.
	
	According to Luminet, to construct a realistic image, it is necessary to consider some assumptions~\cite{Luminet1979}. First, the BH has a disk with negligible self-gravity laying in a static, spherically symmetric, uncharged, and asymptotically flat space-time geometry. In our case, the space-time geometry is that of a non-rotating EOS BH. Moreover, the disk must be geometrically thin but opaque. Secondly, we assume that, on average, the gas particles in the disk orbit the BH in nearly circular geodesic paths that are close to the equatorial plane. Finally, we neglect effects such as possible absorption by distant diffuse clouds surrounding the BH and those from the secondary heating of the disk by reabsorption of some of its light.

	In the Hamiltonian formalism, the equations of motion are \begin{equation}
\label{SVe1}
\begin{array}{ccc}
\dot{k}_\alpha=-\frac{\partial\mathcal{H}}{\partial x^\alpha}&\text{and }&\dot{x}^{\alpha}=\frac{\partial\mathcal{H}}{\partial k_\alpha},
\end{array}
\end{equation} 
where the Hamiltonian is given by Eq.\eqref{SIIe2}. Note that we use  $k^\mu$ to denote the four-momentum of the photon instead of $p^\mu$ (used for massive particles in Sec.\ref{SecIV}). The system of equations has two constant of motion $k_t=-E$ ($\dot{k}_t=0$) and $k_\varphi = L$ ($\dot{k}_\varphi=0$). Hence, the system of equaitons \eqref{SVe1} reduces to
\begin{equation}
\label{SVe2a}
\begin{aligned}
\dot{t}&=\left(1-\frac{2M e^{-\ell M/r}}{r}\right)^{-1}E,\\
\dot{r}&=\left(1-\frac{2M e^{-\ell M/r}}{r}\right)k_r,\\
\dot{\theta}&=\frac{k_\theta}{r^2},\\
\dot{\varphi}&=\frac{L}{r^2\sin^2\theta},\\
\end{aligned}
\end{equation} 
and 
\begin{equation}
\label{SVe2b}
\begin{aligned}
\dot{k}_\theta&=\frac{L^2\cos\theta}{r^2\sin^3\theta},\\\\
\dot{k}_r&=\frac{k^2_\theta}{r^3}-\frac{M e^{-\ell M/r}(r-\ell M)}{r^3}k^2_r+\frac{L^2}{r^3\sin^2\theta}\\\\
&+\frac{Me^{\ell M/r}(\ell M -r)}{r(re^{\ell M/r}-2M)^2}E^2.
\end{aligned}
\end{equation}
For $\ell=0$, the system reduces to that of Schwarzschild.  

	To solve the system of equations \eqref{SVe2a} and \eqref{SVe2b}, we need to set the initial conditions for the photon, i.e., the initial values for both the coordinates, $x^\mu_0$, and the covariant components of the four-momentum $(k_\mu)_0$. In the former case, we have that $[t_0=0,r_0,\theta_0,\phi_0]$. The initial conditions for the covariant four-momentum, $[(k_t)_0,(k_r)_0,(k_\theta)_0,(k_\varphi)_0]$, on the other hand, can be calculated by considering the initial conditions of the contravariant components of the four-momentum by $[k^t_0,k^r_0,k^\theta_0,k^\varphi_0]$ and then using the relation $g_{\mu\nu}k^\nu_0=(k_\mu)_0$.  
\begin{figure}[h!]
\centering
\includegraphics[scale=0.15]{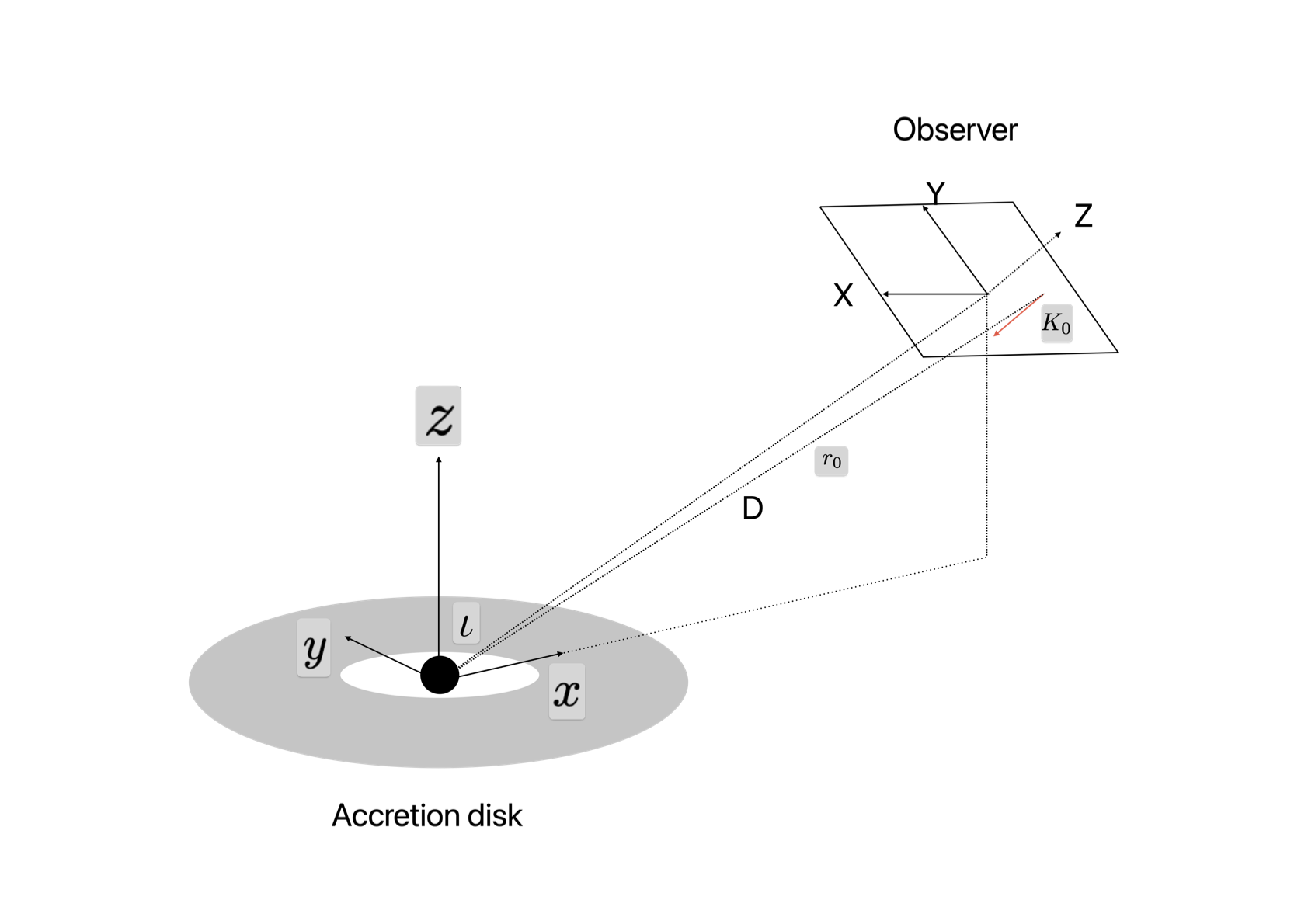}
\caption{The Cartesian coordinates $(x, y, z)$ are centered at the black hole. The Cartesian coordinates $(X, Y, Z)$ are the coordinate system of the distant observer. $D$ is the distance between the black hole and the distant observer. $\iota$ is the inclination angle between $z$-axis and the line of sight of the distant observer.}
\label{Fig13A}
\end{figure}

\begin{figure*}[t]
\centering
\includegraphics[scale=0.58]{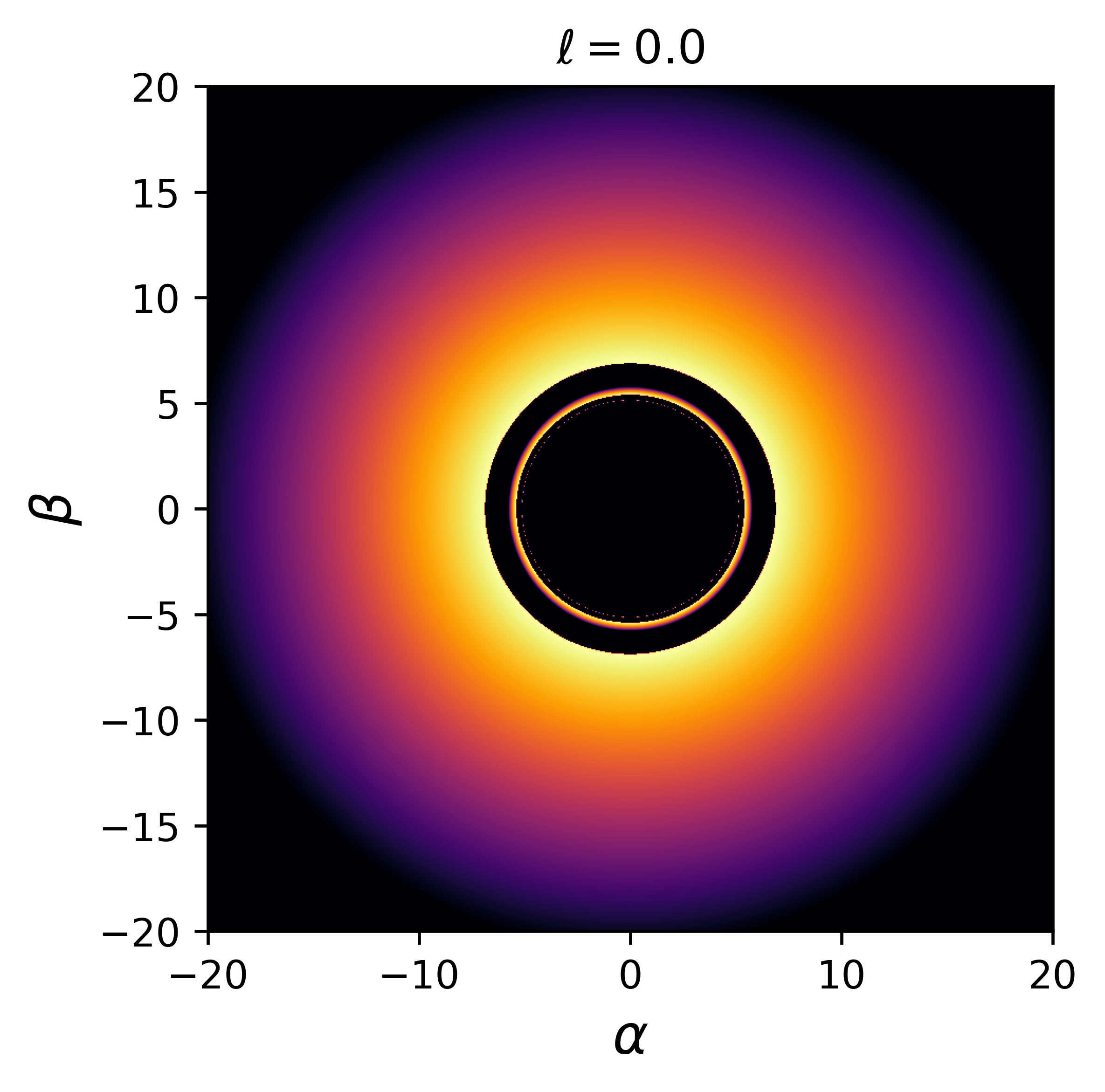}
\includegraphics[scale=0.58]{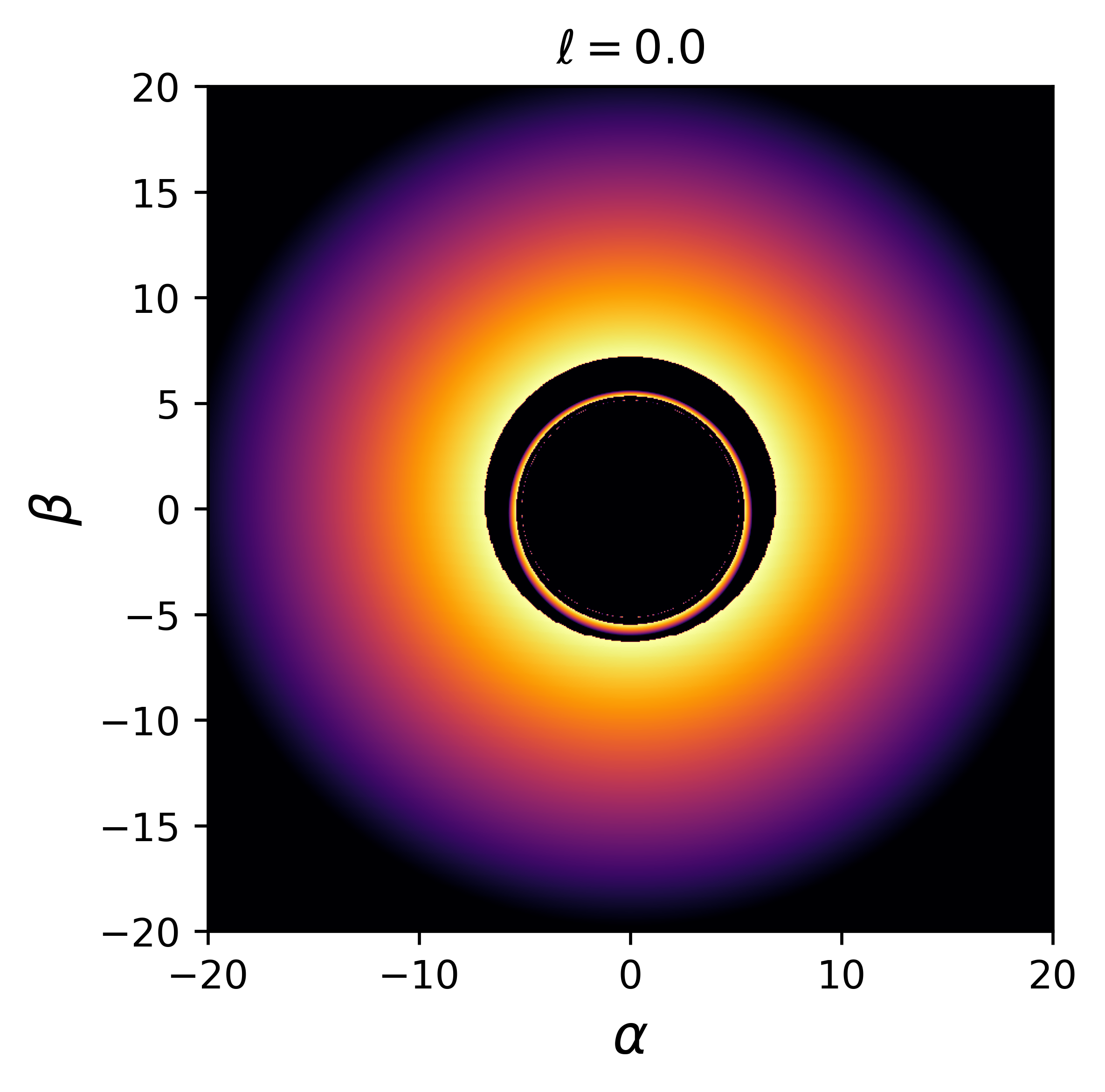}
\includegraphics[scale=0.58]{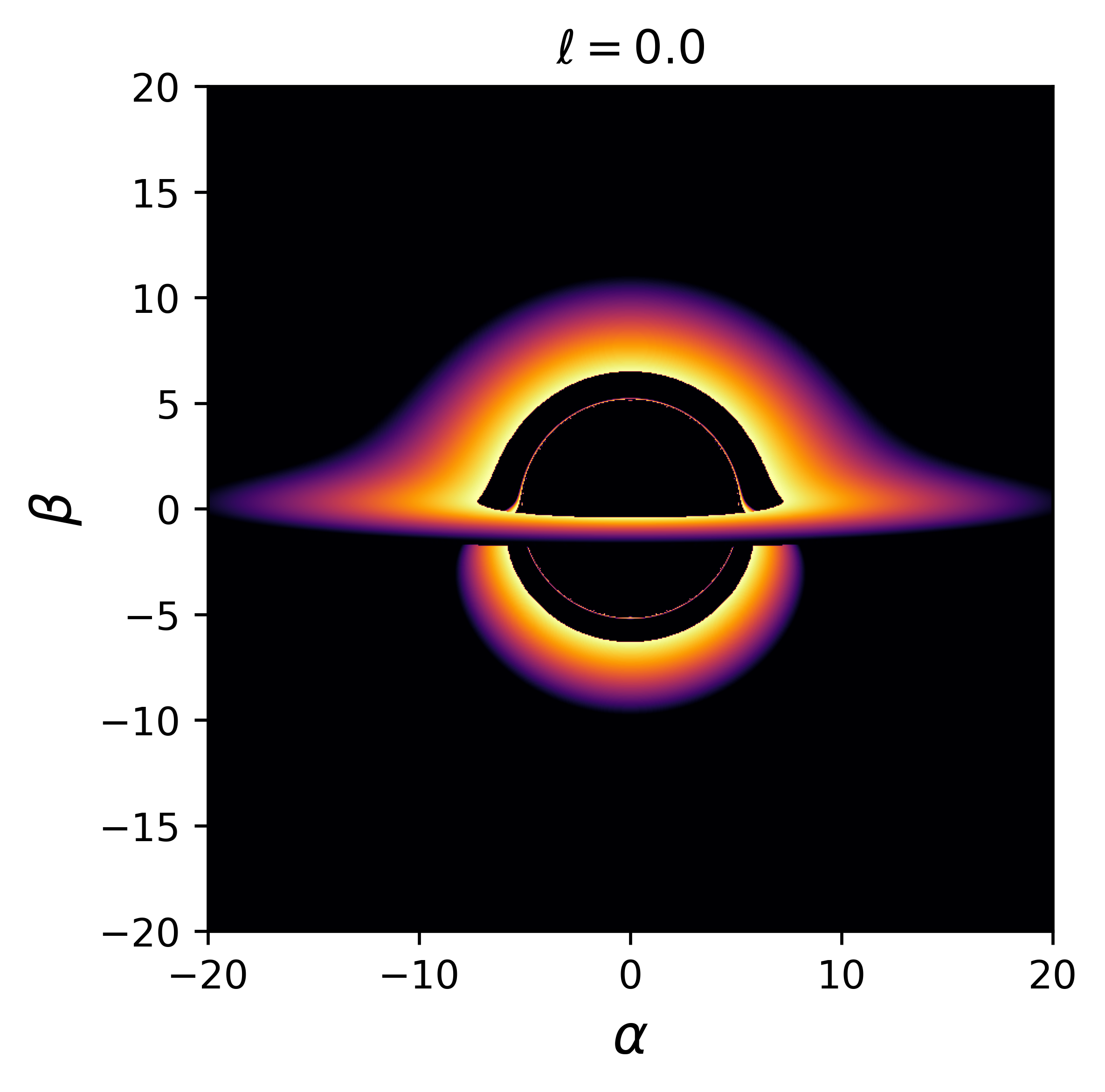}\\
\includegraphics[scale=0.58]{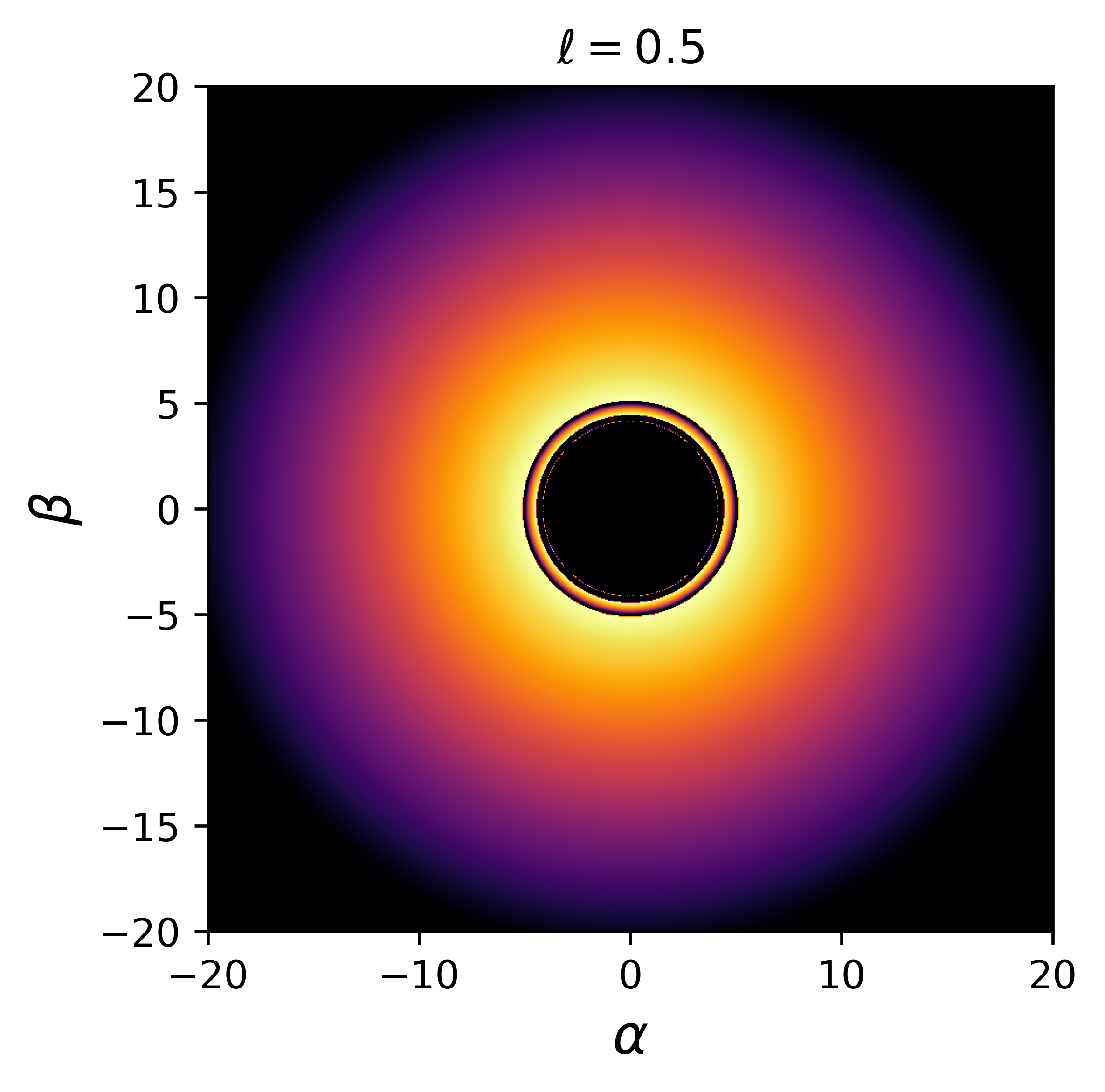}
\includegraphics[scale=0.58]{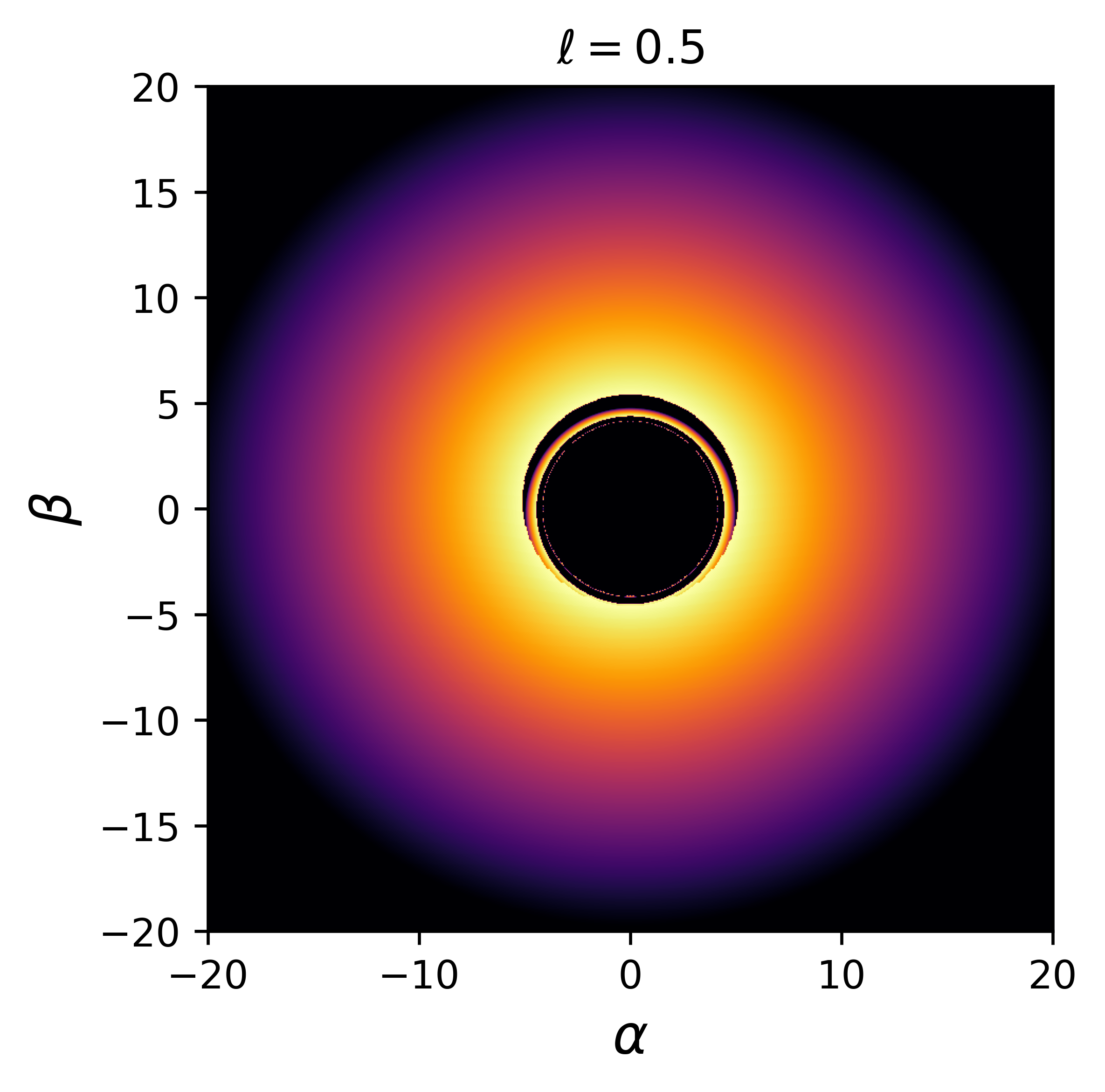}
\includegraphics[scale=0.58]{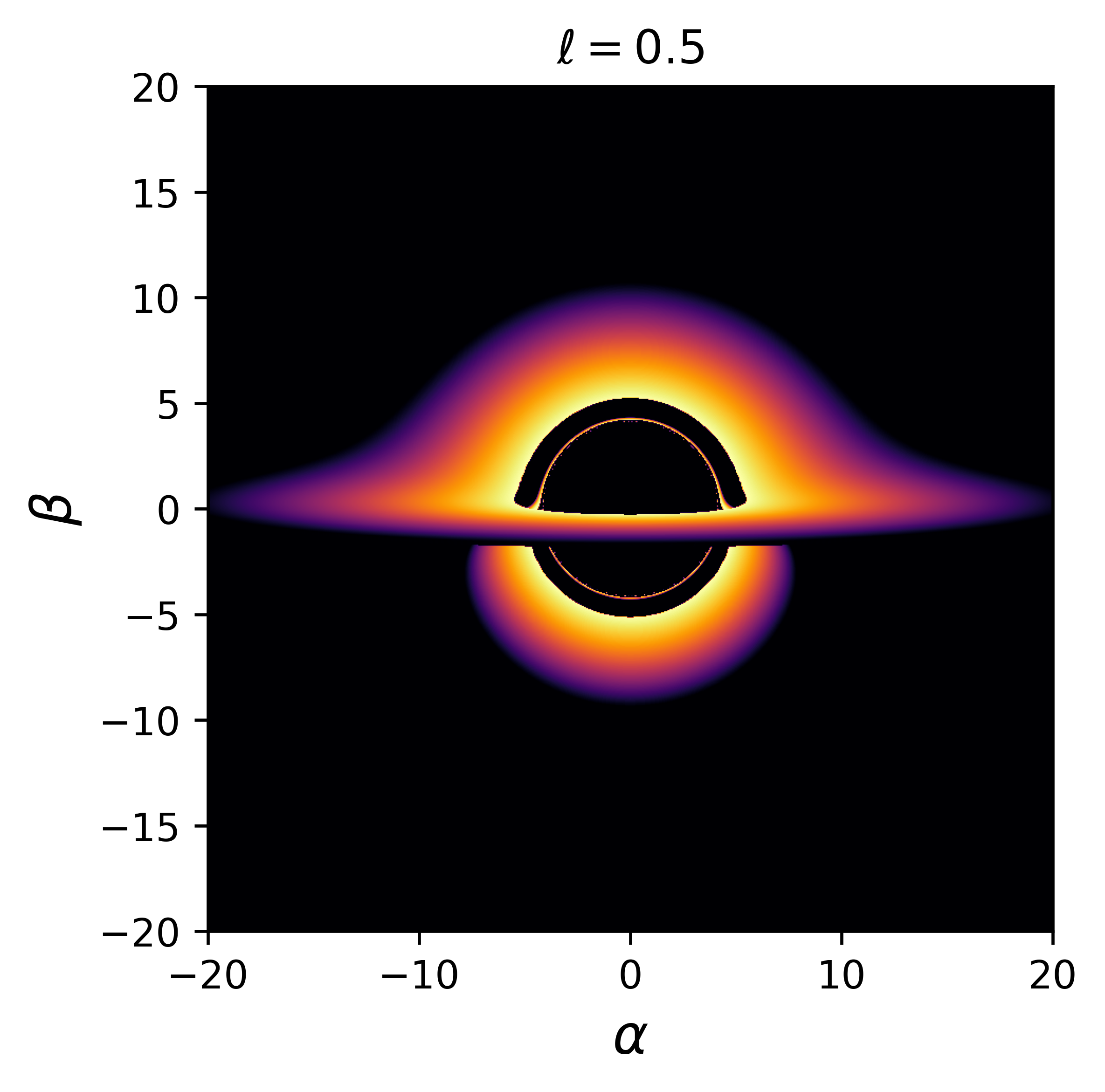}\\
\includegraphics[scale=0.58]{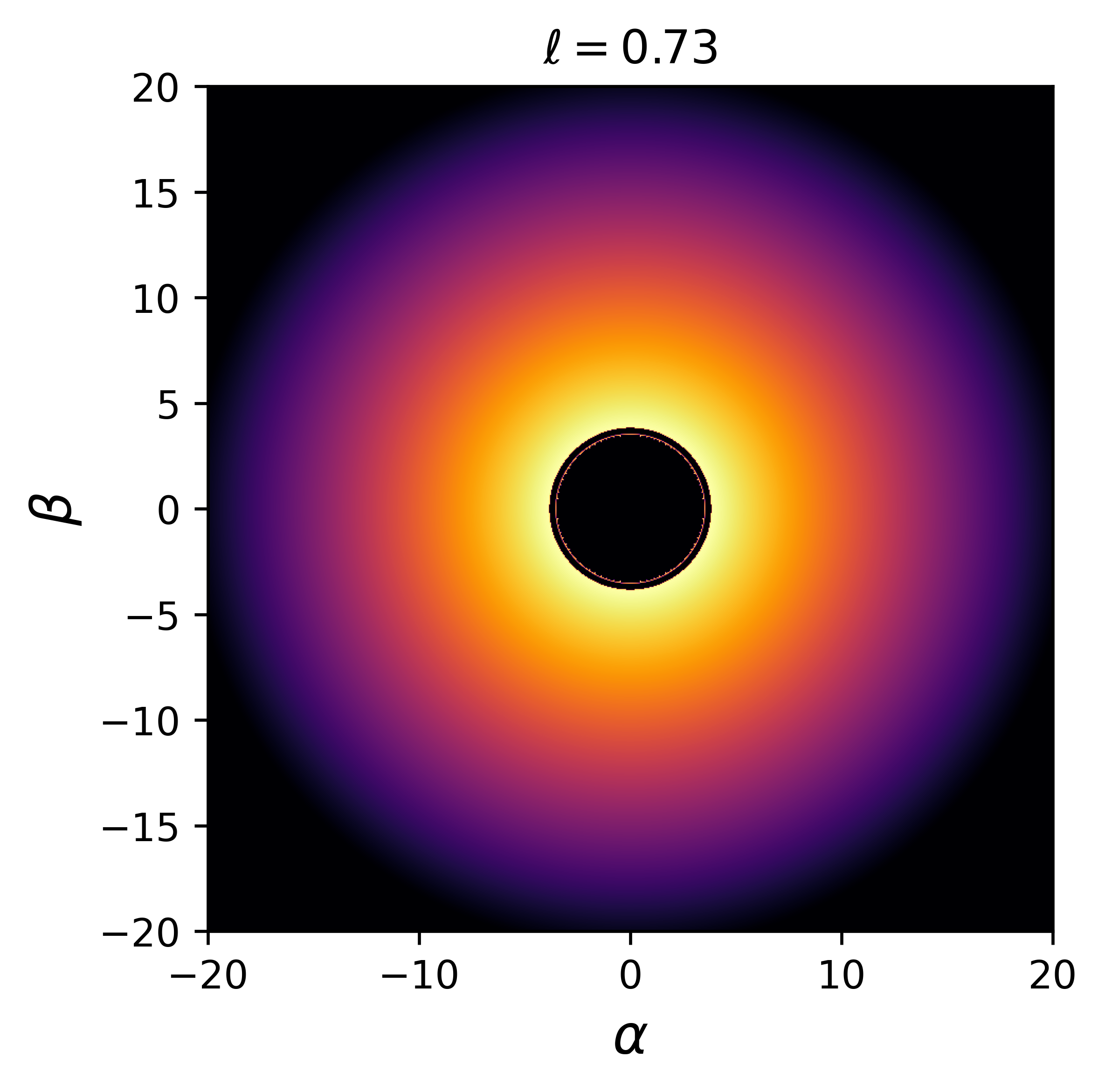}
\includegraphics[scale=0.58]{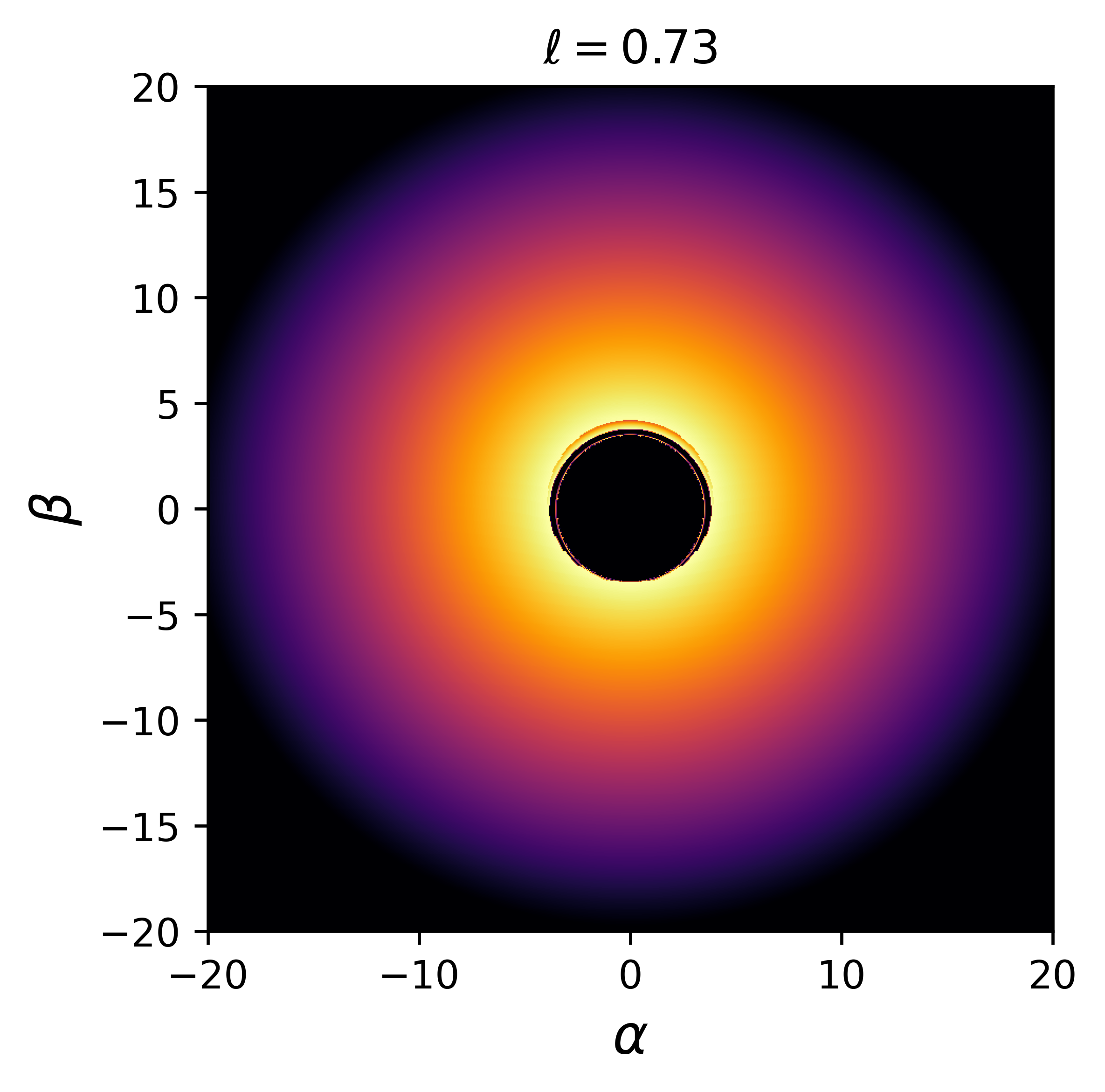}
\includegraphics[scale=0.58]{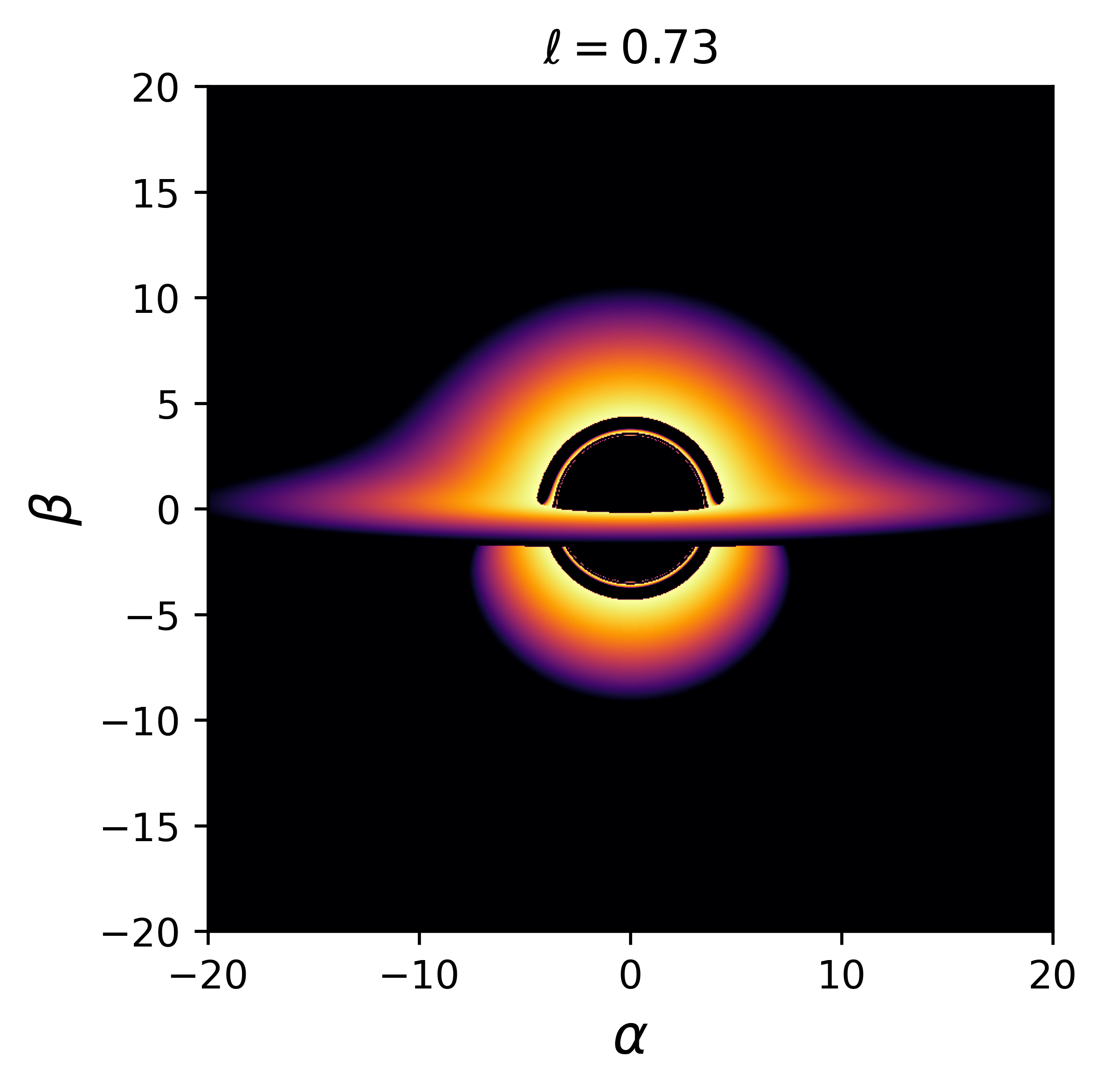}\\
\caption{Image of the EOS BH for different values of $\ell$. The first, second and third colunms show the BH image when $\iota=0^o$, $\iota=15^o$ and $\iota=85^o$. The colors show how close the photon is to the BH center. We use $M=1$.}
\label{Fig13B}
\end{figure*}
	To generate the image, we assume an observer sufficiently far from the BH at a distance $D$ and with an angle of inclination $\iota$ relative to the $z$-axis (See Fig.~\ref{Fig13A}), where one can define a new coordinate system $(X, Y, Z)$. Hence, the photons traveling from the accretion disk will generate an image on the $(X, Y)$ plane, which we redefine as the image plane $(\alpha,\beta)$ once we consider the initial conditions. The relation between the coordinate system $(X,Y,Z)$ and the spherical coordinates $(r,\theta,\varphi)$ is given by the following relations~\cite{Bambi:2024hhi}
\begin{equation}
\label{SVe3a}
\begin{aligned}
x&=(D+Z)\sin\iota-Y\cos\iota\\
y&=X\\
z&=(D+Z)\cos\iota+Y\sin\iota.
\end{aligned}
\end{equation}
The line element \eqref{SIIIe1} is in spherical-like coordinates $(t,r,\theta,\varphi)$. However, far from the BH, the spacial coordinates $(r,\theta,\varphi)$ behave as the usual spherical coordinates in flat space-time (recall the asymptotic flatness assumption mentioned above). Therefore,
\begin{equation}
\label{SVe3b}
\begin{aligned}
r&=\sqrt{x^2+y^2+z^2},\\
\theta&=\arccos\left(\frac{z}{r}\right),\\
\varphi&=\arctan\left(\frac{y}{x}\right).
\end{aligned}
\end{equation}
For the distant observer, a photon has an initial position of the form $(X_0=\alpha, Y_0= \beta, 0)$. Hence, the initial position of the photon is~\cite{Bambi:2024hhi}
\begin{equation}
\label{SVe3c}
\begin{aligned}
t_0&=0,\\
r_0&=\sqrt{\alpha^2+\beta^2+D^2},\\
\theta_0&=\arccos\left(\frac{\beta\sin\iota+D\cos\iota}{r_0}\right),\\
\varphi_0&=\arctan\left(\frac{\alpha}{D\sin\iota-\beta\cos\iota}\right).
\end{aligned}
\end{equation}
It is important to remark that we define the initial conditions on the image plane $(\alpha,\beta)$ so that the photons arriving from the accretion disk intersect the image plane perpendicularly; this means that the 3-momentum of the photon for the distance observer has the form $(0,0,-K_0)$. 
	
	Onth other hand, to obtain the initial 4-momentum of the photon $k^\mu_0$, we use the coordinate transformation $k^\mu_0=\frac{\partial x^\mu}{\partial X^\nu}K^\nu_0$, where $\{x^\mu\}=(t,x,y,z)$ and $\{X^\mu\}=(t,X,Y,Z)$. Since the cartesian coordinates of the four-momentum, relative to the distant observer, is $K^\mu_0=\left(K_0,0,0,-K_0\right)$, we obtain the following expressions for $k^\mu_0$~\cite{Bambi:2024hhi}
\begin{equation}
\label{SVe4}
\begin{aligned}
k^r_0&=-\frac{D}{r_0}K_0,\\
k^\theta_0&=\frac{\left[\cos\iota-\left(\beta\sin\iota+D\cos\iota\right)\frac{D}{r_0}\right]}{\sqrt{\alpha^2+(D\sin\iota-\beta\cos\iota)^2}}K_0,\\
k^\varphi_0&=\frac{\alpha\sin\iota}{\alpha^2+(D\sin\iota-\beta\cos\iota)^2}K_0,
\end{aligned}
\end{equation}
and, from the null condtion $g_{\mu\nu}k^\mu k^\nu=0$,  the component $k^t_0$ takes the form~\cite{Bambi:2024hhi}
\begin{equation}
\label{SVe4a}
\begin{aligned}
k^t_0=\sqrt{(k^r_0)^2+r^2_0(k^\theta_0)^2+r^2_0(k^\varphi_0)^2}.
\end{aligned}
\end{equation}

	Using the initial conditions in Eqs.~\eqref{SVe3c}, \eqref{SVe4}, and \eqref{SVe4a}, we can integrate the system of equations \eqref{SVe2a} and \eqref{SVe2b} in the spherical-like coordinates $(t, r, \theta, \varphi)$ backward in time from a point $(\alpha,\beta)$ in the image plane of the distant observer to its emission point on the accretion disk. In Fig.~\ref{Fig13B}, we show the image of the non-rotating EOS BH for different inclination angles.
\section{Discussion and conclusion\label{SecVII}}

	In this work, we use the LVK and the EHT observations to constrain the free parameter $\ell$ of the non-rotating EOS BH solution. This solution was proposed independently by Ghosh and Simpson-Visser and describes a regular BH, i.e., a singularity-free BH. According to Simpson and Visser, this BH is astrophysically viable and helpful for testing GR observationally.
	
	The first part of the manuscript was devoted to GW observations. We start by calculating the leading-order deviation to the Hamiltonian of a BS in a quasi-circular orbit within the PN approximation for the non-rotating case of the EOS BH solution. Then, we compute the leading-order deviation to the GWs emitted by such a binary system in the frequency domain, where we assume a purely Einstein radiation reaction. Finally, we compare this model to the LIGO-Virgo-KAGRA collaboration gravitational wave detection, placing constraints on the free parameter $\ell$.
	
	Before continuing with the discussion, it is crucial to clarify one point. In the method described in Sec.~\ref{SecIV}, the main idea is to map the two-body problem to an effective one-body problem. Therefore, the free parameter $\ell$ obtained by fitting the data corresponds to an effective $\ell$; i.e. $\ell_\text{eff}$. A more realistic scenario would be considering two different values, $\ell_1$ and $\ell_2$, associated with each BH of the BS, but this would require a deeper analysis of the problem, which is out of the scope of this paper since our goal is to obtain an estimate using two different observation methods. Nevertheless, since we assume small deviations ($\ell<<1$), we expect that the range of $\ell_\text{eff}$ is closer to those of $\ell_1$ and $\ell_2$.
	
	Using the individual redshifted mass of the binaries and the parameter $\delta\varphi_2$ from the posterior samples, we calculate the samples for $\ell$ taking into account only those values for which $0\leq\ell<0.73$. We show the results in Table~\ref {table1}, where we report the constraints obtained from the two waveforms models, i.e., IMPRPhenomPv2 and SEOBNRv4P, with uncertainties at the $95\%$ confidence. We point out that all the events in GWTC-3 use the SEOBRv4 model, and in the case of the GWTC-2 catalog, we only report those events for which the IMPRPhenomPv2 and SEOBNRv4 are available. Moreover, in the case of the GWTC-3 catalog, the event GW200115-042309 corresponds to a neutron star–black hole (NSBH) merger candidate~\cite{LIGOScientific:2021sio}; therefore, we do not consider this event in the analysis. 
	
	To compute the samples for the free parameter $\ell$, we use Eq.~\eqref{SIIIe29}. Nevertheless, since we have obtained this relation by considering the leading order in $\ell$, this equation is valid only if $\ell<<1$. In this sense, only the constraints satisfying this condition are consistent with our approximation. Hence, from all the samples, those events with * in Table~\ref{table1} satisfy this requirement. Our results show that the most stringent constraints on $\ell$ correspond to the events GW191204-171526 (GWTC-3) and GW190924-021846 (GWTC-2). These events are BS with a total redshifted mass of $22.74M_\odot$~\cite{LIGOScientific:2021sio} and $15.5M_\odot$~\cite{LIGOScientific:2020tif}, respectively. In the case of the SEOBNRv4P model, we found the constraints $0.041^{+0.106}_{-0.041}$~\cite{LIGOScientific:2021sio} and $0.05^{+0.165}_{-0.050}$~\cite{LIGOScientific:2020tif} for each event, respectively. Note that the signal-to-noise-ration (SNR) reported for GW191204-171526, $17.5$~\cite{LIGOScientific:2021sio}, is higher than that of GW190924-021846, $11.5$~\cite{LIGOScientific:2020tif}.  
	
	As mentioned in Ref.~\cite{Shashank:2021giy}, assuming the same values for the deformation parameters to combine the constraints will increase the SNR, giving us a stronger constraint. Nevertheless, this possibility depends on the specific theory of gravity or the BH model. The space-time considered in this work violates the no-hair theorem since it contains an additional parameter apart from the mass and spin of the BH. Therefore, as mentioned above, every source could have potentially different values of the parameters (in our case, different values for $M$, $a$, and $\ell$). Since the free parameter $\ell$ could be interpreted as associated with the BH mass distribution, one could assume a BS formed by two BH with different total masses but with the same mass distribution (same value of $\ell$). Nevertheless, we do not see any physical (and solid) argument to assume that all the events considered in this work may have the same mass distribution. For this reason, we do not combine the constraints in our analysis.  

	In the second part of this work, we constrain the free parameter $\ell$ using the EHT observations. Since we consider $a=0$ in the GW part, we constrain our analysis to the non-rotating case. First, we investigated the effective potential for photons to obtain the radius of the photon sphere for different values of $\ell$. We found that $r_\text{ph}$, decreases as the $\ell$ increases. We found a similar behavior in the case of the ISCO. In this sense, the black hole shrinks for $0<\ell<\ell_\text{extremal}$. This behavior has important repercussions on the photon trajectories, particularly when considering those trajectories related to direct, lensed ring and photon ring emissions. Therefore, as the BH shrinks,  our results show that the intervals for lensed and ring emissions increase while the interval for direct emissions decreases. Hence, in the case of ring emissions, for example, more photons participate in this kind of emission (This can be seen clearly in the second row of Fig.~\ref{Fig6}). 
	
	We also investigated the radius of the shadow $b_\text{ph}$, which is crucial for constraining the free parameter $\ell$ via the angular diameter and EHT observations. Our analysis shows that it decreases its value as $\ell$ increases. This behavior extends when considering the angular diameter of the shadow. To constrain the values of $\ell$, we take into account two BHs: Sgr A*, located at a distance $D=8.127 \text{ kpc}$ with a mass of $M = 4.14 \times 10^6 M_\odot$; and M87*, located at a distance $D=16.8 \text{ Mpc}$ with a mass of $M = 6.2 \times 10^9 M_\odot$. According to the EHT collaboration~\cite{EHT1, EHT2}, the angular diameter of the shadows reported for these two BHs is $51.8 \pm 2.3~\text{$\mu$as}$ and $42 \pm 3~\text{$\mu$as}$, respectively. In the case of Sgr A*,  the predicted constraint is $0<\ell<0.148$, while for M87*, we found that $0<\ell<0.212$. Both cases with a $68\%$ of confidence~\cite{EHT1, EHT2}. 
	
	It is worth mentioning that these results agree with those obtained when considering the GW observations. In particular for those events in which $\ell<<1$. For example, in the case of GWTC-1, events such as GW151226, GW170104, and GW170608; the events GW190630-185205, GW190707-093326, GW190708-232457, GW190720-000836, GW190728-064510, GW190828-065509, and GW190924-021846, in the case of GWTC-2; and GW191129-134029, GW191204-171526, GW191216-213338, GW200129-065458, GW200202-154313, GW200225-060421, and  GW200316-215756 for GWTC-3 agree well with the predicted constraints for both Sgr A* and M87* BHs.    
	
	To investigate the optical signatures of the non-rotating EOS BH, we consider the photon ring structure using three different scenarios. First, we focus on static spherical accretion. Secondly, the infalling spherical accretion. Finally, the thin disk accretion, where we investigated the photon trajectories in detail. In the first two scenarios, the observed intensity $I_o$ increases when the free parameter $\ell$ goes from 0 (Schwarzschild case) to 0.73 (near the extremal case $\ell_\text{extremal}$). The behavior of $I_o$ as a function of the impact parameter is similar between the two scenarios; i.e., for values of $|b|<b_c$, the observed intensity increases and diverges when $b=b_c$ (at the photon sphere), then it decreases its value for $|b|>b_c$. However, when comparing the values of $I_o$ (for all the values of $\ell$), we found that it is smaller in the case of infalling spherical accretion. Moreover, in the central region ($|b|<b_c$) $I_o$ is deeper when compared with the outer side of the shadow. 
	
	In the third scenario: thin disk accretion, we analyzed the photon trajectories by classifying them using three categories, depending on the number of intersections with the accretion disk. Hence, we define the direct emissions to the photons that intersect once ($n<0.75$), the lensed emissions for those photons intersecting twice ($0.75<n<1.25$), and finally, the photon ring emissions for more than three intersections of the accretion disk ($n>1.25$). We found that more photos will follow photon ring trajectories as the free parameter $\ell$ increases from $0$ to $0.73$. We found a similar behavior in the case of lensed emissions. This means that the impact parameter interval becomes broader for both lensed and photon ring emissions as the free parameter approaches the naked regular compact object limit, $\ell_\text{extremal}$, which is a direct consequence of the ``shrinking'' property of the EOS BHs.
	
	The fact that photon trajectories intersect the thin accretion disk once, twice, or more has a direct impact on $I_o$ since the light from the emission becomes brighter each time the photon intersects the accretion disk. In this sense, we investigated the ring structure using different models for the total emission intensity, $I_\text{emit}$. In the case of the first model, we found three peaks associated with photon ring, lensed, and direct emissions. The observed intensity of the photon ring emission peak is smaller than the other two. Therefore, it becomes difficult to see when plotting the intensity distribution.
	
	In the second model, we identified two peaks, with the observed intensity of the inner one smaller than that of the outer one. The latter is related to the photon ring emissions. We also found two peaks in the third model; nevertheless, these are closer to each other when contrasted with those of the second model. The models' profile does not change notably when the free parameter $\ell$ changes. However, when we consider the case $\ell=0.73$, we found some changes in the observed intensity. For example, in the first model, the middle peak merges with the outer peak (associated with direct emissions). Eventually, this peak disappears for $\ell>\ell_\text{extremal}$. We also observed an additional peak in the intensity profile for the second model. We believe this phenomenon is associated with the ``shrinking" behavior of the EOS space-time as it transitions towards the naked compact object regime (the ring structure already discussed in Ref.~\cite{Guerrero:2022msp}). We found no notable differences in the third model.
	
	Finally, we plan to extend our work by considering X-ray observations; this will allow us to constrain the free parameter $\ell$ by considering the spin of the EOS BH since, in the present work, we constrain our analysis to the non-rotating case. This work is already in progress.

\section*{Acknowledgements} \label{sec:acknowledgements}
LIGO Laboratory and Advanced LIGO are funded by the United States National Science Foundation (NSF) as well as the Science and Technology Facilities Council (STFC) of the United Kingdom, the Max-Planck-Society (MPS), and the State of Niedersachsen/Germany for support of the construction of Advanced LIGO and construction and operation of the GEO600 detector. Additional support for Advanced LIGO was provided by the Australian Research Council. Virgo is funded, through the European Gravitational Observatory (EGO), by the French Centre National de Recherche Scientifique (CNRS), the Italian Istituto Nazionale di Fisica Nucleare (INFN) and the Dutch Nikhef, with contributions by institutions from Belgium, Germany, Greece, Hungary, Ireland, Japan, Monaco, Poland, Portugal, Spain. The construction and operation of KAGRA are funded by Ministry of Education, Culture, Sports, Science and Technology (MEXT), and Japan Society for the Promotion of Science (JSPS), National Research Foundation (NRF) and Ministry of Science and ICT (MSIT) in Korea, Academia Sinica (AS) and the Ministry of Science and Technology (MoST) in Taiwan. Unless otherwise specified, the contents of this release are licensed under the Creative Commons Attribution 4.0 International License. To view a copy of this license, visit http://creativecommons.org/licenses/by/4.0/ or send a letter to Creative Commons, PO Box 1866, Mountain View, CA 94042, USA. 

This work is supported by the Chinese Ministry of Science and Thecnology of China (grant No.~2020SKA0110201) and the National Science Foundation of China (grant No.~11835009). We thank Alejandro Cárdenas-Avendaño for useful discussions. C.A.B.G. thanks Prof. Larranaga for his seminar on computational astrophysics, which was of great help while writing the Python code used for the BH images in Sec.~\ref{SecVI} of this manuscript. 



	\end{document}